\documentclass[twocolumn]{aastex61}
\usepackage{natbib}
\usepackage{multirow}
\usepackage{color}
\usepackage{bm}
\usepackage{soul}

\usepackage[normalem]{ulem}
\usepackage{amsmath}
\usepackage{amssymb}

\newcommand\aastex{AAS\TeX}

\received{\today}
\revised{}
\accepted{}
\submitjournal{\apjs}

\shorttitle{\aastex\ Atomic data of Pr II - Gd~II ions}
\shortauthors{Rad\v{z}i\={u}t\.{e} et al.}

\begin{document}

\title{Extended calculations of energy levels and transition rates for singly ionized lanthanide elements I: Pr - Gd}

\correspondingauthor{Laima Rad\v{z}i\={u}t\.{e}}
\email{Laima.Radziute@tfai.vu.lt}

\author{Laima Rad\v{z}i\={u}t\.{e}}
\affil{Institute of Theoretical Physics and Astronomy, 
Vilnius University, Saul\.{e}tekio Ave. 3, Lithuania}

\author{Gediminas Gaigalas}
\affil{Institute of Theoretical Physics and Astronomy, 
Vilnius University, Saul\.{e}tekio Ave. 3, Lithuania}

\author{Daiji Kato}
\affiliation{National Institute for Fusion Science, 
322-6 Oroshi-cho, Toki 509-5292, Japan}

\author{Pavel Rynkun}
\affil{Institute of Theoretical Physics and Astronomy, 
Vilnius University, Saul\.{e}tekio Ave. 3, Lithuania}

\author{Masaomi Tanaka}
\affiliation{Astronomical Institute, Tohoku University, Sendai 980-8578, Japan}



\begin{abstract}
Lanthanide elements play important roles as an opacity source in the ejected material from neutron star mergers.
Accurate and complete atomic data are necessary to evaluate the opacities and to analyze the observed data.
In this paper, we perform extended, {\it ab-initio} atomic calculations from Pr II (Z=59) to Gd~II (Z=64).
By using multiconfiguration Dirac-Hartree-Fock and
relativistic configuration-interaction methods, 
implemented in the general-purpose relativistic atomic structure package GRASP2K, 
we calculate the energy levels and transition data of electric dipole transitions. 
These computations are based on strategies (with small variations) of Nd II published by \citet{Nd_jonai}.
Accuracy of data is evaluated by comparing computed energy levels with the NIST database or other works.
For the energy levels, we obtain the average relative accuracy of 8\%, 12\%, 6\%, 8\%, and 7\% for Pr~II, Pm~II, Sm~II, Eu~II, and Gd~II ions, respectively as compared with the NIST data. Accuracy of energy transfer to the wavelength as 3\%, 14\% and 11\% for Pr~II, Eu~II and Gd~II.
 Our computed E1 type transition probabilities 
are in good agreement with experimental values presented by other authors
especially for strong transitions.
\end{abstract}



\section{Introduction} 
\label{sec:intro}

Atomic opacities of heavy elements have a wide impact to astrophysics.
In particular, recent observations of gravitational waves
and electromagnetic waves from a neutron star merger 
(GW170817, \citealt{abbott17MMA}) highlight the needs for heavy-element opacities.
In optical and infrared wavelengths,
the electromagnetic counterpart of GW170817 shows characteristics of kilonova,
emission powered by radioactive decays of newly synthesized $r$-process
(or rapid neutron capture process) nuclei.
To study the $r$-process nucleosynthesis from the observed emission,
we need to accurately understand the opacities of lanthanide elements
since properties of kilonova are mainly governed by bound-bound
opacities of $r$-process elements
and lanthanide elements give the largest contributions
(\citealt{kasen13}; \citealt{barnes13}; \citealt{tanaka13}).

Several works have been done to study the properties and opacities
of lanthanide elements
(\citealt{kasen13,fontes17,tanaka18,tanaka19}).
However, atomic calculations to evaluate the total opacities
are not necessarily accurate enough to give a wavelength and
a transition probability of each transition \citep{tanaka19}.
Recently, \citet{watson19} reported identification of Sr in the spectra
of kilonova associated with GW170817.
In principle, other elements can also be identified in the spectra.
However, the line list used for astrophysics
is not neccesarily complete even for strong transitions, in particular, in infrared wavelengths.
By these reasons, it is still not straightforward to fully decode the
spectra of kilonova.
Accurate atomic calculations of lanthanide elements, therefore,
play an important role as a benchmark to give accurate atomic data
\citep{Nd_jonai}.

There are many semi-empirical works which provide accurate atomic data of the lanthanide elements. 
In these works, the Racah-Slater parametric method  is used \citep{Racha-Slater}.
This method is known to give an excellent agreement between calculated energies using fitted radial parameters and available experimental energies. 
However, correct level identification of experimental spectra is needed, 
which is not always available. On the other hand, {\it ab-initio} methods 
can provide complete atomic data set without any empirical parameter. 
Nevertheless, there are few applications of such {\it ab-initio} methods 
for lanthanide with spectroscopic accuracy. 
This is because systematic improvement of subtle correlation effects 
in complicated atomic structures of open-$4f$ shell is not studied thoroughly.

In our previous paper \citep{Nd_jonai}, we have performed accurate calculations for Nd ions. In this paper, 
we extend our calculations to Pr~II, Pm~II, Sm~II, Eu~II, and Gd~II.
Namely, we perform energy spectrum computations for states of the 
following configurations:  
[Xe]$4f^N \{6s,5d,6p\}$ and [Xe]$4f^{N-1} \{5d6s,5d6p,6s6p, 5d^2\}$
for $N=3,5,6,7,8$. 
We also perform energy spectrum computations for states of [Xe]$4f^{N+1}$ configuration for Sm~II and Eu~II,
and [Xe]$4f^{N-1} 6s^2$ configuration for Gd~II. 
 Levels up to 10 eV are computed since such low-lying energy levels
play dominant roles in the opacities in the neutron star merger ejecta at 
typical temperature of 5,000 K \citep{Nd_jonai}. 
Using these results, electric dipole (E1) transitions data were computed between these states. 
 In this paper, we aim at providing complete atomic data
with the overall accuracy of about 10\%.
This accuracy is not high enough to directly compare with spectroscopic experiments,
but it is adequate to evaluate the opacities ("opacity accuracy" rather than "spectroscopic accuracy", \citealt{Nd_jonai}).
In fact, typical accuracy of complete atomic calculations \citep{kasen13,tanaka18} is much lower than the accuracy presented in this paper.

The calculations are done using multiconfiguration 
Dirac-Hartree-Fock (MCDHF) and relativistic configuration-interaction
(RCI) methods \citep{grant,topical_rev}, which are implemented 
in the general-purpose relativistic atomic structure
package GRASP2K \citep{graspV3}.  
We employ a strategy similar to the \citet{Nd_jonai} including electron correlation, 
which is suitable for series of rare earth ions.
 For low lying levels, higher accuracy can be achieved
 using computational schemes including more electron correlations as in \citet{ErIII}.
In addtion, there is an advantage in the computation since large computational tasks can be split in to smaller tasks by using this method.

In Section \ref{sec:method}, we describe our method and strategy of calculations. 
Then, we show results of energy level structure and transition probabilities in Sections \ref{sec:evaluation} 
and \ref{sec:transitions}, respectively. Finally we give summary in Section \ref{sec:summary}.

\section{Methods} 
\label{sec:method}

\subsection{Computational procedure}
The computational methods used in this paper follow the methods 
used in \citet{Nd_jonai}. 
Therefore, we briefly outline the methods in this section.
We refer the reader to \citep{topical_rev} for further details.
We use the MCDHF method, based on Dirac-Coulomb Hamiltonian, in this work. 
The atomic state functions (ASFs) are expressed by a linear combination of symmetry adapted configuration state functions (CSFs).
The CSFs are built from products of one-electron Dirac orbitals. 
The radial parts of the Dirac orbitals and the 
expansion coefficients are optimized to self-consistency 
in the relativistic self-consistent field procedure.

The spin-angular approach \citep{Gaigalas_1996,Gaigalas_1997} is used in these computations.
The approach is based on the second quantization 
in a coupled tensorial form, on the angular momentum theory in
the orbital, spin, and quasispin spaces 
and on the reduced coefficients of fractional parentage. 
It allows us to study configurations with open $f$-shells without any restrictions.

\begin{deluxetable}{c c c c c c c c c }[ht!!]
\tablecaption{\label{summary} Summary of Computed Levels and Active Space Size.}
\tablehead{
Ion   &\multicolumn{2}{c}{Number of levels} &&  \multicolumn{2}{c}{N$_{CSFs}$} \\ 
\cline{2-3} \cline{5-6} 
      & Even & Odd && Even & Odd}
\startdata
Pr~II &   927 & 1~218 && 29~129    &    45~045\\
Nd II*& 3~270 & 2~813 && 188~357   &   113~900\\
Pm~II & 5~206 & 4~568 && 380~588   &   518~957\\	
Sm~II & 3~153 & 5~240 && 1~272~634 & 2~133~183\\
Eu~II & 1~306 & 1~241 && 1~501~949 & 2~201~859\\
Gd~II & 2~035 & 2~335‬ && 3~033~793 & 1~721~371\\
\hline
\enddata 
\tablecomments{* Nd II data are published in \citep{Nd_jonai}.}
\end{deluxetable}

In the following RCI calculations, the Breit interaction
is included in the Hamiltonian. 
In the RCI calculation, 
the leading quantum electrodynamics corrections (QED),  
self-interaction and vacuum polarization are also included.

The label of the ASF is the same as the label of the dominating CSF.
The ASFs are obtained as expansions over $jj$-coupled CSFs. 
 To provide the ASFs in the $LSJ$ labeling system, transformation from a $jj$-coupled CSF basis to an $LSJ$-coupled CSF basis has been done \citep{jj2lsj_atoms}.  
Review on all these methods and on the GRASP2K package can be found in \citet{topical_rev}.

\subsection{Computation of transition probabilities}

For electric dipole transitions there are two forms of the transition operator:
 the length (Babushkin) and velocity (Coulomb) forms.  
Although the exact solutions of the Dirac-equation should give the same value of the 
transition moment \citep{gauge}, they do not necessarily agree in numerical calculations. 
The quantity $dT = | A_{\rm l} - A_{\rm v} | / {\rm max} (A_{\rm l}, A_{\rm v})$ \citep{ekman} 
defines the accuracy of the computed transition rates, where $A_{\rm l}$ and $A_{\rm v}$ are 
the transition rates in length and velocity forms, respectively.

 The calculation of the transition moment breaks down in the task of 
summing up reduced matrix elements between different CSFs. 
Using standard techniques, by assuming
that both left and right hand CSFs are formed from the same orthonormal set of spin-orbitals,
the reduced matrix elements can be evaluated. 
This constraint is severe, since a high-quality and compact
wave function requires orbitals optimized
for a specific electronic state (see for example \citealt{SF}). 
To avoid the problems of having a single orthonormal set of spin-orbitals, 
the wave-function representations of the two states
are transformed in a way that the orbital sets became biorthonormal \citep{biotra}.
To evaluate the matrix elements of the transformed CSFs,
standard methods as in \citet{topical_rev} are used. 

\subsection{Computational Schemes} 
\label{sec:scheme}
To compute singly ionized lanthanide elements,
the \textbf{strategy C} by \cite{Nd_jonai} is used.
Details of this strategy and extension of it are given below.   
Active space method is used for computation of energy levels and E1 transitions.
The configuration space is increased step by step,
by increasing the number of layers (L), that is, a set of virtual orbitals. 
The virtual orbitals of the increased layer are optimized in the relativistic self-consistent field procedure, while all orbitals of inner layers are fixed.
The scheme used to increase the active spaces of the CSF’s is presented below: \\
AS$_{0L}$ = $\{6s, 6p, 5d\}$, \\
AS$_{1L}$ = AS$_{0L}$ + $\{7s, 7p, 6d, 5f\}$, \\
AS$_{2L}$ = AS$_{1L}$ + $\{8s, 8p, 7d, 6f, 5g\}$.\\
The number of computed levels and CSFs in the final even and odd state expansions 
are presented in Table \ref{summary}.

Computations are performed for each configuration separately (single reference method). 
This method allows to split the large computations into several tasks. In each task, the wave function expansion for a single reference configuration is constructed by substitution of one and two electrons from the reference configuration. 
For configurations $4f^N 6s$, $4f^N 6p$ and $4f^N 5d$,
single and/or double (SD) substitutions are allowed from $4f^N nl$ shells ($l=s,p,d$) 
to $AS_{0L,1L}$ and single (S) substitutions are allowed to $AS_{2L}$.
For configurations $4f^{N-1} 5d6s$, $4f^{N-1} 5d6p$, $4f^{N-1} 6s6p$, and $4f^{N-1} 5d^2$, only S substitutions are allowed. 
For Sm~II and Eu~II ions, a new configuration $4f^{N+1}$, 
which was not taken into account in \textbf{the strategy C} of \cite{Nd_jonai},
is computed.  
For this configuration, 
single, double, and triple (SDT) substitutions 
are allowed from $4f^{N+1}$ shell to $AS_{0L,1L}$ and SD substitutions are allowed to $AS_{2L}$. 
For configuration $4f^{N-1} nl n'l'$, two electrons are excited from $4f$ orbital, and for $4f^N nl$, only one electron is excited from $4f$ orbital. Therefore, to include compensated correlations, we need to make less excitations from the first configuration and more excitations from the second one. For example, if we do SD substitutions for $4f^{N-1} nl n'l'$ configuration, we need to make SDT substitutions for $4f^N nl$ configuration.


To compute energy levels, it is important to have correct core radial wave functions, that is, initial Dirac-Fock (DF) computations.
Correct selection of the core stabilizes solution of self-consistent field computation. 
We find that core radial wave functions [Xe]$4f$ from
the ground configuration [Xe]$4f^N6s$ are the best solution. 
Radial wave functions up to $4f, 5s, 5p$ 
orbital are taken from the ground configuration for these configurations $4f^{N-1} 5d6s$, $4f^{N-1} 5d6p$, $4f^{N-1} 6s6p$, and $4f^{N-1} 5d^2$. 
Meanwhile, the radial wave functions were computed for each configurations $4f^N nl$ ($l=s,p,d$) separately.  

For neutral atoms and ions of lanthanide elements with different ground configurations, 
we suggest that their ground configuration radial wave functions are used as common core.  
For example, for neutral lanthanides, radial wave functions of the ground configurations [Xe]$4f^N6s^2$ can be used as common core.

\begin{figure}
\includegraphics[width=0.47\textwidth]{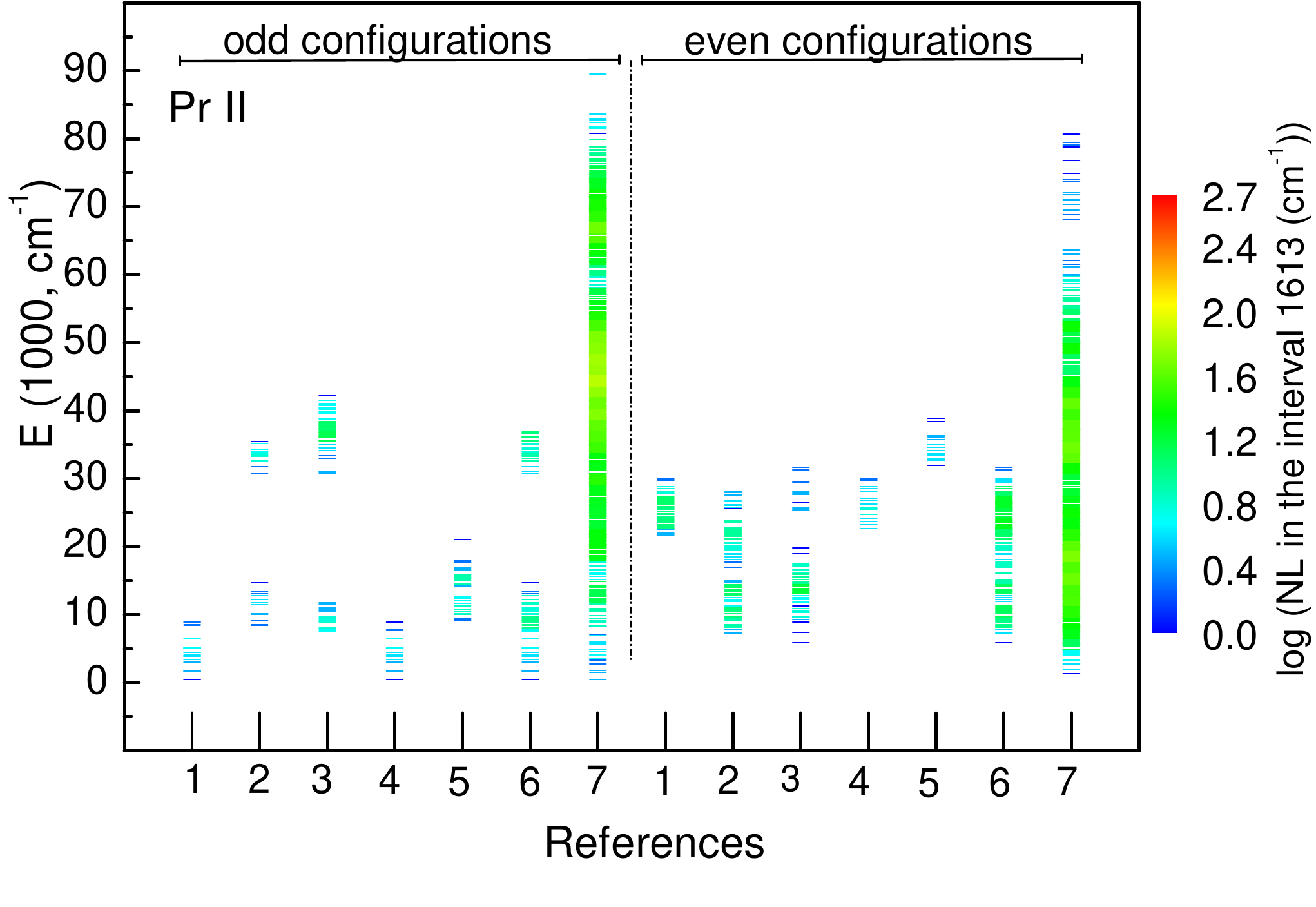}
	\caption{Comparison of energy levels for odd and even configurations of Pr~II with experimental values. 
	References 
	(1) \citet{Rosen_Pr}; 
	(2) \citet{Blaise_Pr} and \citet{Blaise_Pr_b}; 
	(3) \citet{Ginibre_Pr_1989_a};
	(4) \citet{Ivarsoon_Pr};
	(5) \citet{Furmann_Pr_a}, \cite{Furmann_Pr_b} and \cite{Furmann_Pr_c};
	(6) \citet{Akhtar_Pr};
	(7) Our computed levels. NL is the number of levels.
	\label{Pr_II_su_kitais_autoriai}}
\end{figure}

For Eu~II and Gd~II, wave function is investigated differently 
due to the rapid increase of the number 
of configuration state functions in the active space (see Table \ref{summary}).
For these ions, self-consistent field computations are 
performed not for all $J$ values but only for one $J$ value.
Then, using computed radial wave functions, RCI computations are performed. 
For example, for the configuration of Eu,
$4f^5 6s$ atomic states only with $J=4$ are computed and 
it is later used in the RCI computation for $J =0-13$. 
For all configurations, the lowest $J$ values are selected 
for computation of the radial wave functions. 
 This computational method demands less computational resources.

\begin{figure*}
\includegraphics[width=0.50\textwidth]{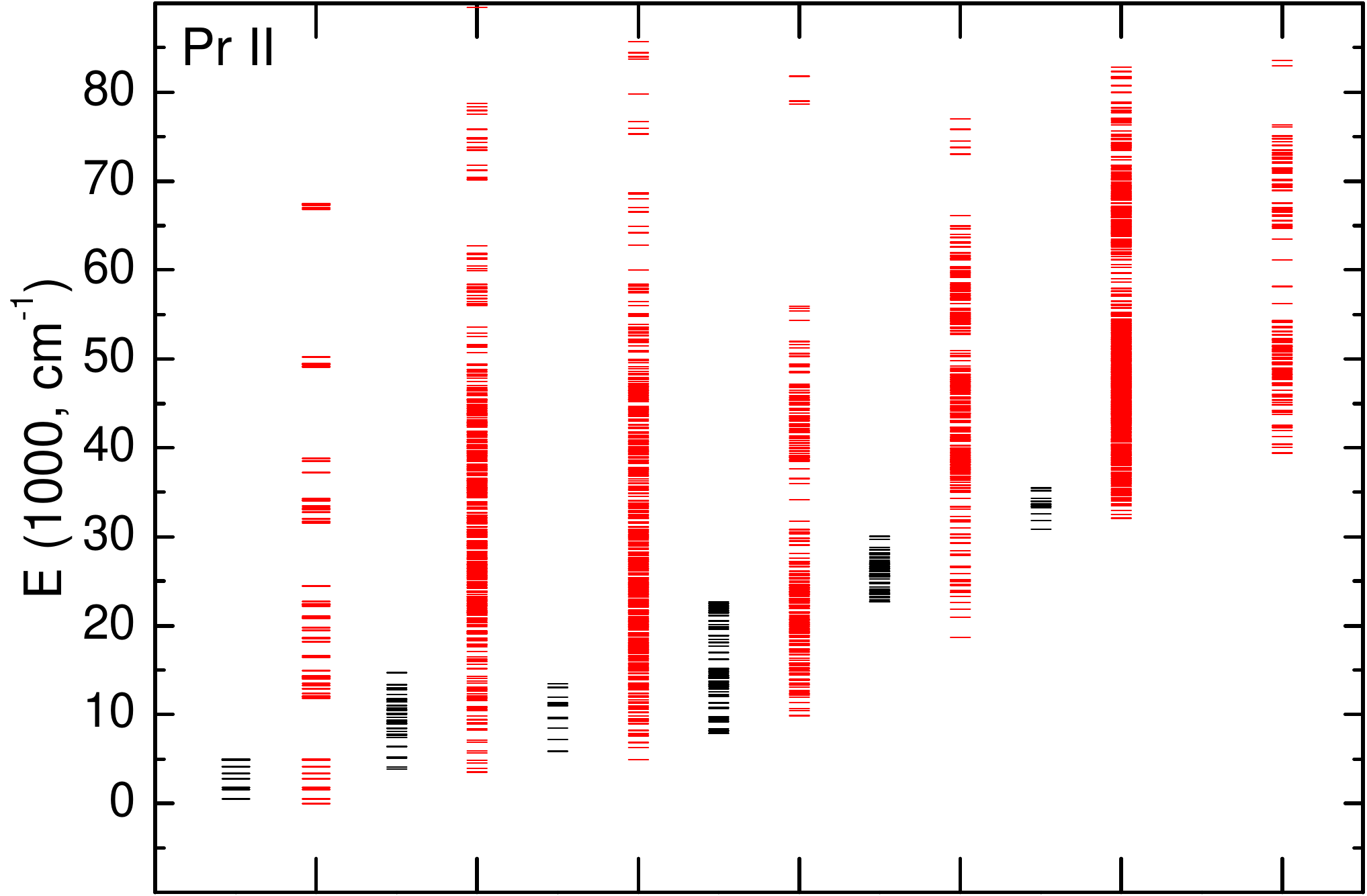}
\includegraphics[width=0.50\textwidth]{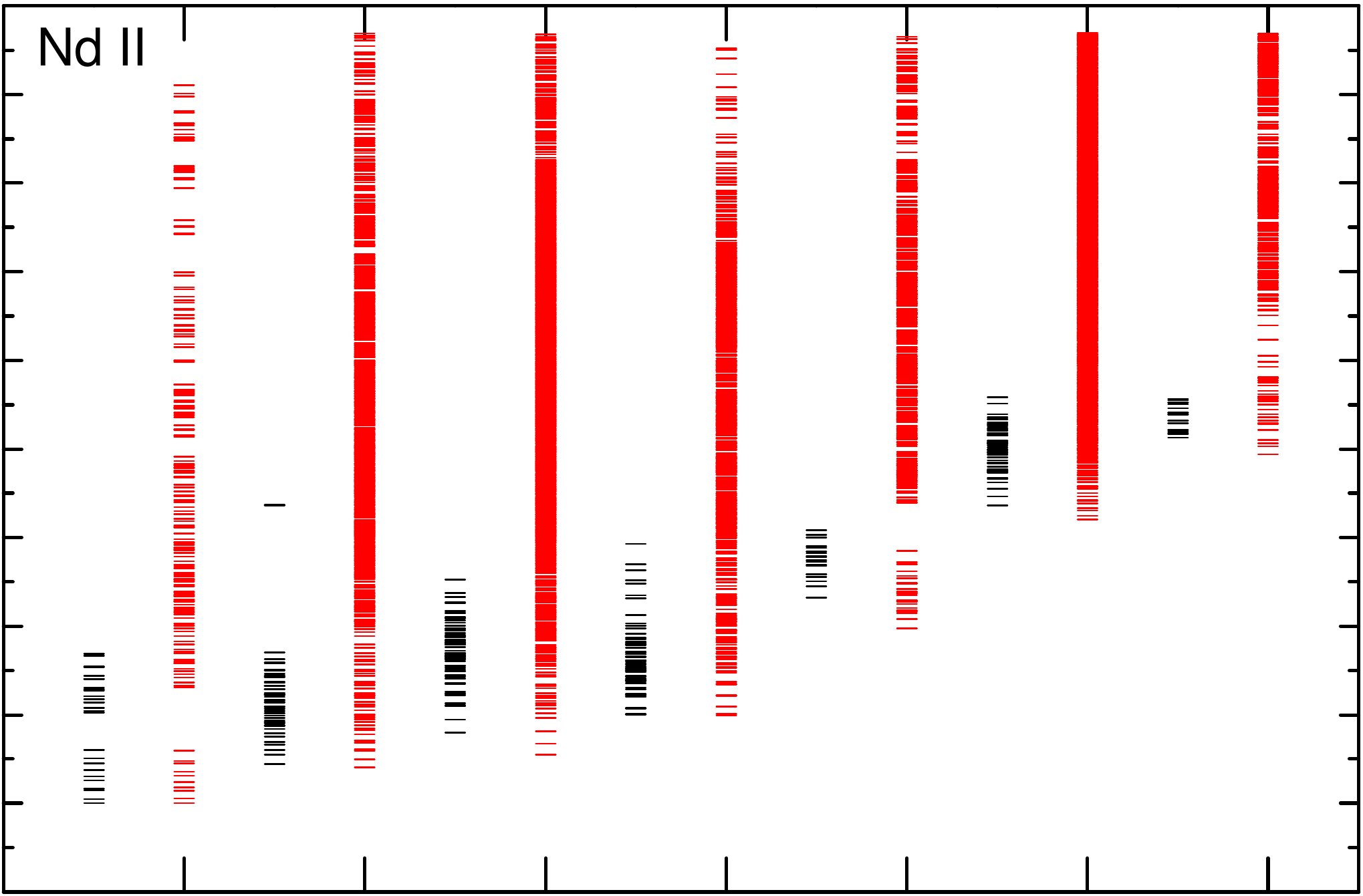}
\includegraphics[width=0.50\textwidth]{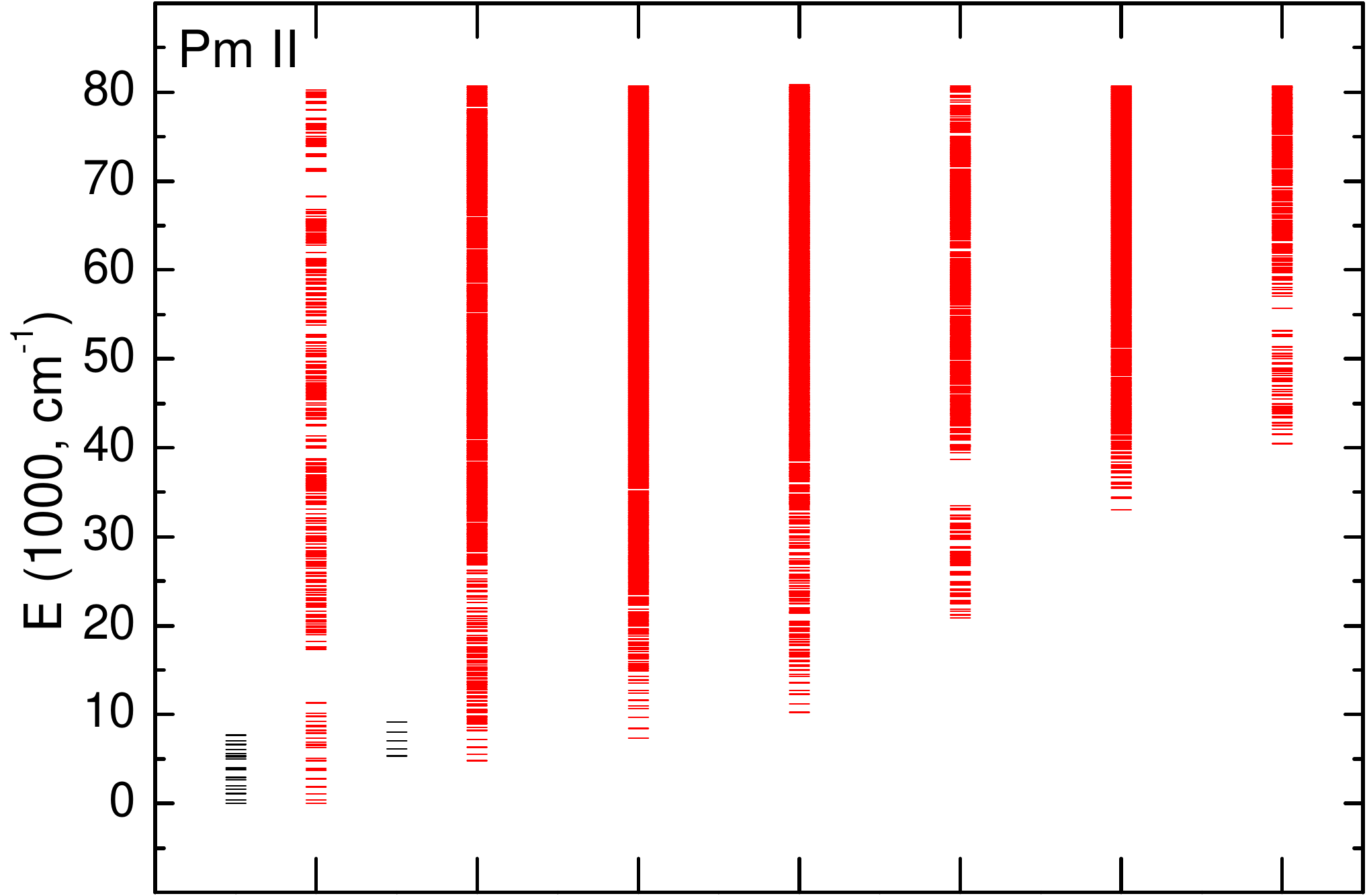}
\includegraphics[width=0.50\textwidth]{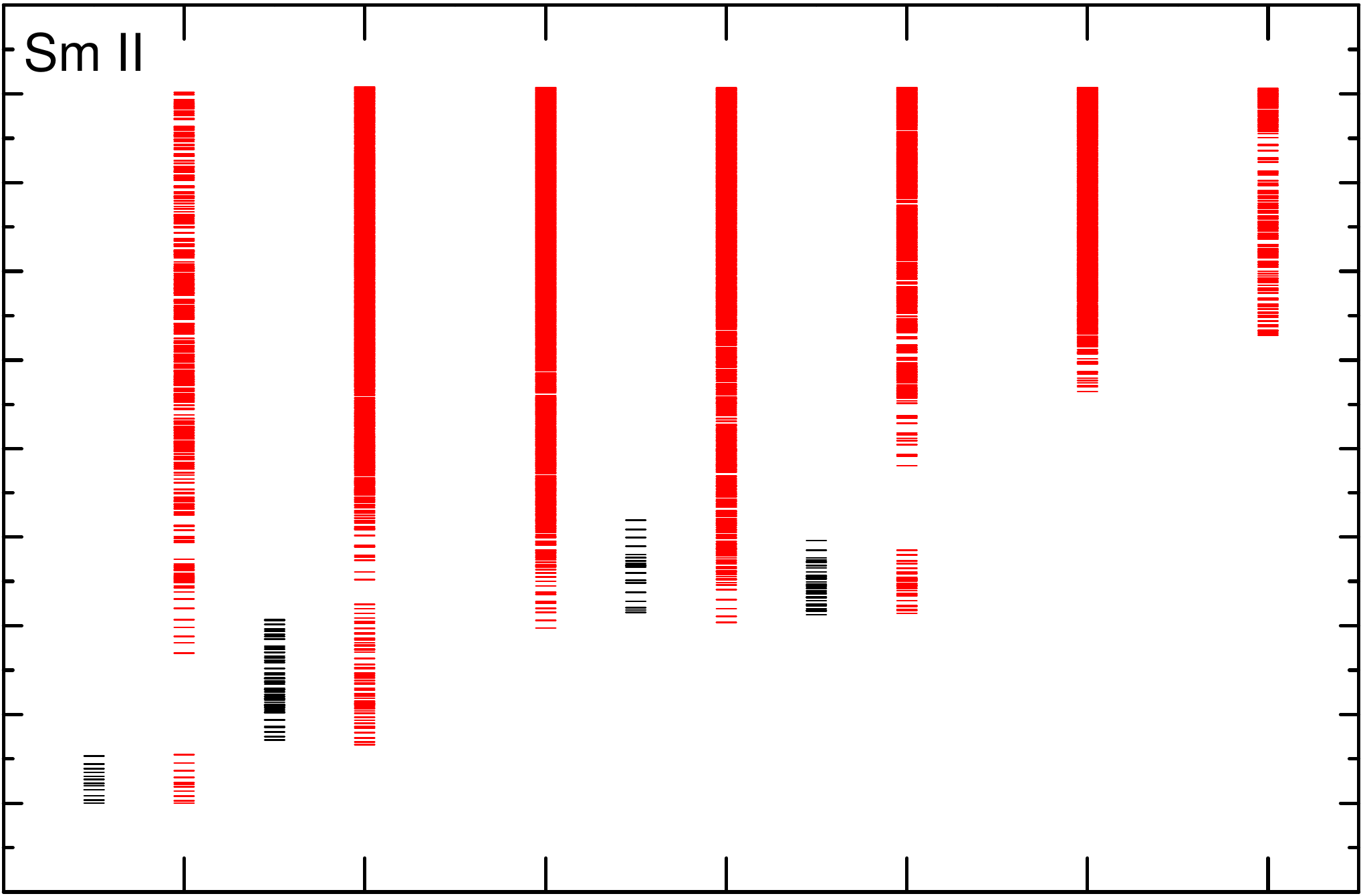}
\includegraphics[width=0.50\textwidth]{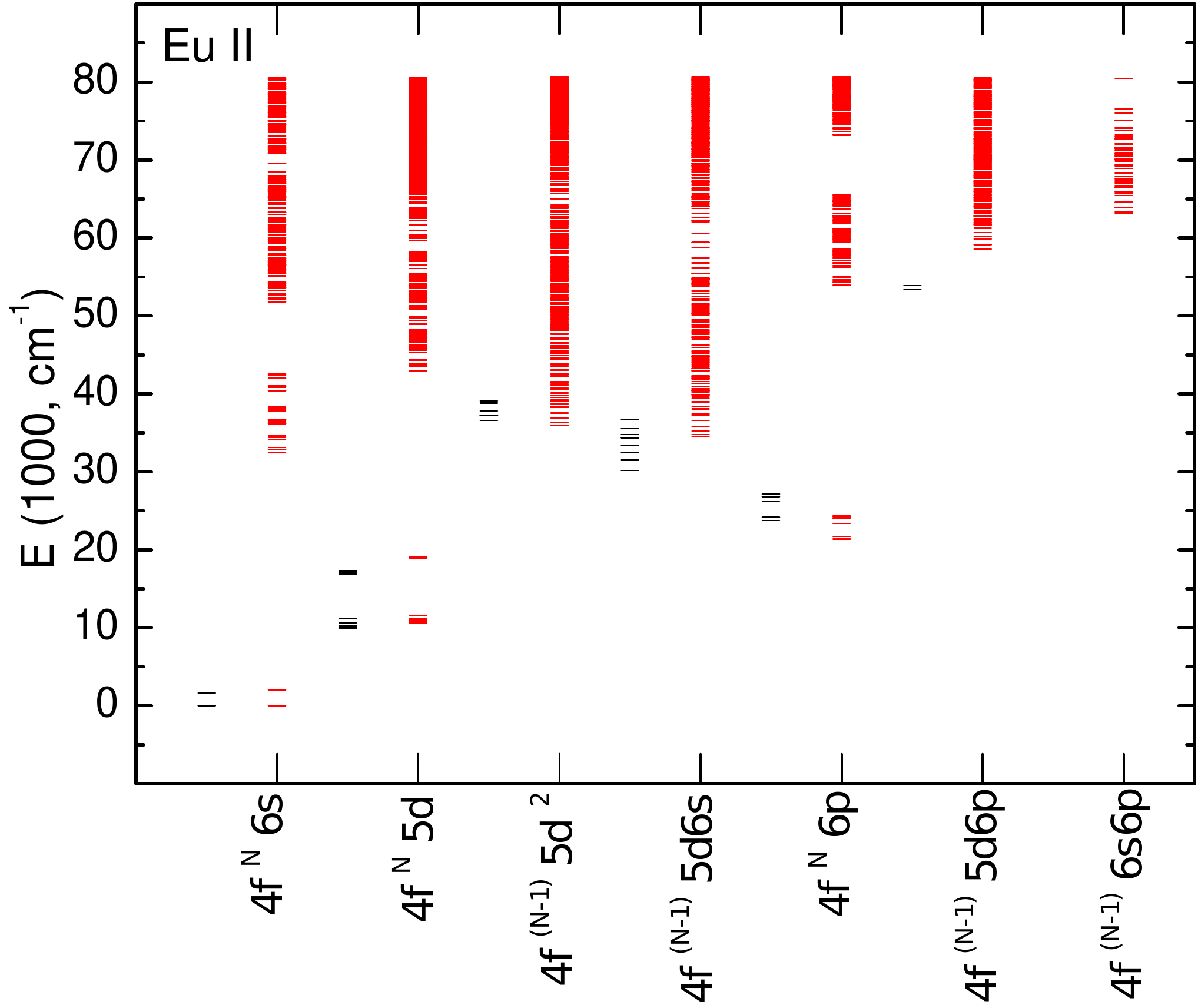}
\includegraphics[width=0.50\textwidth]{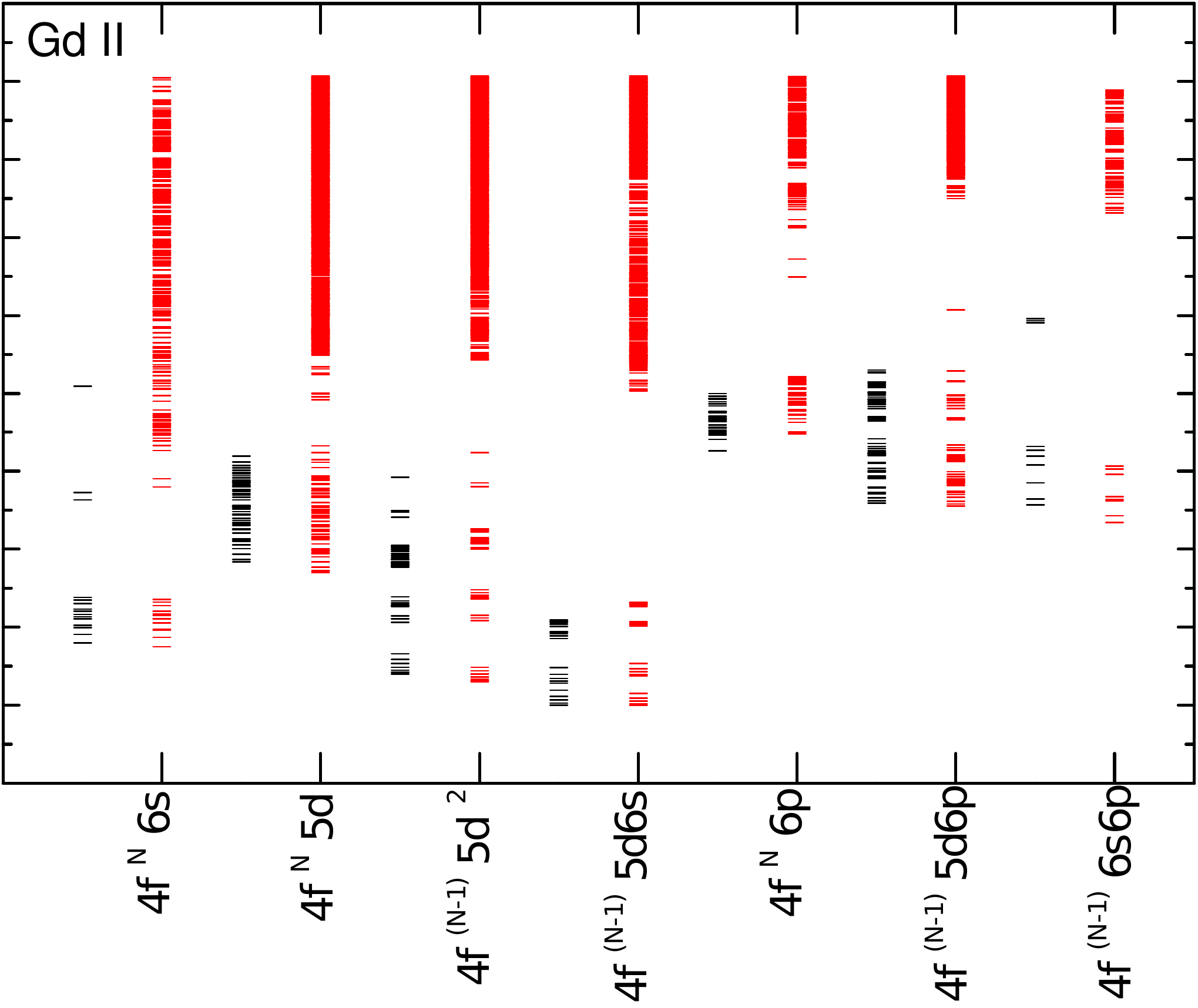}
	\caption{Energy levels for configurations of Pr~II - Gd~II are compared with data in the NIST database. Energy levels with questionable identification
	 in the NIST database also included. Black lines represent the NIST data
	 while red lines represent all computed energy levels. Nd II data are published in \citep{Nd_jonai}.
	$N$ is the number of electrons in the $4f$ shell ($N=3 - 8$ for Pr~II, Nd II, Pm~II, Sm~II, Eu~II and Gd~II respectively).
	\label{Pr_II_pilnas}}
\end{figure*}

\begin{figure*}
 \includegraphics[width=0.50\textwidth]{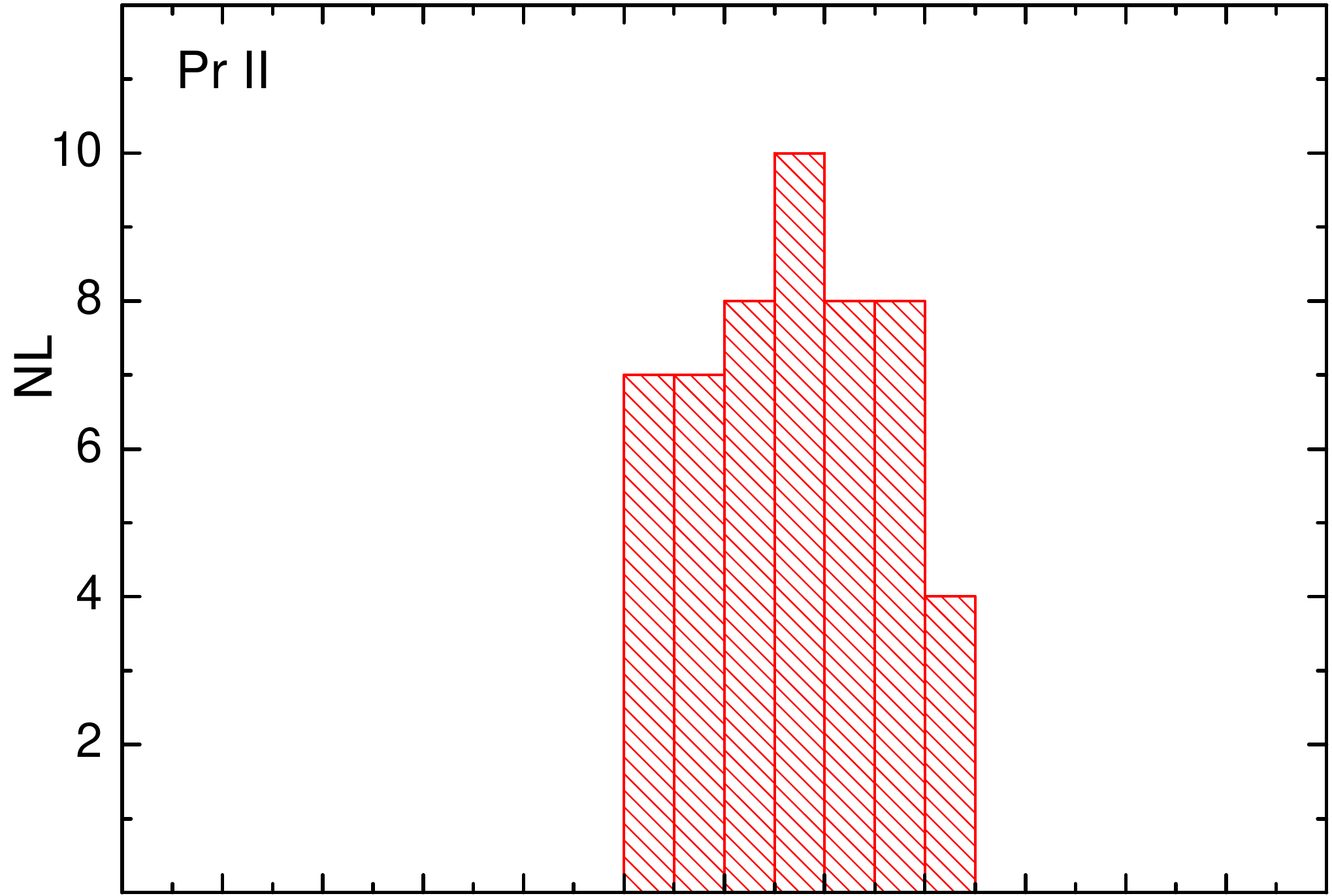}
 \includegraphics[width=0.50\textwidth]{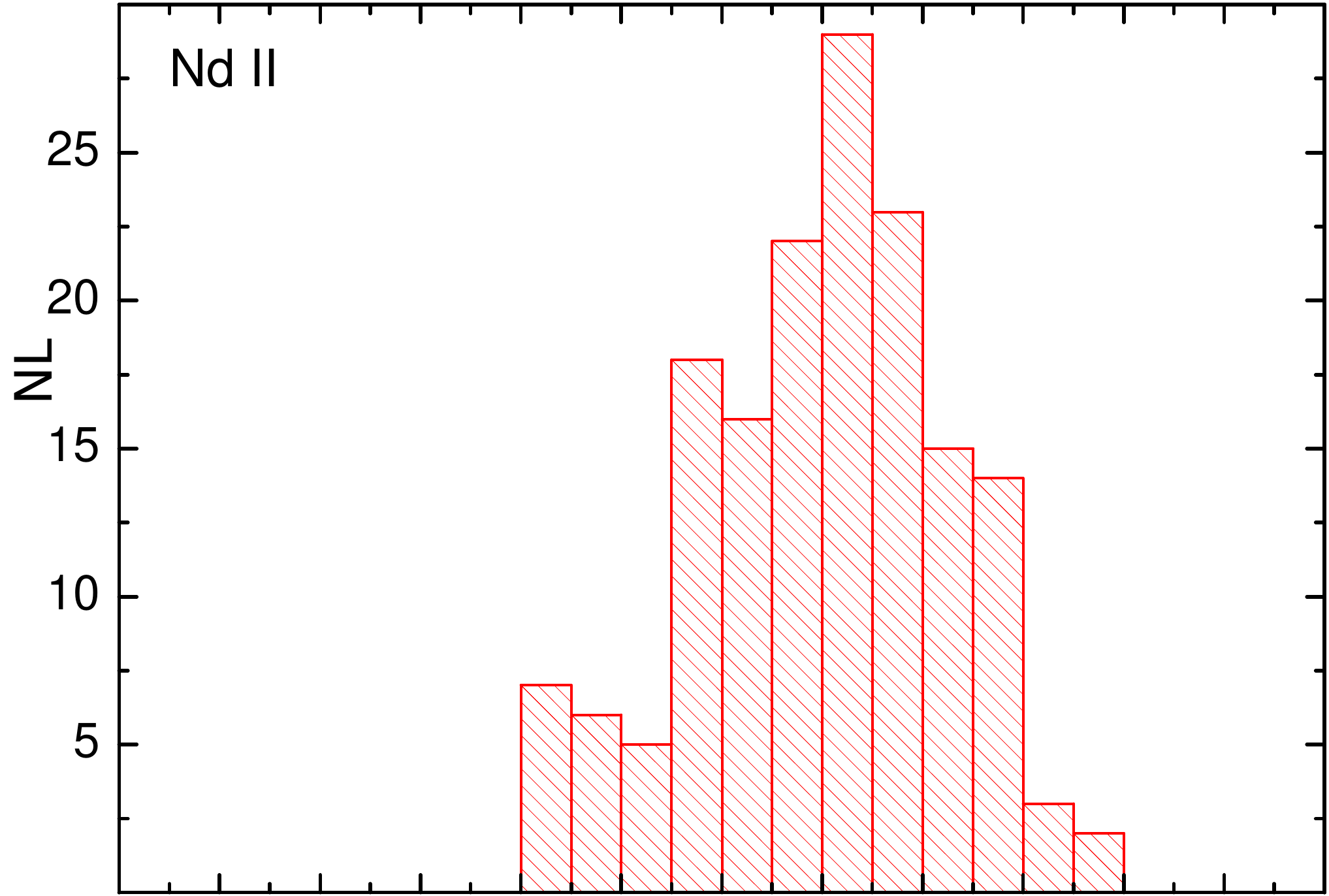}
 \includegraphics[width=0.50\textwidth]{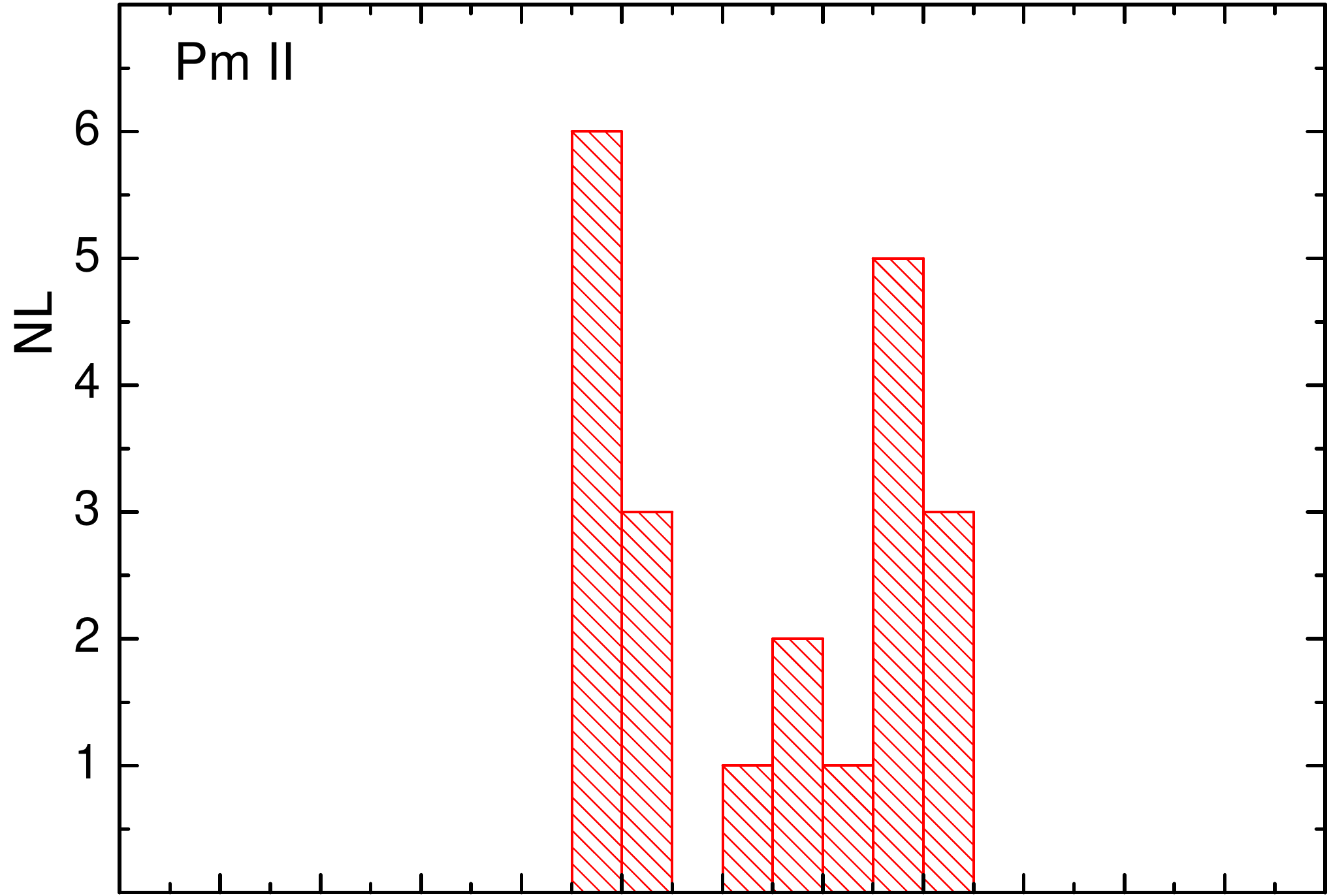}
 \includegraphics[width=0.50\textwidth]{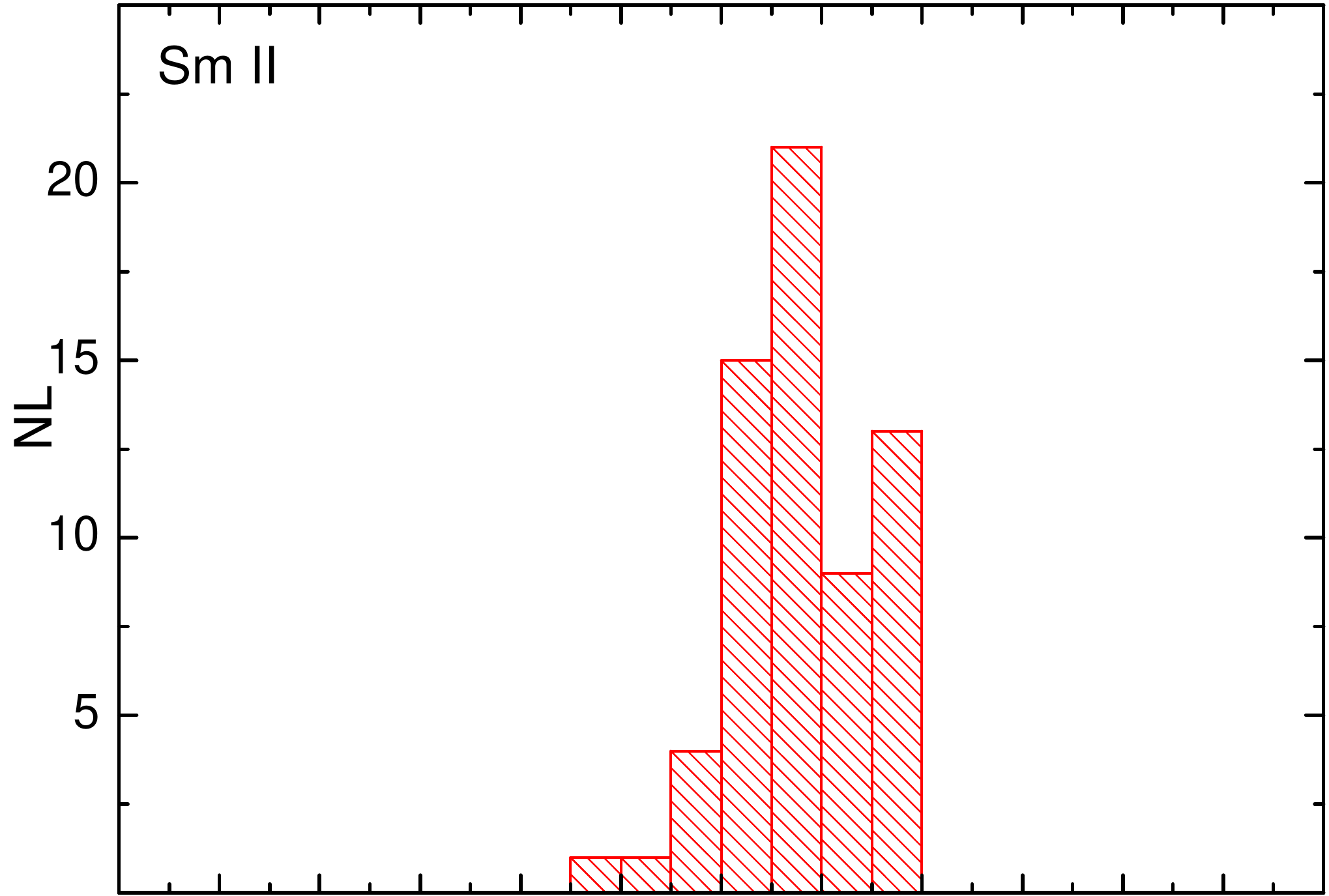}
 \includegraphics[width=0.50\textwidth]{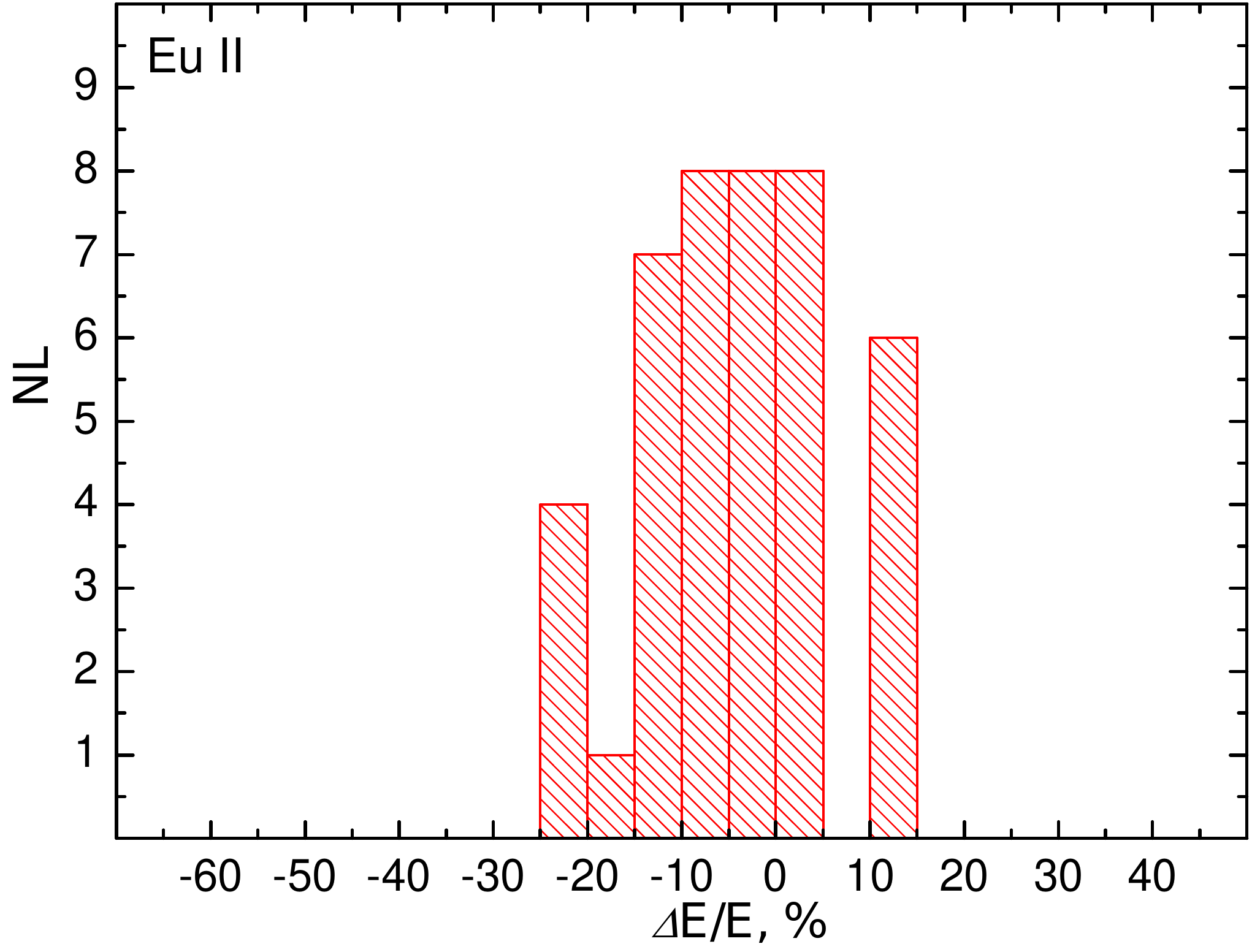}
 \includegraphics[width=0.50\textwidth]{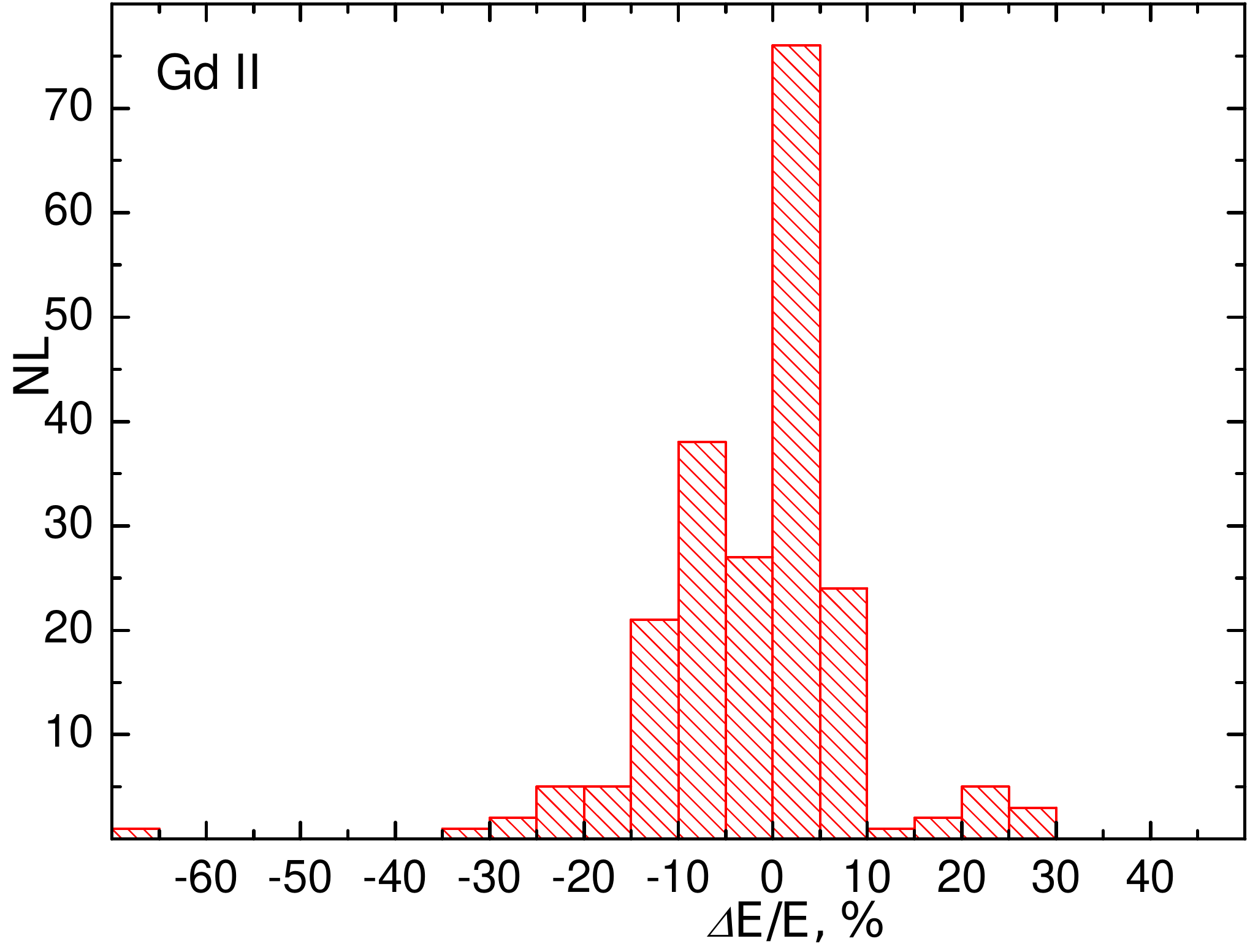}
	\caption{	
Histogram of the relative accuracy with respect to the NIST data ($\Delta E_{i}/E_{\rm NIST} =  (E_{\rm NIST} - E_{i}) / E_{\rm NIST} \times 100\%$) for Pr~II - Gd~II. Nd II data are published in \citet{Nd_jonai}. NL is number of levels.
	\label{Statiska}}
\end{figure*}

\begin{deluxetable*}{c c c c c c c c c cc ccccccc}[ht!!]
\tablecaption{\label{summary_accuracy} Comparison of Energy Levels with the NIST Database, $\overline{\Delta E/E}$ (in \%) and 
the Number of Compared Levels (NL). }
\tablehead      {
 &\multicolumn{2}{c}{Pr~II}
&&\multicolumn{2}{c}{Nd~II}
&&\multicolumn{2}{c}{Pm~II}
&&\multicolumn{2}{c}{Sm~II}
&&\multicolumn{2}{c}{Eu~II}
&&\multicolumn{2}{c}{Gd~II}\\
                 \cline{2-3}\cline{5-6}\cline{8-9}\cline{11-12}\cline{14-15}\cline{17-18}
                 &$\overline{\Delta E/E}$ &NL 
								&&$\overline{\Delta E/E}$ &NL
								&&$\overline{\Delta E/E}$ &NL
								&&$\overline{\Delta E/E}$ &NL
								&&$\overline{\Delta E/E}$ &NL
								&&$\overline{\Delta E/E}$ &NL}
\startdata
$4f^N 6s$        & 6  & 7    && 15 & 27&&  13& 17&&  8 & 12&& 23 & 1 && 4 &13\\ 
$4f^N 6p$        & 4  &12    && 13 & 23&&   -&  -&&  - &  -&& 10 & 6 && 6 &30\\ 
$4f^N 5d$        &10  &33    &&  8 & 47&&  10&  5&&  5 & 52&&  9 &10 && 3 &56\\ 
$4f^{N-1} 5d6s$  & -  & -    &&  2 & 14&&   -&  -&& -  &  -&& 15 & 8 &&19 &19\\ 
$4f^{N-1} 5d6p$  & -  & -    &&  6 & 12&&   -&  -&& -  &  -&&  - & - && 2 &49\\ 
$4f^{N-1} 6s6p$  & -  &-     &&  4 & 13&&   -&  -&& -  &  -&&  - & - && 8 & 6\\ 
$4f^{N-1} 5d^2$  &15  &1     && 15 & 22&&   -&  -&& -  &  -&& 13 & 3 &&13 &39\\ 
$4f^{N-1} 6s^2$  &    &      &&    &   &&    &   &&    &   &&    &   &&66 & 1\\ 
$4f^{N+1}$       &    &      &&    &   &&    &   && 14 &  1&&  - & - &&   &  \\ 
$4f^N 7s$        &    &      &&    &   &&    &   &&    &   &&  3 & 2 &&   &  \\ 
$4f^N 8s$        &    &      &&    &   &&    &   &&    &   &&  9 & 1 &&   &  \\ 
$4f^N 6d$        &    &      &&    &   &&    &   &&    &   &&  2 &10 &&   &  \\ 
all              & 8  &53    &&10  &158&&  12&22 && 6  & 65&&  8 &41 &&7  &213\\ 
\hline
\enddata 
\tablecomments{* Nd II data are published in \citet{Nd_jonai}. Levels with unquestionable identification are included in to the comparison.}
\end{deluxetable*}

In addition, some states of Rydberg  series (up to 10 eV) 
are computed for Eu~II. 
This includes 38 levels from configurations $4f^6\{7s,8s,6d,7d,7p,8p\}$. 
Radial wave functions for configurations $4f^6\{7s,8s\}$ up to $4f$ are 
taken from the ground configuration ($4f^6 6s$). 
For the rest configurations, radial wave functions are computed in the same manner as in the configurations $4f^6\{5d,6p\}$. 
This means that each configuration from $4f^6\{6d,7d,7p,8p\}$ has 
different radial wave functions.
Active space generated in a similar manner as for the configurations $4f^6\{6s,5d,6p\}$. 
For example, active space for the configuration are generated by SD substitutions 
from $4f^6~8s$ to AS$_{0L}$ = $\{6p, 5d\}$ and AS$_{1L}$ = AS$_{0L}$ + $\{6s, 7p, 6d, 5f\}$ 
and by S substitutions to AS$_{2L}$ = AS$_{1L}$ + $\{7s, 8p, 7d, 6f, 5g\}$. 

For Gd~II, radial wave function is generated also for only one $J$ value. 
Radial wave functions of $4f^7 5d^2$, $4f^7 6s^2$, $4f^7 5d6s$, and $4f^8 6p$ are computed together, using radial wave function of configuration $4f^8 6s$ up to $4f$. 
Rest of configurations are computed in the same manner as for Eu~II. 
The MCDHF calculations are then followed by RCI calculations
by including the Breit interaction and leading QED effects.
 The same active space (AS$_{2L}$) is used for the RCI computations as well as for MCDHF computations.

\section{Energy levels}
\label{sec:evaluation}

All levels for $Z=59 - 64$ ions are given in Figure \ref{Pr_II_pilnas},
and the energy data computed for Pr~II, Pm~II, Sm~II, Eu~II, and Gd~II are given in machine-readable format in Tables
\ref{Pr_II_Energies}, \ref{Pm_II_Energies}, \ref{Sm_II_Energies}, \ref{Eu_II_Energies}, and \ref{Gd_II_Energies},
respectively. 
This includes label, $J$ and $P$ values, and energy value. 
Levels are given in $LS$-coupling, 
although it is suitable only for the lowest states of configurations and 
determination the configuration is complicated for higher states \citep{Cowan}. 
 For the labels, we use notation $4f^N~^{(2S+1)}_{Nr} L ~n'l' ~^{(2S'+1)}L'$. 
Intermediate quantum numbers define parent levels $4f^N~^{(2S+1)}_{Nr} L$, 
where $N$ is electron number in $4f$ shell, $(2S+1)$ is multiplicity, 
$Nr$ is a sequential index number representing the group labels $nWU$ for the term, 
and $L$ is orbital quantum number
(see \citealt{senioriy_2} for more about $Nr$). More complicated configurations are presented in the similar way.

To evaluate the accuracy of our calculations, comparison with critically evaluated data is necessary.
In this section, we first summarize the available data for energy levels of Pr~II, Pm~II, Sm~II, Eu~II, Gd~II in the NIST database. Then, we compare calculated energy levels with these available data. 

\subsection{Available data}
\label{sec:nist}

\subsubsection{Pr~II}
\label{Pr_II_energies}
\citet{Ginibre_Pr_1989_a} have investigated 105 odd and 187 even experimental energies based on Fourier transform (FT) 
spectroscopy in range 2~783 - 27~920 cm$^{-1}$.
Also, the large amount levels were investigated by \citet{Rosen_Pr}, \citet{Blaise_Pr}, and \citet{Blaise_Pr_b}.
They performed semi-empirical fitting procedure to assign for some levels labels in $LS$-coupling \citep{Ginibre_Pr_1989_b}. 
Later, \cite{Ivarsoon_Pr} presented improved 39 energy levels using FT spectroscopy in 2~800 - 8~000~\AA~region. 
\cite{Furmann_Pr_a,Furmann_Pr_b,Furmann_Pr_c} investigated 31 odd and 14 even levels, 
using laser-induced fluorescence spectroscopy (LIF) in a hollow cathode discharge lamp.
More recently, \citet{Akhtar_Pr} have redetermined energy values of 227 levels (74 having odd and 153 even parity) and hyperfine structures of 477 transitions in the range of 3~260 - 11~700 \AA. They corrected the energy levels from the works of  
\citet{Ginibre_Pr_1989_a} and \cite{Ivarsoon_Pr}.

All of these levels are measured/reanalyzed in high accuracy. 
However, 
each work presents energy levels in a narrow range as shown in Figure \ref{Pr_II_su_kitais_autoriai}.
Therefore, the transitions between measured energy levels give
too small amount of lines needed for computation of opacities in neutron star mergers.
Data of these authors are summarized by \cite{Martin}.
Since the NIST database \citep{NIST} includes the work by \citep{Martin},
we only give comparison with the NIST database here.

\subsubsection{Pm~II}
\label{Pm_II_energies}
Pm~II is one of the ions whose spectrum is not well investigated. 
Energy levels of two configurations $4f^5 6s$ and $4f^5 5d$ were investigated by \cite{Martin}.
Five new levels of $4f^5 5d$ configuration were measured by \cite{Otto_Pm} 
with the collinear laser ion beam spectroscopy (CLIBS) method 
and were identified using Hartree-Fock method.

\subsubsection{Sm~II}
\cite{Albertson_Sm} have assigned terms of 40 even levels of the $4f^6 6s$ and $4f^6 5d$ configurations based on the Zeeman patterns of over 300 lines.
\cite{Spector_Sm_Gd} have done semi-empirical computation of energy values and $LS$-composition of 55 levels for $4f^6(^7F) 5d$ configuration.
Also, a large amount of work for energy levels was done by \cite{Blaise_Sm}:
325 levels for Sm~II were obtained from the Zeeman effect measurement in the visible and the ultraviolet spectrum. 
Then, these energy levels were re-evaluated by \cite{Martin}.
Attempt of identification of odd configurations for some levels was done by \cite{Rao_Sm} using isotope shifts data, which was carried out on a recording Fabry-Perot spectrometer. 
The hyperfine structure and isotope shift were also measured by collinear fast ion beam laser spectroscopy.
These data were used to assign configurations to the 13 odd upper levels by 
\cite{Villemoes_Sm}. 
Note that some of them do not have identification by \cite{Martin}. 

\subsubsection{Eu~II}
156 levels of configurations $4f^7\{6s,7s,8s,5d,6d,6p\}$ and $4f^6 5d6s$, $4f^6 5d^2$ 
were resolved with the spark spectrum of arc by \cite{Russel}. 
This work is the extension of the analysis by \cite{Albertson} 
on 9 levels of $4f^7\{6s,5d,6p\}$ configurations.
Then these energy levels were re-evaluated by \cite{Martin}.
More recently, 13 new energy levels of $4f^7 6s$ configuration were suggested from hyperfine constant and isotope shift measurements \citep{Furmann_Eu}.

\begin{figure}[ht!!]
 \includegraphics[width=0.47\textwidth]{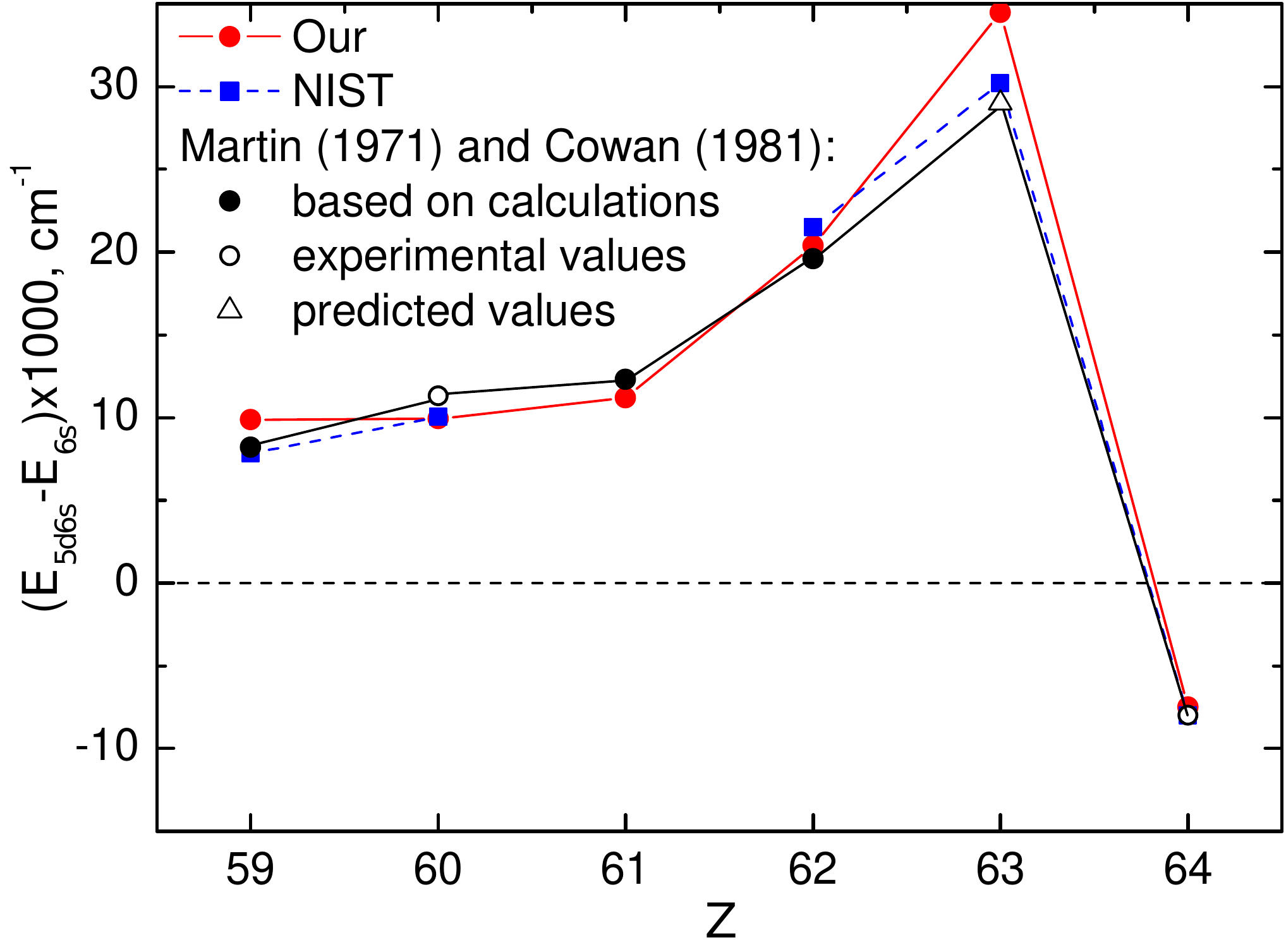}
	\caption{The difference between the lowest level of 
$4f^{N-1} 5d6s$ and the lowest level of $4f^{N} 6s$ 
for singly ionized lanthanide elements. 
Red circles are our computed theoretical values,
while
  blue squares are values recommended by the NIST database.
Black symbols indicate data from  \cite{Martin_SD} and \cite{Cowan}:
closed circles are predicted values, open circles are experimental values,
and triangles are estimated data based on incomplete experimental data.
	\label{Kowan_type}}
\end{figure}

\begin{figure*}
\includegraphics[width=0.47\textwidth]{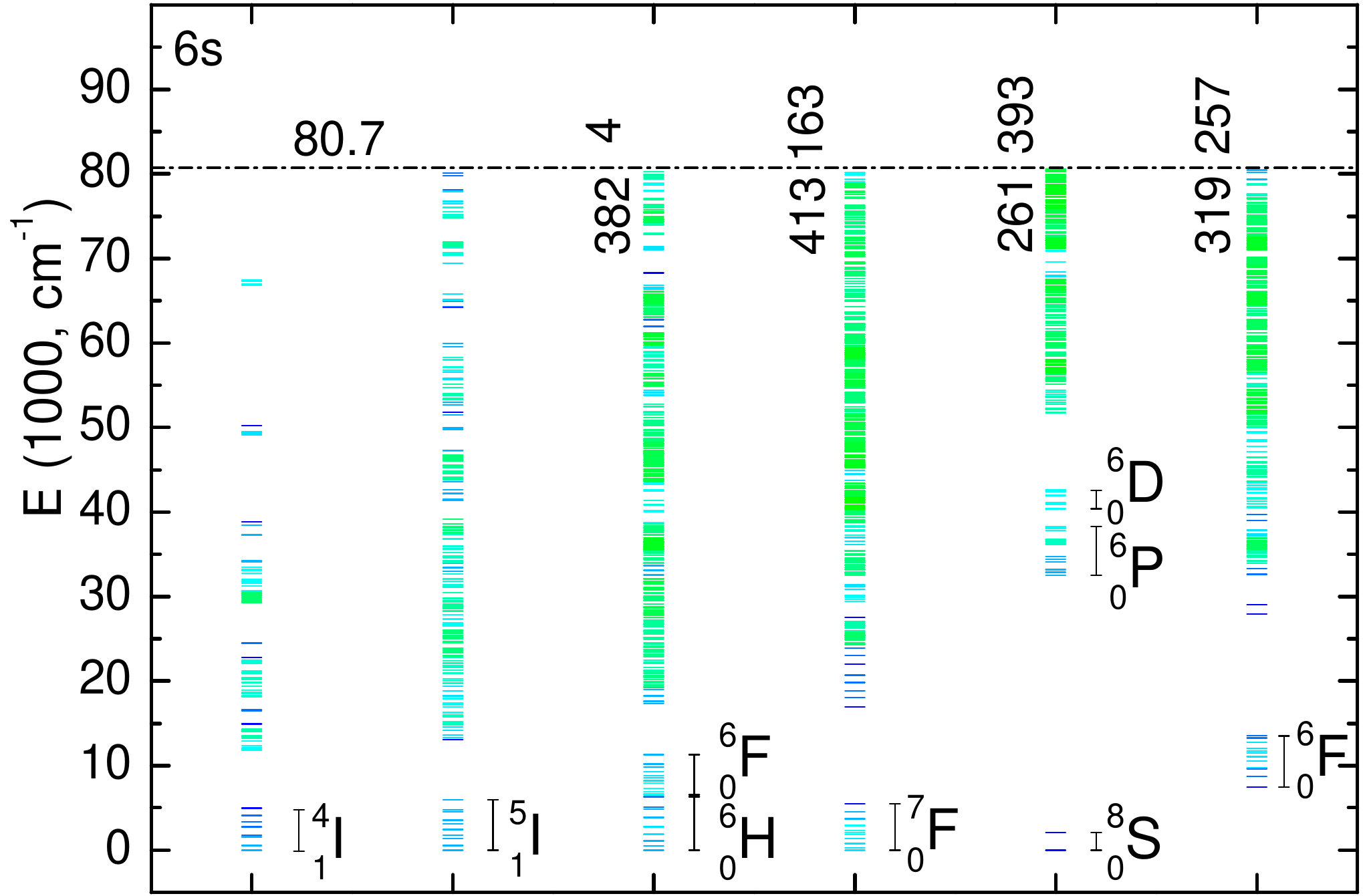}
\includegraphics[width=0.47\textwidth]{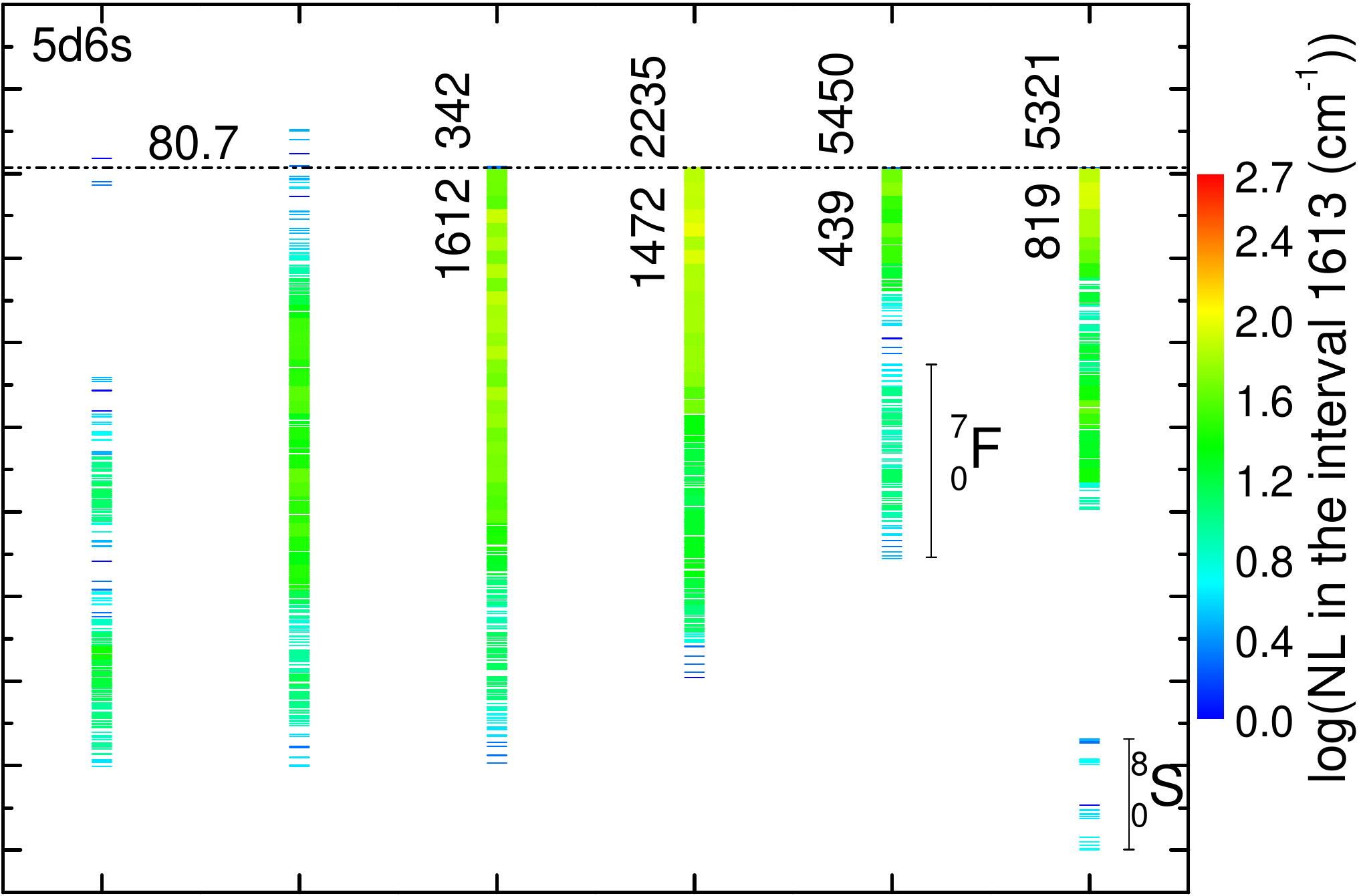}
\includegraphics[width=0.47\textwidth]{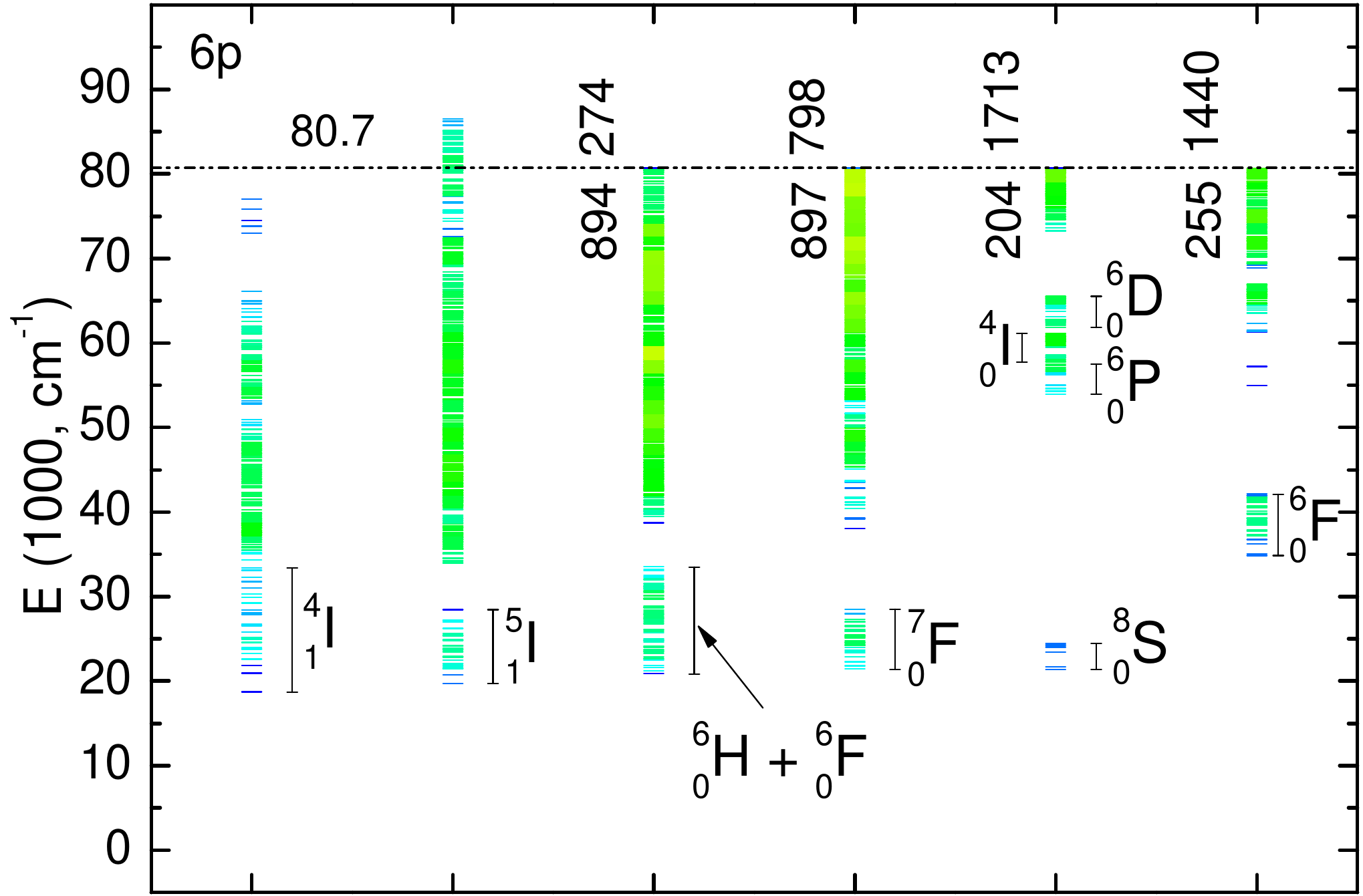}
\includegraphics[width=0.47\textwidth]{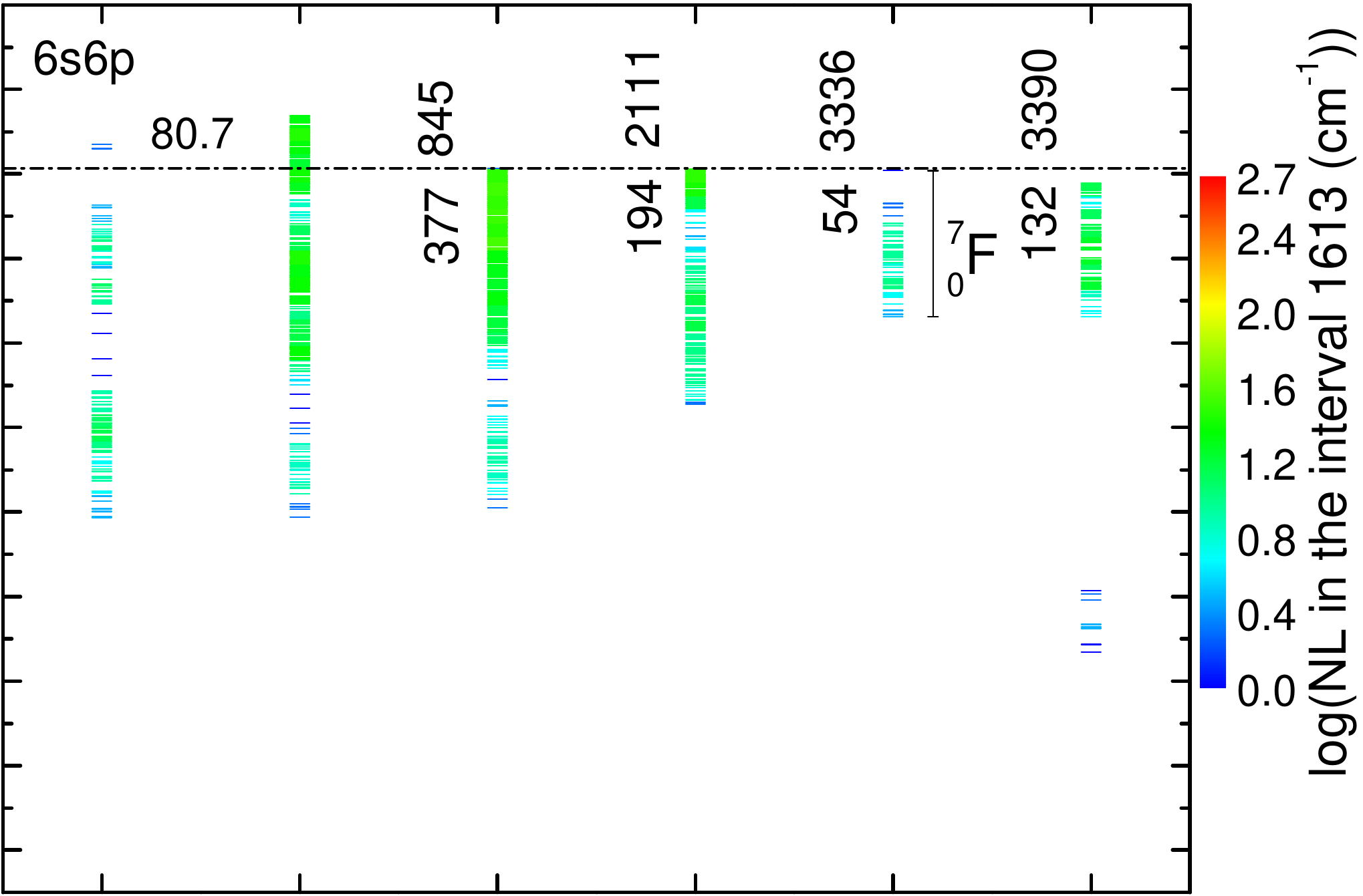}
\includegraphics[width=0.47\textwidth]{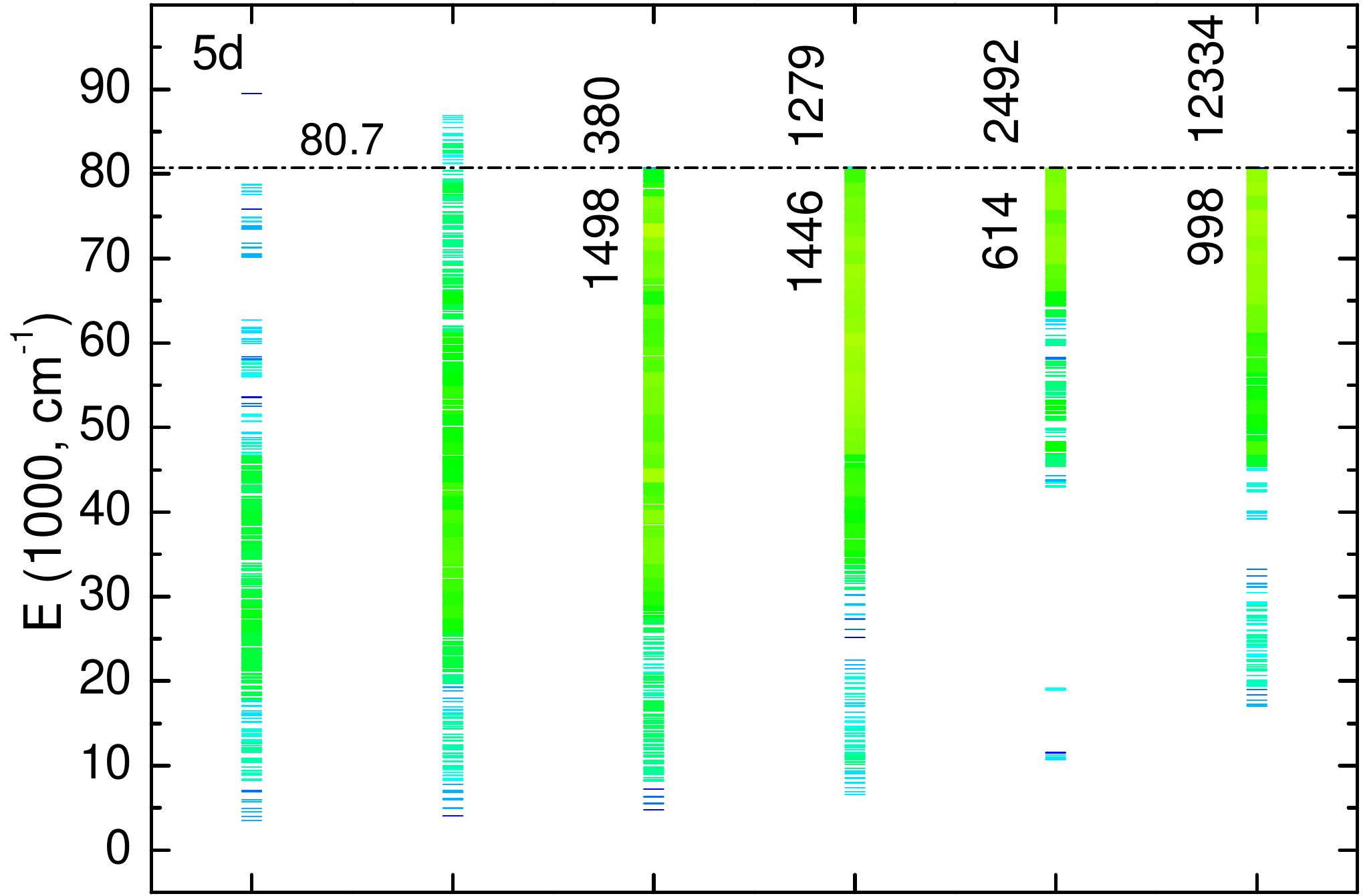}
\includegraphics[width=0.47\textwidth]{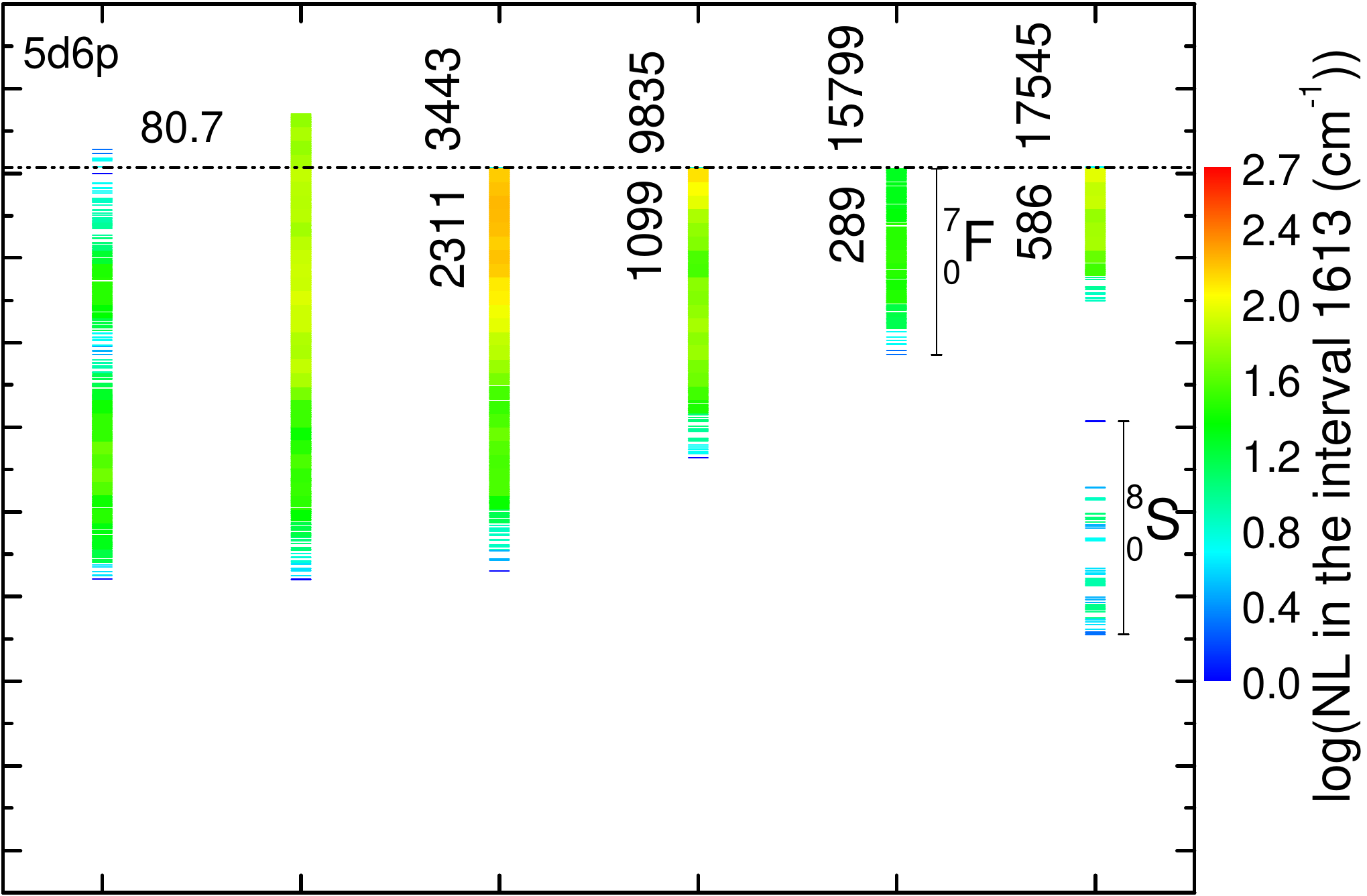}
\includegraphics[width=0.47\textwidth]{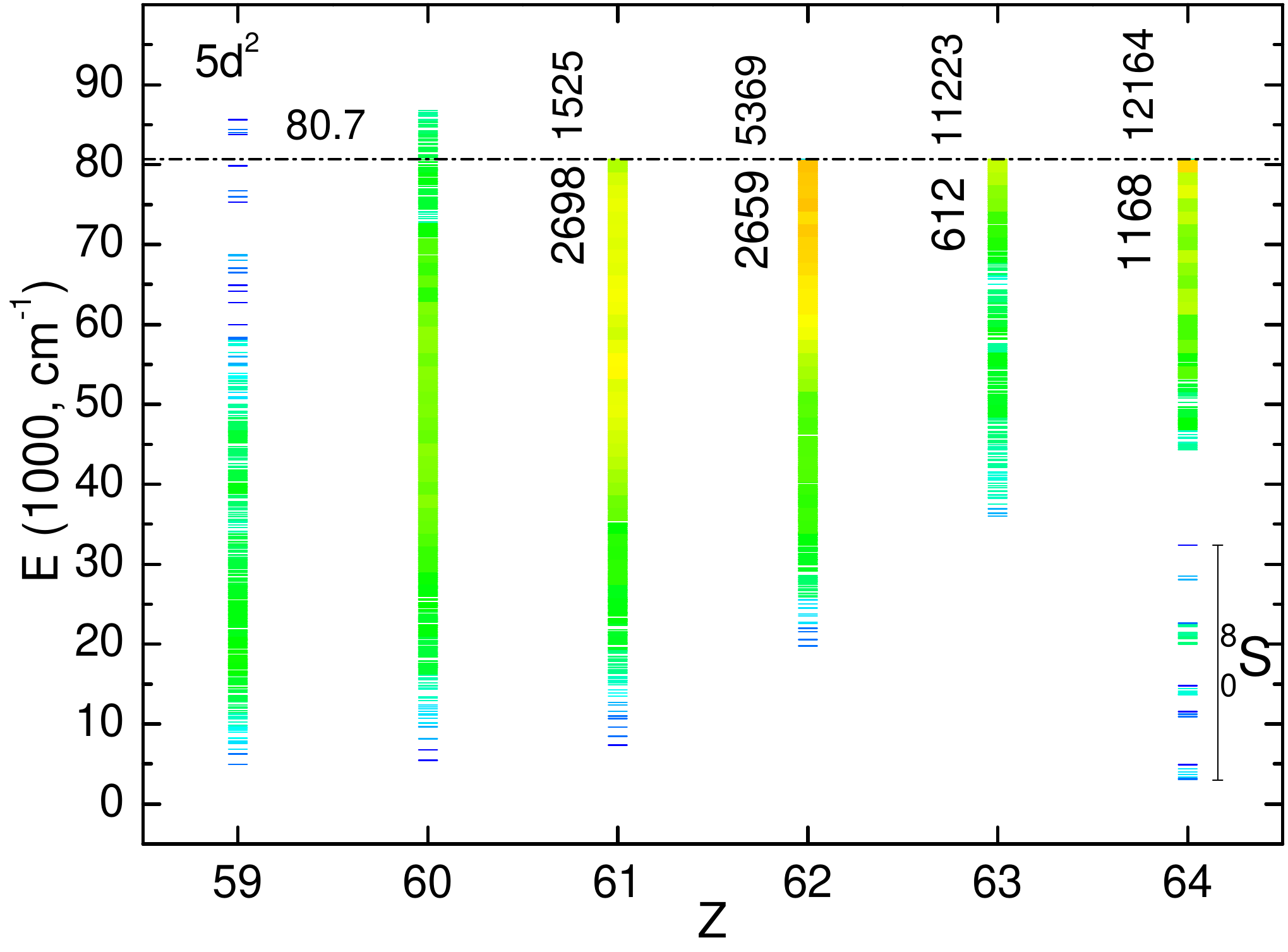}~~~~~~~~~~~
\includegraphics[width=0.47\textwidth]{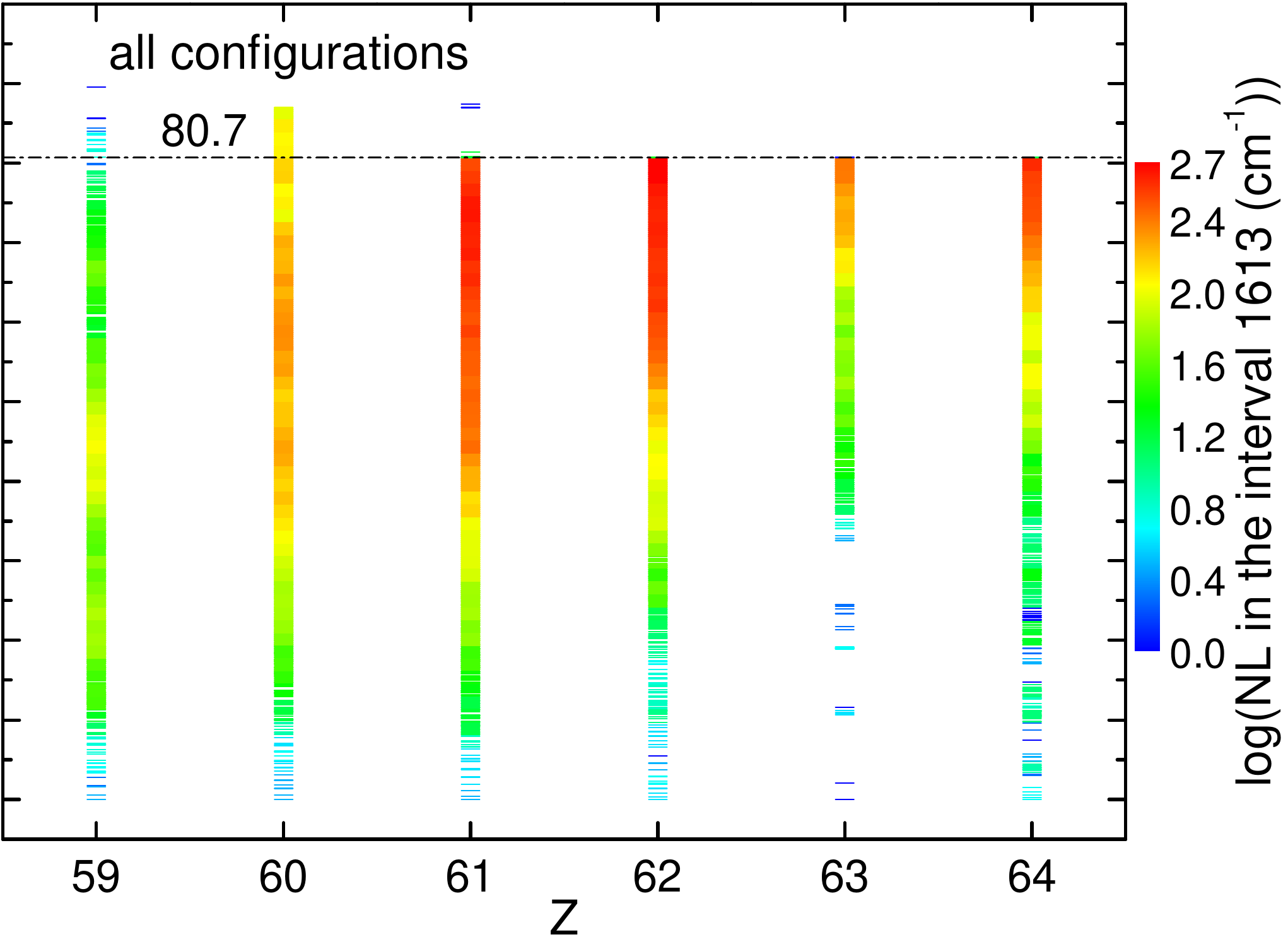}
\caption{Energy level density and structure of 
$4f^N \{6s,6p,5d\}$, 
$4f^{N-1} \{5d6s,5d6p,6s6p\}$, and 
$4f^{N-1} 5d^2$ configurations and all levels for the ions with $Z=59 - 64$.  
The blocks of levels and the corresponding parent levels are also given. NL is the number of levels. 
The horizontal lines show our energy threshold (10 eV), and the numbers above/below the lines
show the number of levels above/below this threshold.
}
\label{all_configurations_energies}
\end{figure*}

\subsubsection{Gd~II}
\cite{Albertson_Gd} have investigated 9 odd and 11 even energy levels,
have deduced quantum numbers from Zeeman effect pattern, 
and have established the ground configurations to be $4f^75d6s$.
\cite{Venugopalan_Gd} and \cite{Venugopalan_II_Gd} have measured isotope 
shift of 33 spectroscopic lines, 
using photoelectric recording Fabry-Perot spectrometer.
They suggested new configuration identification of 4 high energy levels (lying above 35 000 cm$^{-1}$):
35~362.630 cm$^{-1}$ ($J = 13/2$) as $4f^7 5d6s$;
35~822.697 cm$^{-1}$  ($J = 9/2$) as mix of two configurations $4f^75d6s$+$4f^86p$;
37~831.032 cm$^{-1}$ ($J = 11/2$) 
and 38~010.603 cm$^{-1}$ ($J = 11/2$) as $4f^86p$.
\cite{Blaise_Gd} have done the analysis of the spark spectrum of Gd~II of 178 new levels.
Total 30 levels were ascribed to $4f^8 (^7F) 6p$ configuration by their strong transitions 
with the levels on the $4f^8 (^7F) 6s$ and $4f^8(^7F) 5d$ sub-configurations.
\cite{Spector_Sm_Gd} have done semi-empirical computation of energy values and $LS$-composition of 57 levels for $4f^8(^7F) 5d$ configuration.
164 odd and 150 even parity energy levels of Gd~II are listed by \citet{Martin}.
\cite{Spector_Gd} have done extended analysis on levels of the configurations $4f^8(^7F)\{6s,6p,5d\}$
and measured new levels of $4f^8(^7F)\{6s,5d\}$ configurations and new odd levels.

\begin{deluxetable}{c c c c c c c c c }[ht!!]
\tablecaption{\label{radius} Singly Ionized Atom Ground State Orbitals Radii (a.u.) for Lanthanide $Z = 59 - 64$ Compared with those by \citet{Indelicato}[I07].}
\tablehead{
Ion& $<4f\_>$ & $<4f>$ & $<5d\_>$ & $<5d>$ & $<6s>$}
\startdata
Pr~II & 1.0833& 1.0986 & 2.5036   & 2.5484 & 4.3130\\
 I07  & 1.0589& 1.0667 &          &        & 4.2924\\
Nd II*& 1.0291& 1.0440 & 2.4607   & 2.5085 & 4.2522\\
 I07  & 1.0054& 1.0190 &          &        & 4.2252\\
Pm~II & 0.9832& 0.9981 & 2.4223   & 2.4717 & 4.1948\\
 I07  & 0.9624& 0.9796 &          &        & 4.1608\\	
Sm~II & 0.9442& 0.9590 & 2.3892   & 2.4400 & 4.1402\\
 I07  & 0.9249& 0.9392 &          &        & 4.1012\\
Eu~II & 0.9098& 0.9256 & 2.3642   & 2.4272 & 4.0870\\
 I07  & 0.8920& 0.8999 &          &        & 4.0438\\
Gd~II & 0.8797& 0.8929 & 2.3991   & 2.5041 & 3.6878\\
 I07  & 0.8218& 0.8221 & 2.4547   & 2.4846 & 3.7930\\
\hline
\enddata 
\tablecomments{* Orbital radii of Nd II were computed during MCDHF computations in \citet{Nd_jonai},
but have not been unpublished.}
\end{deluxetable}

\subsection{Comparison of the energy levels}
The energy levels for each configuration are compared with
those in the NIST database in Figure \ref{Pr_II_pilnas}. 
Only the common configurations for Pr~II - Gd~II are presented in the figure. 
Although the energy levels in the NIST database sometimes include 
questionable identification of the configuration, this figure includes all levels.

To analyse the accuracy of our calculations as compared with the NIST data,
we use an expression
 $\Delta E_{i}/E_{\rm NIST} =  (E_{\rm NIST} - E_{i}) / E_{\rm NIST} \times 100\% $ . 
For the indicator of the accuracy for many levels, we use a value
$\overline{\Delta E/E} = \sum \frac{|\Delta E_{i}/E_{\rm NIST}|}{N}$, 
where $N$ is the number of the compared levels. 
Summary of the accuracy for each configuration is given in Table \ref{summary_accuracy}.
 Levels with unquestionable identification are included in to the comparison.
Empty space in Table \ref{summary_accuracy} means that configuration is not computed
while a mark with "-" means that data are missing in the NIST database
(or there is only one level).
The last line (all) of the table presents averaged accuracy with unquestionable identification 
between our results and the NIST database.

Overall, we find that our calculations give good accuracy: 
8\%, 12\%, 6\%, 8\%, and 7\% for Pr~II, Pm~II, Sm~II, Eu~II, and Gd~II ions, respectively.
There is no clear trend with the atomic number $Z$. 
The accuracy depends on the configurations. For example, 
the degree of agreement for $4f^N 6s$, $4f^N 5d$ and $4f^N 6p$ configurations 
slightly differ.
These variations of the differences is mainly caused by the different number of levels 
used for comparison.
Note that the biggest deviation is found for level $^8S_{7/2}$ of configuration $4f^7~6s^2$
(66\% difference for this 1 level).



\begin{deluxetable}{rlrrr}[ht!!]
	\tabletypesize{\footnotesize}
	\setlength{\tabcolsep}{10pt}
	\tablecaption{
Energy Levels (in cm$^{-1}$) Relative to the Ground State for the States of Pr~II. \label{Pr_II_Energies}}
	\tablehead{ \colhead{No.} & \colhead{label}  & \colhead{$J$} & \colhead{P} & \colhead{$E$} }
	\startdata
   1 & $4f^{3}(^4_1I)~6s~^{5}I          $  & 4  &  $-$ &    0.00 \\
   2 & $4f^{3}(^4_1I)~6s~^{5}I          $  & 5  &  $-$ &  511.12 \\
   3 & $4f^{3}(^4_1I)~6s~^{5}I          $  & 6  &  $-$ & 1558.26 \\
   4 & $4f^{3}(^4_1I)~6s~^{3}I          $  & 5  &  $-$ & 1772.15 \\
   5 & $4f^{3}(^4_1I)~6s~^{5}I          $  & 7  &  $-$ & 2773.88 \\
   6 & $4f^{3}(^4_1I)~6s~^{3}I          $  & 6  &  $-$ & 3337.17 \\
   7 & $4f^{3}(^4_1I)~5d~^{5}L          $  & 6  &  $-$ & 3506.00 \\
   8 & $4f^{3}(^4_1I)~5d~^{5}K          $  & 5  &  $-$ & 3948.78 \\
   9 & $4f^{3}(^4_1I)~6s~^{5}I          $  & 8  &  $-$ & 4104.37 \\
  10 & $4f^{3}(^4_1I)~5d~^{5}L          $  & 7  &  $-$ & 4553.26 \\
  11 & $4f^{3}(^4_1I)~5d~^{5}K          $  & 6  &  $-$ & 4898.65 \\
  12 & $4f^{3}(^4_1I)~6s~^{3}I          $  & 7  &  $-$ & 4937.25 \\
  13 & $4f^{2}(^3_1H)~5d^{2}(^3_2F)~^5L $  & 6  &  $+$ & 4948.47 \\
  14 & $4f^{3}(^4_1I)~5d~^{5}L          $  & 8  &  $-$ & 5703.77 \\
  15 & $4f^{3}(^4_1I)~5d~^{5}K          $  & 7  &  $-$ & 5941.26 \\
  16 & $4f^{2}(^3_1H)~5d^{2}(^3_2F)~^5L $  & 7  &  $+$ & 6265.05 \\
  17 & $4f^{2}(^3_1H)~5d^{2}(^3_2F)~^5I $  & 4  &  $+$ & 6861.72 \\
  18 & $4f^{3}(^4_1I)~5d~^{5}L          $  & 9  &  $-$ & 6944.26 \\
  19 & $4f^{3}(^4_1I)~5d~^{5}K          $  & 8  &  $-$ & 7067.68 \\
  20 & $4f^{2}(^3_1H)~5d^{2}(^3_2F)~^5K $  & 5  &  $+$ & 7591.81 \\
			\enddata
\tablecomments{ Table~\ref{Pr_II_Energies} is published in its entirety in the machine-readable format. Part of the values are shown here for guidance regarding its form and content.}
\end{deluxetable}

As mentioned in Section \ref{sec:scheme}, 
computations of Eu~II are performed in a slightly different manner:
radial wave functions are computed only for one $J$ symmetry of the lowest ASF. 
To test the influence of such splitting, 
we compute configurations $4f^7{6s}$, $4f^7{5d}$, $4f^6 5d6s$, $4f^6 5d6p$, and $4f^6 5d^2$ 
in both ways.
We find that the differences between two methods are small:
the maximum averaged difference of energy levels per configuration is 0.5\%  for $4f^7{5d}$ configuration (614 levels)
and the minimum difference is 0.02\% for $4f^7{6s}$ (261 levels).
 Levels of Rydberg state of the configurations $4f^7\{7s,8s,6d\}$ for Eu~II are also compared in Table \ref{summary_accuracy}. 
There is a good agreement for levels of configurations $4f^7\{7s,6d\}$ obtained in this research with values from NIST database.

Figure \ref{Statiska} show the histogram of the relative difference compared to the NIST for all computed ions. 
This figure includes only the levels with the exact identification.
Note that the number of the available energy levels has a large variation as summarized in 
Section \ref{sec:nist}.
The biggest numbers of levels are available for the Nd II and the Gd~II in the NIST database,
and thus, the distribution is close to the normal distribution for these ions. 

The accuracy of our calculations can also be evaluated using Figure \ref{Kowan_type}, 
which shows the energy difference between the lowest levels of $4f^{N-1} 5d6s$ 
and the lowest levels of $4f^{N} 6s$ 
configurations for singly ionized lanthanides.
As shown in the figure, the overall agreement is very good.
Our results and those in the NIST database give smaller energy differences
than those in \cite{Martin_SD} and \cite{Cowan} for Nd II and Pm~II ions.
The increase of the energy difference is observed for Eu~II by all the works, 
but our result shows a bigger increase than in \citet{Martin_SD}, \cite{Cowan} and the NIST data.

Here it should be noted that, for the cases of Pr~II and Sm~II, 
the identification of $4f^{N-1} 5d6s$ configurations are questionable in the NIST database.
More detailed investigation was done by \cite{Brewer_Sm} (see their Figure 1).
They have estimated energies for lowest levels of configurations involving $4f$, $5d$, $6p$, and $6s$ shells
for singly-triply ionized lanthanides and actinides. 
Their computations are based on the thermodynamic data of the metals.
In a similar manner, the energy differences were also analysed by \cite{Vander_Sm}. 
In fact, our results are very close to the data of these authors.  

\begin{deluxetable}{rlrrr}
	\tabletypesize{\footnotesize}
	\setlength{\tabcolsep}{10pt}
	\tablecaption{
Energy Levels (in cm$^{-1}$) Relative to the Ground State for the States of Pm~II. \label{Pm_II_Energies}}
	\tablehead{ \colhead{No.} & \colhead{label}  & \colhead{$J$} & \colhead{P} & \colhead{$E$} }
	\startdata
   1 &$ 4f^{5} (^6_0H)~6s~^7H  $  & 2 &  $-$ &    0.00 \\ 
   2 &$ 4f^{5} (^6_0H)~6s~^7H  $  & 3 &  $-$ &  441.78 \\ 
   3 &$ 4f^{5} (^6_0H)~6s~^7H  $  & 4 &  $-$ & 1073.15 \\ 
   4 &$ 4f^{5} (^6_0H)~6s~^5H  $  & 3 &  $-$ & 1850.51 \\ 
   5 &$ 4f^{5} (^6_0H)~6s~^7H  $  & 5 &  $-$ & 1858.38 \\ 
   6 &$ 4f^{5} (^6_0H)~6s~^7H  $  & 6 &  $-$ & 2765.09 \\ 
   7 &$ 4f^{5} (^6_0H)~6s~^5H  $  & 4 &  $-$ & 2830.31 \\ 
   8 &$ 4f^{5} (^6_0H)~6s~^7H  $  & 7 &  $-$ & 3765.53 \\ 
   9 &$ 4f^{5} (^6_0H)~6s~^5H  $  & 5 &  $-$ & 3914.91 \\ 
  10 &$ 4f^{5} (^6_0H)~5d~^7K  $  & 4 &  $-$ & 4799.46 \\ 
  11 &$ 4f^{5} (^6_0H)~6s~^7H  $  & 8 &  $-$ & 4836.02 \\ 
  12 &$ 4f^{5} (^6_0H)~6s~^5H  $  & 6 &  $-$ & 5078.38 \\ 
  13 &$ 4f^{5} (^6_0H)~5d~^7K  $  & 5 &  $-$ & 5507.08 \\ 
  14 &$ 4f^{5} (^6_0H)~6s~^5H  $  & 7 &  $-$ & 6298.26 \\ 
  15 &$ 4f^{5} (^6_0H)~5d~^7K  $  & 6 &  $-$ & 6322.77 \\ 
  16 &$ 4f^{5} (^6_0F)~6s~^7F  $  & 0 &  $-$ & 6567.93 \\ 
  17 &$ 4f^{5} (^6_0F)~6s~^7F  $  & 1 &  $-$ & 6678.52 \\ 
  18 &$ 4f^{5} (^6_0F)~6s~^7F  $  & 2 &  $-$ & 6917.29 \\ 
  19 &$ 4f^{5} (^6_0H)~5d~^7K  $  & 7 &  $-$ & 7232.66 \\ 
  20 &$ 4f^{5} (^6_0F)~6s~^7F  $  & 3 &  $-$ & 7320.95 \\ 
				\enddata
\tablecomments{ Table~\ref{Pm_II_Energies} is published in its entirety in the machine-readable format. Part of the values are shown here for guidance regarding its form and content.}
\end{deluxetable}

\subsection{Energy level distribution for each configuration}
\label{E_l_d_e_c}

Identification of energy levels is 
a complicated task for lanthanides due to a mix of configurations. 
Even assigning particular configuration labeling to some levels is complicated. 
The discussion below should give enlightenment on the inner structure of the energy spectrum.
Energy levels have formed groups around parent level of $4f^N$ or $4f^{N-1}$ configurations with the same term of $f$ shell.
 Levels belonging to the different groups are separated by energy gaps. 
Below are given more details about these groups for each configuration.

Energy level structures for states of 
$4f^N \{6s,6p,5d\}$, $4f^{N-1} \{5d6s,5d6p,6s6p\}$, and $4f^{N-1} 5d^2$ 
configurations are presented in Figure \ref{all_configurations_energies}. 
Cut off line 80 700 cm$^{-1}$ (10 eV) is given by the horizontal lines. 
For the Pr~II and Nd II, computations are done up to ionizations limits: 
it is 85 745 cm$^{-1}$ for Pr~II and 86 970 cm$^{-1}$ for Nd II according to the NIST database.
 The number of computed levels 
are displayed below the line and the number of levels above the line are left uncomputed. 
The sum of these numbers comprise a possible 
number of levels in $jj$-coupling. 
We find that the increase of the nuclear charge has a small effect on 
the positions of first level relative to the ground state for the configurations $4f^N 6p$ and $4f^N 5d$.  
The energy level structures of these configurations are influenced by the structure of core [Xe]$4f^N$. 
Similar to the system difference analysed by \cite{Cowan},
the increase of the energy of first level relative to the ground state is found
for the configurations of $4f^{N-1} 5d6s$, $4f^{N-1} 5d6p$, $4f^{N-1} 6s6p$, and $4f^{N-1} 5d^2$
(see Figure \ref{Kowan_type} for $4f^{N-1} 5d6s$).
The highest density of the energy levels are found for $4f^{N-1} 5d^2$ and $4f^{N-1} 5d6p$ configurations.

\begin{deluxetable}{rlrrr}
	\tabletypesize{\footnotesize}
	\setlength{\tabcolsep}{10pt}
	\tablecaption{
Energy Levels (in cm$^{-1}$) Relative to the Ground State for the States of Sm~II. \label{Sm_II_Energies}}
	\tablehead{ \colhead{No.} & \colhead{label}  & \colhead{$J$} & \colhead{P} & \colhead{$E$} }
	\startdata
    1 &$ 4f^{6} (^7_0F)~6s~^8F $   &   1/2 & $ + $ &    0.00 \\
    2 &$ 4f^{6} (^7_0F)~6s~^8F $   &   3/2 & $ + $ &  296.55 \\
    3 &$ 4f^{6} (^7_0F)~6s~^8F $   &   5/2 & $ + $ &  765.67 \\
    4 &$ 4f^{6} (^7_0F)~6s~^8F $   &   7/2 & $ + $ & 1372.98 \\
    5 &$ 4f^{6} (^7_0F)~6s~^6F $   &   1/2 & $ + $ & 1853.48 \\
    6 &$ 4f^{6} (^7_0F)~6s~^8F $   &   9/2 & $ + $ & 2084.20 \\
    7 &$ 4f^{6} (^7_0F)~6s~^6F $   &   3/2 & $ + $ & 2285.57 \\
    8 &$ 4f^{6} (^7_0F)~6s~^8F $   &  11/2 & $ + $ & 2870.66 \\
    9 &$ 4f^{6} (^7_0F)~6s~^6F $   &   5/2 & $ + $ & 2918.77 \\
   10 &$ 4f^{6} (^7_0F)~6s~^6F $   &   7/2 & $ + $ & 3688.53 \\
   11 &$ 4f^{6} (^7_0F)~6s~^8F $   &  13/2 & $ + $ & 3709.70 \\
   12 &$ 4f^{6} (^7_0F)~6s~^6F $   &   9/2 & $ + $ & 4548.72 \\
   13 &$ 4f^{6} (^7_0F)~6s~^6F $   &  11/2 & $ + $ & 5467.12 \\
   14 &$ 4f^{6} (^7_0F)~5d~^8H $   &   3/2 & $ + $ & 6571.55 \\
   15 &$ 4f^{6} (^7_0F)~5d~^8H $   &   5/2 & $ + $ & 6913.83 \\
   16 &$ 4f^{6} (^7_0F)~5d~^8H $   &   7/2 & $ + $ & 7375.65 \\
   17 &$ 4f^{6} (^7_0F)~5d~^8H $   &   9/2 & $ + $ & 7942.79 \\
   18 &$ 4f^{6} (^7_0F)~5d~^8D $   &   3/2 & $ + $ & 8488.11 \\
   19 &$ 4f^{6} (^7_0F)~5d~^8H $   &  11/2 & $ + $ & 8600.85 \\
   20 &$ 4f^{6} (^7_0F)~5d~^8D $   &   5/2 & $ + $ & 9058.38 \\
				\enddata
\tablecomments{ Table~\ref{Sm_II_Energies} is published in its entirety in the machine-readable format. Part of the values are shown here for guidance regarding its form and content.}
\end{deluxetable}

The lowest levels of $4f^N 6s$ and $4f^N 6p$ configuration 
form blocks of energy levels around the parent levels of $4f^N$ 
($^4_1I^*$), 
($^5_1I^*$),  
($^6_0H$ and $^6_0F^*$), 
($^7_0F^*$), 
($^8_0S^*$, $^6_0P$, $^6_0I^*$, and $^6_0D^*$), and 
($^7_0F^*$) 
for $Z=59 - 64$, respectively. 
After the levels with core configuration marked by "$^*$" above,
there is an energy gap, except for $4f^3 6p$ of Pr~II ion.
Levels with the specific parent levels do not mix with others,
except for the parent level states of $4f^5 6p$ of Pm~II ion ($4f^5~^6_0H$ mix between $4f^5~^6_0F^*$). 
For $4f^N 5d$ configuration, the situation is different 
because of the strong interaction between $4f$ and $5d$ (Figure \ref{all_configurations_energies}). 

\begin{deluxetable}{rlrrr}
	\tabletypesize{\footnotesize}
	\setlength{\tabcolsep}{10pt}
	\tablecaption{
Energy Levels (in cm$^{-1}$) Relative to the Ground State for the States of Eu~II. \label{Eu_II_Energies}}
	\tablehead{ \colhead{No.} & \colhead{label}  & \colhead{$J$} & \colhead{P} & \colhead{$E$} }
	\startdata
   1 &$ 4f^7 (^8_0S)~6s~^9S  $&  4  &$ - $&     0.00  \\ 
   2 &$ 4f^7 (^8_0S)~6s~^7S  $&  3  &$ - $&  2057.90  \\ 
   3 &$ 4f^7 (^8_0S)~5d~^9D  $&  2  &$ - $& 10657.96  \\ 
   4 &$ 4f^7 (^8_0S)~5d~^9D  $&  3  &$ - $& 10784.18  \\ 
   5 &$ 4f^7 (^8_0S)~5d~^9D  $&  4  &$ - $& 10964.35  \\ 
   6 &$ 4f^7 (^8_0S)~5d~^9D  $&  5  &$ - $& 11212.47  \\ 
   7 &$ 4f^7 (^8_0S)~5d~^9D  $&  6  &$ - $& 11551.69  \\ 
   8 &$ 4f^7 (^8_0S)~5d~^7D  $&  5  &$ - $& 18922.41  \\ 
   9 &$ 4f^7 (^8_0S)~5d~^7D  $&  4  &$ - $& 18964.72  \\ 
  10 &$ 4f^7 (^8_0S)~5d~^7D  $&  3  &$ - $& 19032.93  \\ 
  11 &$ 4f^7 (^8_0S)~5d~^7D  $&  2  &$ - $& 19095.86  \\ 
  12 &$ 4f^7 (^8_0S)~5d~^7D  $&  1  &$ - $& 19143.88  \\ 
  13 &$ 4f^7 (^8_0S)~6p~^9P  $&  3  &$ + $& 21378.10  \\ 
  14 &$ 4f^7 (^8_0S)~6p~^9P  $&  4  &$ + $& 21708.29  \\ 
  15 &$ 4f^7 (^8_0S)~6p~^9P  $&  5  &$ + $& 23385.62  \\ 
  16 &$ 4f^7 (^8_0S)~6p~^7P  $&  4  &$ + $& 23999.79  \\ 
  17 &$ 4f^7 (^8_0S)~6p~^7P  $&  3  &$ + $& 24276.27  \\ 
  18 &$ 4f^7 (^8_0S)~6p~^7P  $&  2  &$ + $& 24446.93  \\ 
  19 &$ 4f^7 (^6_0P)~6s~^7P  $&  4  &$ - $& 32530.49  \\ 
  20 &$ 4f^7 (^6_0P)~6s~^7P  $&  3  &$ - $& 32852.51  \\ 
				\enddata
\tablecomments{ Table~\ref{Eu_II_Energies} is published in its entirety in the machine-readable format.
 Part of the values are shown here for guidance regarding its form and content.}
\end{deluxetable}

For $4f^{N-1} 5d6s$ configuration, groups of energy levels are formed around 
the lowest parent levels for only two elements i.e., Eu~II and Gd~II.
These parent levels are $4f^6~^7_0F$ and $4f^7~^8_0S^*$ for Eu~II and Gd~II, respectively.
For $4f^{N-1} 5d6p$ configuration, only for Gd~II has formed a group of energy levels around $4f^7~^8_0S^*$ parent level 
(Figure \ref{all_configurations_energies}). 

Levels of $4f^{N-1} 6s6p$ configuration do not 
 form group of energy levels around the parent levels. 
For Eu~II, all levels of $4f^6 6s6p$ and of $4f^6 5d6p$ 
configurations belong to the parent levels $4f^6~^7_0F$, because of the 10 eV cut off
(Figure \ref{all_configurations_energies}).
For $4f^{N-1} 5d^2$ configuration, groups of energy levels forms around 
the lowest parent levels for Gd~II $4f^8~^8_0S^*$ (Figure \ref{all_configurations_energies}).

\begin{deluxetable}{rlrrr}
	\tabletypesize{\footnotesize}
	\setlength{\tabcolsep}{10pt}
	\tablecaption{
Energy Levels (in cm$^{-1}$) Relative to the Ground State for the States of Gd~II. \label{Gd_II_Energies}}
	\tablehead{ \colhead{No.} & \colhead{label}  & \colhead{$J$} & \colhead{P} & \colhead{$E$} }
	\startdata
   1 & $4f^7 (^8_0S)~5d~^9D 6s~^1D    $ &  5/2 & $ - $ &    0.00 \\
   2 & $4f^7 (^8_0S)~5d~^9D 6s~^1D    $ &  7/2 & $ - $ &  225.02 \\
   3 & $4f^7 (^8_0S)~5d~^9D 6s~^1D    $ &  9/2 & $ - $ &  536.90 \\
   4 & $4f^7 (^8_0S)~5d~^9D 6s~^1D    $ & 11/2 & $ - $ &  959.98 \\
   5 & $4f^7 (^8_0S)~5d~^9D 6s~^1D    $ & 13/2 & $ - $ & 1536.81 \\
   6 & $4f^7 (^8_0S)~5d^2 (^3_2F)~^1F $ &  3/2 & $ - $ & 3026.08 \\
   7 & $4f^7 (^8_0S)~5d^2 (^3_2F)~^1F $ &  5/2 & $ - $ & 3173.49 \\
   8 & $4f^7 (^8_0S)~5d^2 (^3_2F)~^1F $ &  7/2 & $ - $ & 3382.26 \\
   9 & $4f^7 (^8_0S)~5d^2 (^3_2F)~^1F $ &  9/2 & $ - $ & 3654.80 \\
  10 & $4f^7 (^8_0S)~5d~^9D 6s~^8D    $ &  3/2 & $ - $ & 3767.49 \\
  11 & $4f^7 (^8_0S)~5d~^9D 6s~^8D    $ &  5/2 & $ - $ & 3965.84 \\
  12 & $4f^7 (^8_0S)~5d^2 (^3_2F)~^1F $ & 11/2 & $ - $ & 3994.44 \\
  13 & $4f^7 (^8_0S)~5d~^9D 6s~^8D    $ &  7/2 & $ - $ & 4269.62 \\
  14 & $4f^7 (^8_0S)~5d^2 (^3_2F)~^1F $ & 13/2 & $ - $ & 4405.08 \\
  15 & $4f^7 (^8_0S)~5d~^9D 6s~^8D    $ &  9/2 & $ - $ & 4713.57 \\
  16 & $4f^7 (^8_0S)~5d^2 (^3_2F)~^1F $ & 15/2 & $ - $ & 4888.14 \\
  17 & $4f^7 (^8_0S)~5d~^9D 6s~^8D    $ & 11/2 & $ - $ & 5359.77 \\
  18 & $4f^7 (^8_0S)~6s^2~^8S         $ &  7/2 & $ - $ & 5713.88 \\
  19 & $4f^8 (^7_0F)~6s~^8F           $ & 13/2 & $ + $ & 7507.66 \\
  20 & $4f^8 (^7_0F)~6s~^8F           $ & 11/2 & $ + $ & 8691.73 \\
				\enddata
\tablecomments{ Table~\ref{Gd_II_Energies} is published in its entirety in the machine-readable format.
 Part of the values are shown here for guidance regarding its form and content.}
\end{deluxetable}

Radii of the orbitals of the configuration $4f^N 6s$ and $4f^{N-1} 5d6s$ are presented in Table \ref{radius}. 
For higher $Z$, all orbitals contract (see Table \ref{radius}).
The exception is Eu~II and Gd~II: there is no big differences for $<5d\_>$ and $<5d>$ orbitals between Eu~II and Gd~II. 
Indeed, for Gd~II, the radii for orbitals $<5d\_>$ and $<5d>$ show small increase with respect to 
Eu~II. This may be caused by different computation of the radial wave functions (see section \ref{sec:scheme}).
Some of the radii are compared with computations by \citet{Indelicato}. 
Radii by \citet{Indelicato} differ from 1 to 8\% from those computed in this paper.
It is likely that these differences are caused by inclusion of Breit interaction into the self-consistent field procedure 
in the MCDHF computations.

\begin{figure}
 \includegraphics[width=0.47\textwidth]{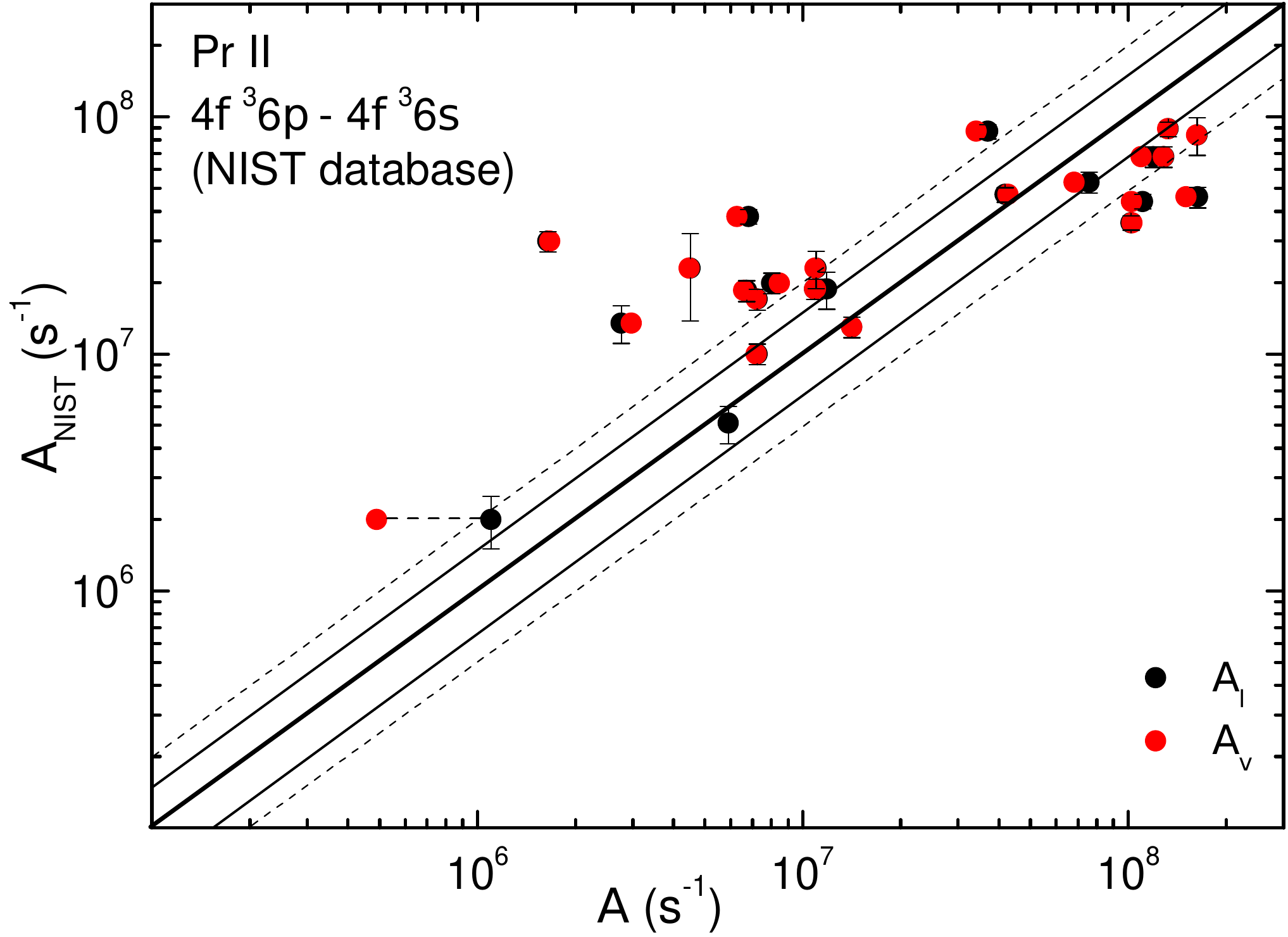}
 \includegraphics[width=0.47\textwidth]{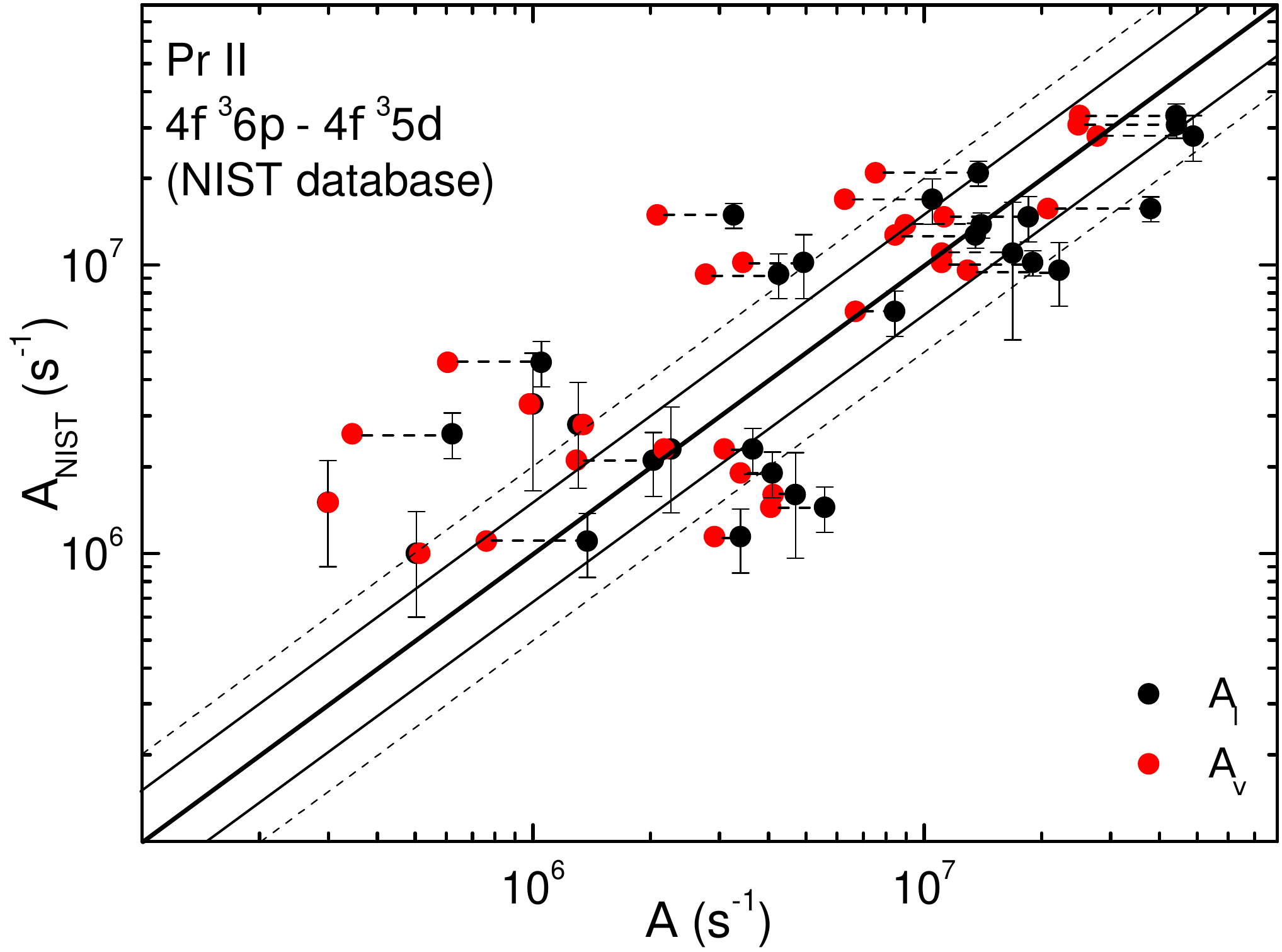}
	\caption{Comparison of transition probability between states of configurations $4f^3~6s$ - $4f^3~6p$ 
	and between states of configurations $4f^3~6p$ and $4f^3~5d$ for Pr~II. The top and bottom
panels show a comparison between our results and
results from the NIST database. The thick line corresponds to perfect agreement, while the thin
solid and dashed lines correspond to deviations by factors of 1.5 and 2.0,
respectively. The black and reds points show the values calculated with the length
(Babushkin) and velocity (Coulomb) forms, respectively.
	\label{Pr_II_6p-5d_transitions}}
\end{figure}

\section{E1 transitions}
\label{sec:transitions}

In this section, we show the results of our calculations of transition probabilities.
The transition data computed for Pr~II, Pm~II, Sm~II, Eu~II, and Gd~II are given in machine-readable format in Tables
\ref{Pr_II_transitions}, \ref{Pm_II_transitions}, \ref{Sm_II_transitions}, \ref{Eu_II_transitions}, and \ref{Gd_II_transitions}. 
The tables include identification of upper and lower levels in $LSJ$ coupling, 
transition energy, wavelength, line strength, weighted oscillator strength, 
and transition probabilities in length form. 
The numbers of transitions are 411~314, 7~104~005, 4~720~626, 467~724 (plus 13~154 transitions with Rydberg states, 480~878 in total), and 1 383 694‬ for Pr~II, Pm~II, Sm~II, Eu~II, and Gd~II, respectively.

In the following sections, we compare the calculated transition probabilities with available data, 
except for Pm and Sm for which enough data are not available
\footnote{For Sm~II, there are transitions probabilities for 7 lines in the NIST database. 
Unfortunately upper levels do not have clear identification of the configuration. 
\cite{7630TP} have performed radiative lifetime measurements with time-resolved laser induced
fluorescence (LIF) techniques for 47 levels and 
have performed relativistic Hartree-Fock (HFR) computations over the energy 
range {21 000} - {36 000} cm$^{-1}$, but again the identification of these levels is unclear. 
Large amount of data (958 lines) have been measured with the same method \citep{Lawler}, 
but all upper levels do not have clear identification.
Lifetimes of 82 levels in range 21 655.420 - 29 591.120 cm $^{-1}$ were investigated beam-laser method and 
transition probabilities were calculated using branching ratios for 35 transitions by \citet{Scholl_sm}.
}.

\subsection{Pr~II}
For Pr~II, rather rich data are available in the NIST database. 
Therefore, it can be used as evaluation of our calculations.
Comparison between the calculated E1 transitions probabilities and those in the NIST database 
is presented in Figure \ref{Pr_II_6p-5d_transitions}. 
Figure \ref{Pr_II_6p-5d_transitions} includes transitions between $4f^3~6s$ and $4f^3~6p$ 
and transitions between $4f^3~6p$ and $4f^3~5d$ with clear level identification.
The same transitions in length and velocity form are connected with dashed lines.
Transitions in the NIST database are based on FT spectroscopy by \citet{Ivarsoon_Pr} and 
measurements of branching fractions with use of a laser/fast-ion-beam
method by \citet{Li_Pr_transitions} and lifetimes determined in a previous study 
with beam-laser method \citep{Li_Pr_transitions_lifetime}. 

We find that transition probabilities calculated in two forms 
agree better for the transitions between $4f^3~6s$ and $4f^3~6p$ than 
those between $4f^3~6p$ and $4f^3~5d$. 
Compared with the data by other authors, 
our transitions in velocity form gives a better agreement in the strong transition area. 
Therefore, hereafter we show transition probabilities computed in velocity form.

\begin{figure}
 \includegraphics[width=0.47\textwidth]{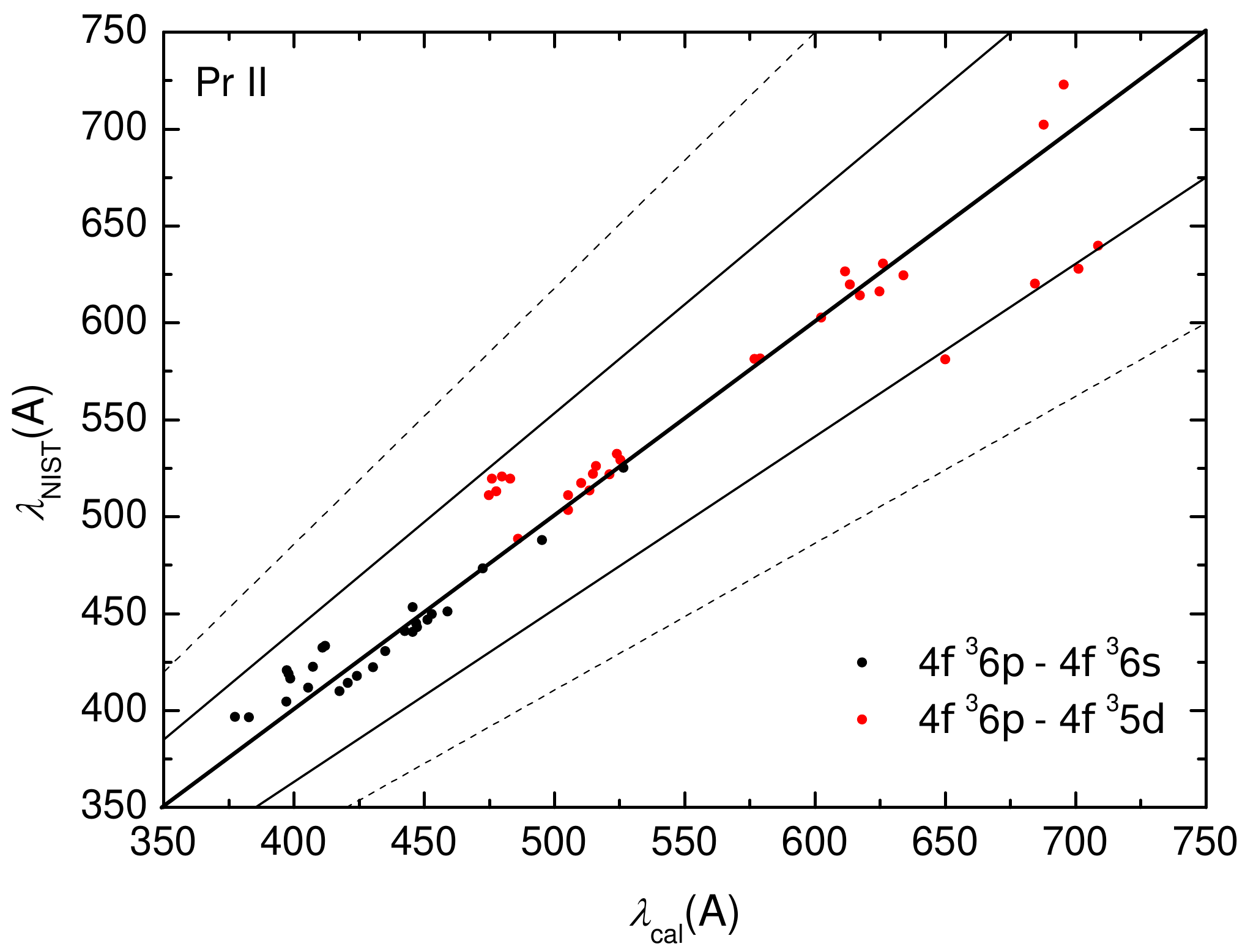}
	\caption{Comparison of transition wavelengths for Pr~II between our results ($\lambda_{cal}$)
	and NIST database recommended values ($\lambda_{NIST}$). The
thick line corresponds to perfect agreement, while thin solid and dashed lines
correspond to 10\% and 20\% deviations.
	\label{Pr_II_lambda}}
\end{figure}

As for the transition wavelength, our calculations give a good agreement with the NIST data. 
Averaged agreement in the transition wavelength is 2\% for 
the transitions between states of configurations $4f^3~6s$ and $4f^3~6p$, and 
4\% for the transitions between states of configurations $4f^3~6p$ and $4f^3~5d$ (see Figure \ref{Pr_II_lambda}). 

\begin{figure}
 \includegraphics[width=0.47\textwidth]{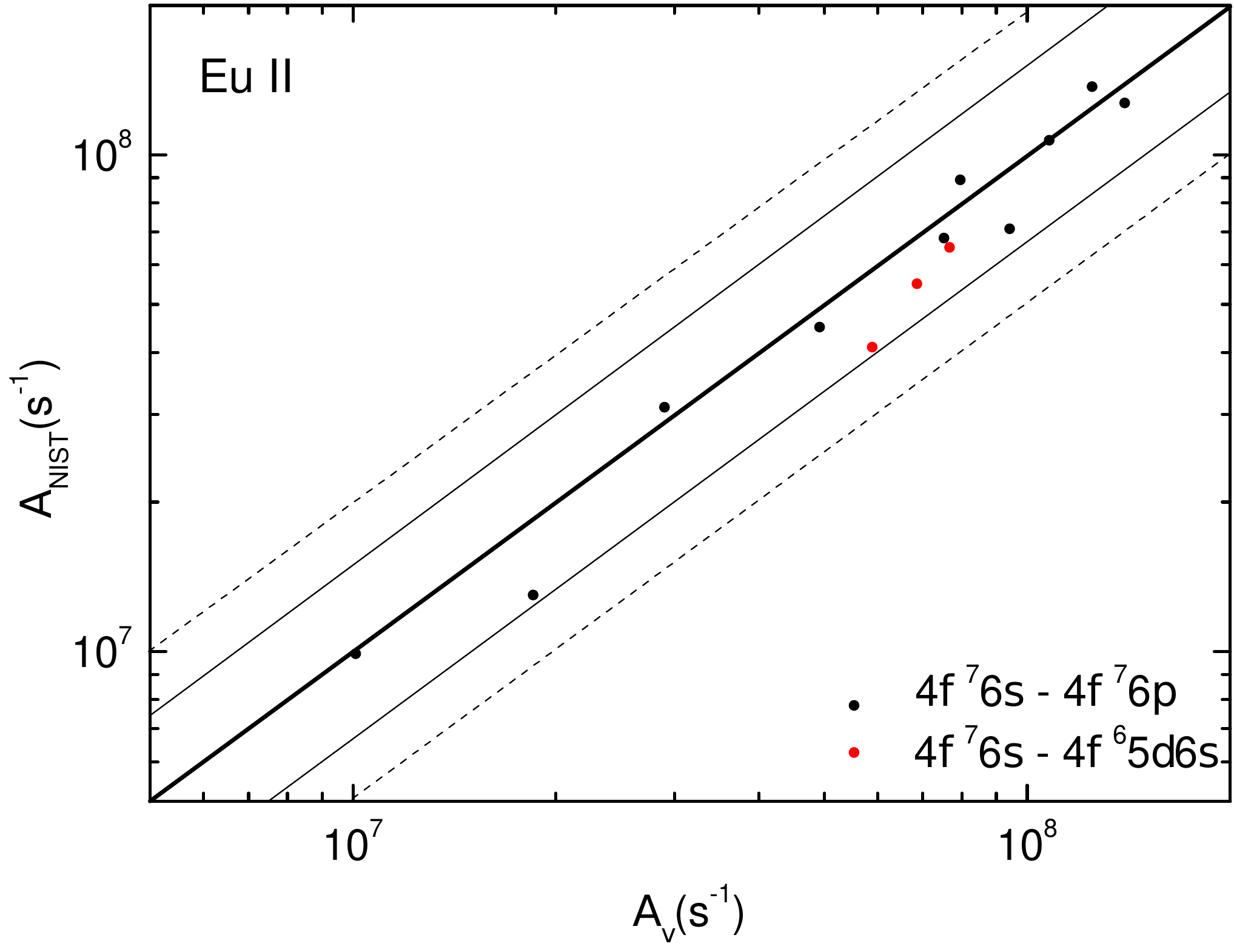}
	\caption{Comparison of transition probability of Eu~II between our results ($A_{v}$) and NIST database.
 The thick line corresponds to perfect agreement, while the thin
solid and dashed lines correspond to deviations by factors of 1.5 and 2.0,
respectively. Transition probabilities are presented in the velocity (Coulomb) form.
	\label{Eu_transitions_A}}
\end{figure}

\begin{deluxetable}{r c c c c }[ht!!]
\tablecaption{\label{summary_tran} Summary of Experiments on Eu~II.}
\tablehead{
References        &N$_{\tau}$    & Method$_{\tau}$& Method$_{BF}$& N$_L$ } 
\startdata
\citet{Biemont}    & --& TR-LIF   & HCL   & 27\\
\citet{Zhang_Eu}   &  9& TR-LIF   & HCL   & 31\\
\citet{Lawer_Eu_A} &  6& TR-LIF   & FTS   & 24\\
\citet{Wang_Eu}    & 30& TR-LIF   & HCL   & 18\\ 
\citet{Than_Eu}    & 11& TR-LIF   & HCL   & 24\\ 
\hline
\enddata 
\end{deluxetable}


\subsection{Eu~II}
The NIST database presents 13 lines with transition probabilities
which are compared with our calculations in Figure \ref{Eu_transitions_A}.
There is a very good agreement of transitions probabilities although the agreement in the transition wavelength
is rather poor, about 14\%.

\begin{figure}
 \includegraphics[width=0.47\textwidth]{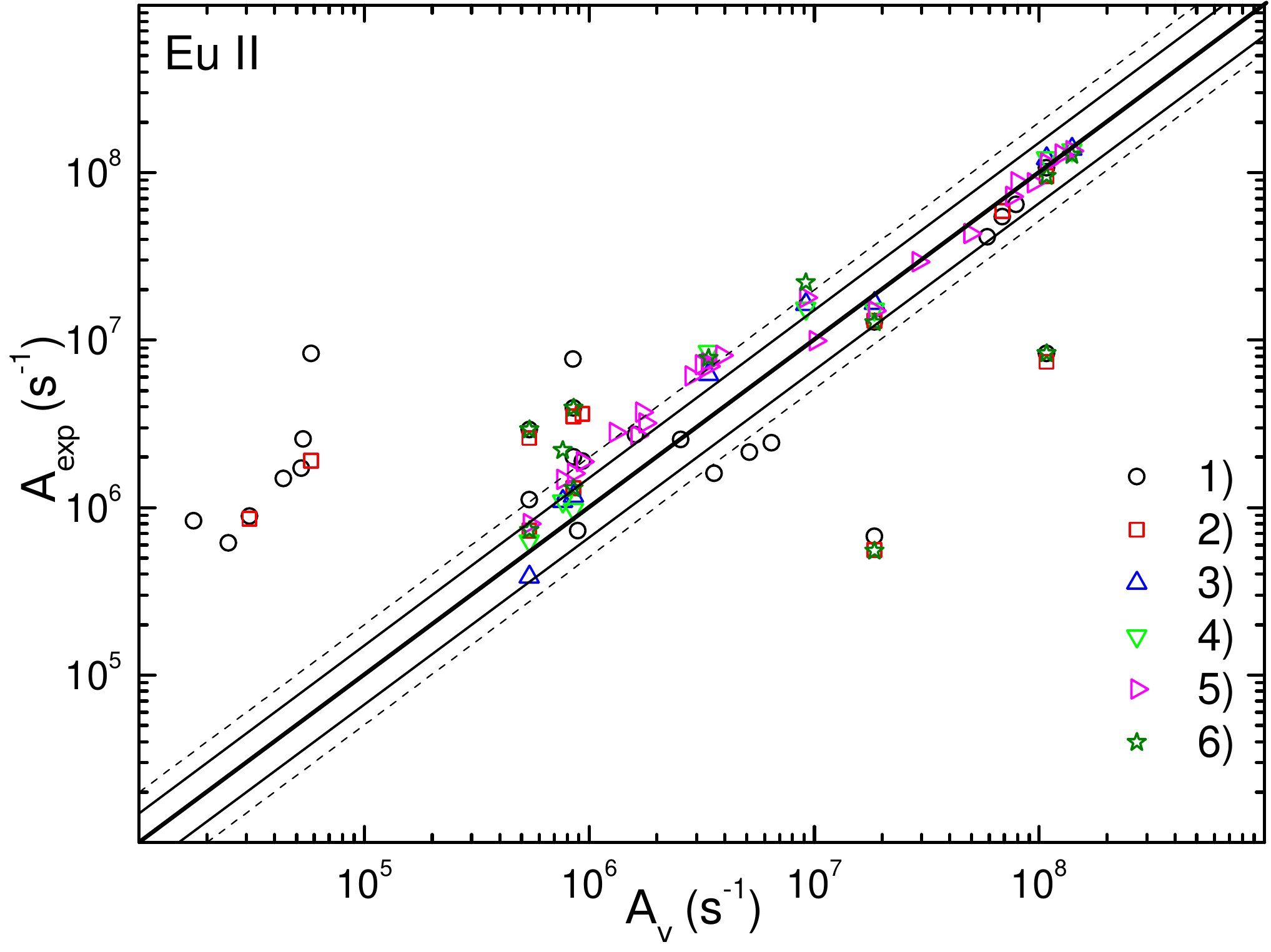}
	\caption{Comparison of transition probability of Eu~II between our results ($A_{v}$) and data of other authors ($A_{exp}$):
1) \citet{Zhang_Eu}; 2) \citet{Komarovskii}; 3) \citet{Than_Eu}; 4) \citet{Wang_Eu}; 5) \citet{Lawer_Eu_A};
6) \citet{Karner} and \citet{Biemont}.
 The thick line corresponds to perfect agreement, while the thin
solid and dashed lines correspond to deviations by factors of 1.5 and 2.0,
respectively. 
Transition probabilities are presented in the velocity (Coulomb) form.
	\label{Eu_transitions_A_authors}}
\end{figure}

It is worth comparing our results with more available measurements although the data are 
not always critically evaluated. Summary of experiments for Eu~II is given in Table \ref{summary_tran}.
Absolute transitions probabilities are measured experimentally
through the measurements of lifetimes ($\tau$) and branching fractions (BF) by other authors. 
Measurements for the lifetime are done using time-resolved laser-induced fluorescence (TR-LIF)
while branching factors are estimated from emission spectra of a hollow-cathode discharge lamp
with Eu powder in the cathode (HCL) or Fourier transform spectrometer (FTS) data. 
Table \ref{summary_tran} includes the methods as well as the number of lifetimes measurements N$_{\tau}$ 
and the number of lines N$_{L}$.


Comparison with these measurement is given in Figure \ref{Eu_transitions_A_authors}. 
In this figure, only the levels with clear identifications are included. 
The most transitions are in the ranges of dashed lines showing the deviation by a factor of 2.0.
However, we observe a relatively large deviation
in the weak transitions: our calculations give a much smaller transition probabilities than 
those estimated from the experiments. This may suggest that our strategy of computations is not good enough
for weak transitions. Another possible reason is that transitions other than E1, 
which we do not include in our calculations, may contribute to these weak lines.


\begin{figure}
 \includegraphics[width=0.47\textwidth]{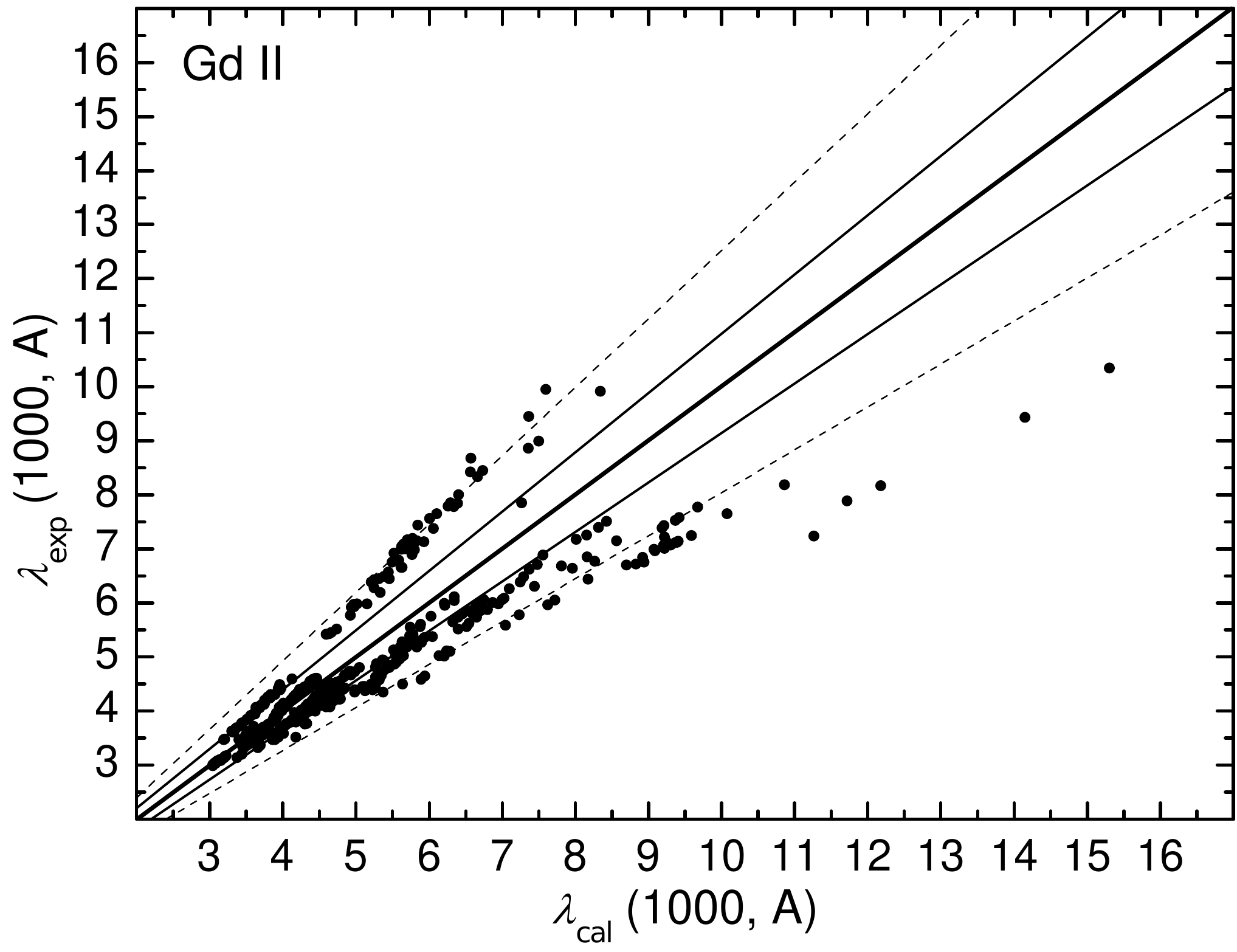}
	\caption{Comparison of transition wavelengths for Gd~II between our results 
	and experimental data by \citet{Hartog}. The
thick line corresponds to perfect agreement, while thin solid and dashed lines
correspond to 10\% and 20\% deviations.
	\label{Gd_II_ambda}}
\end{figure}	

\begin{figure}
\includegraphics[width=0.47\textwidth]{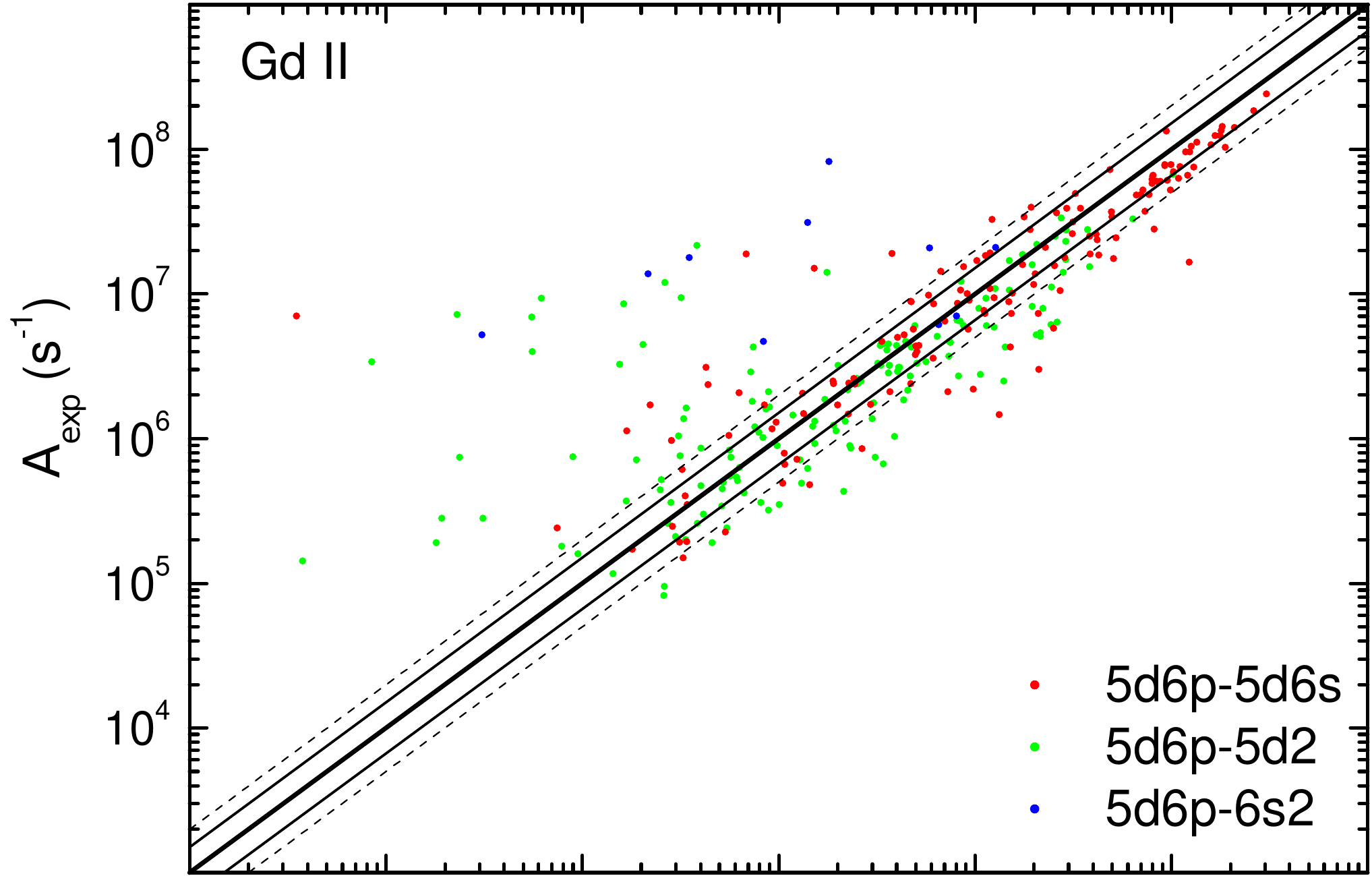}
\includegraphics[width=0.47\textwidth]{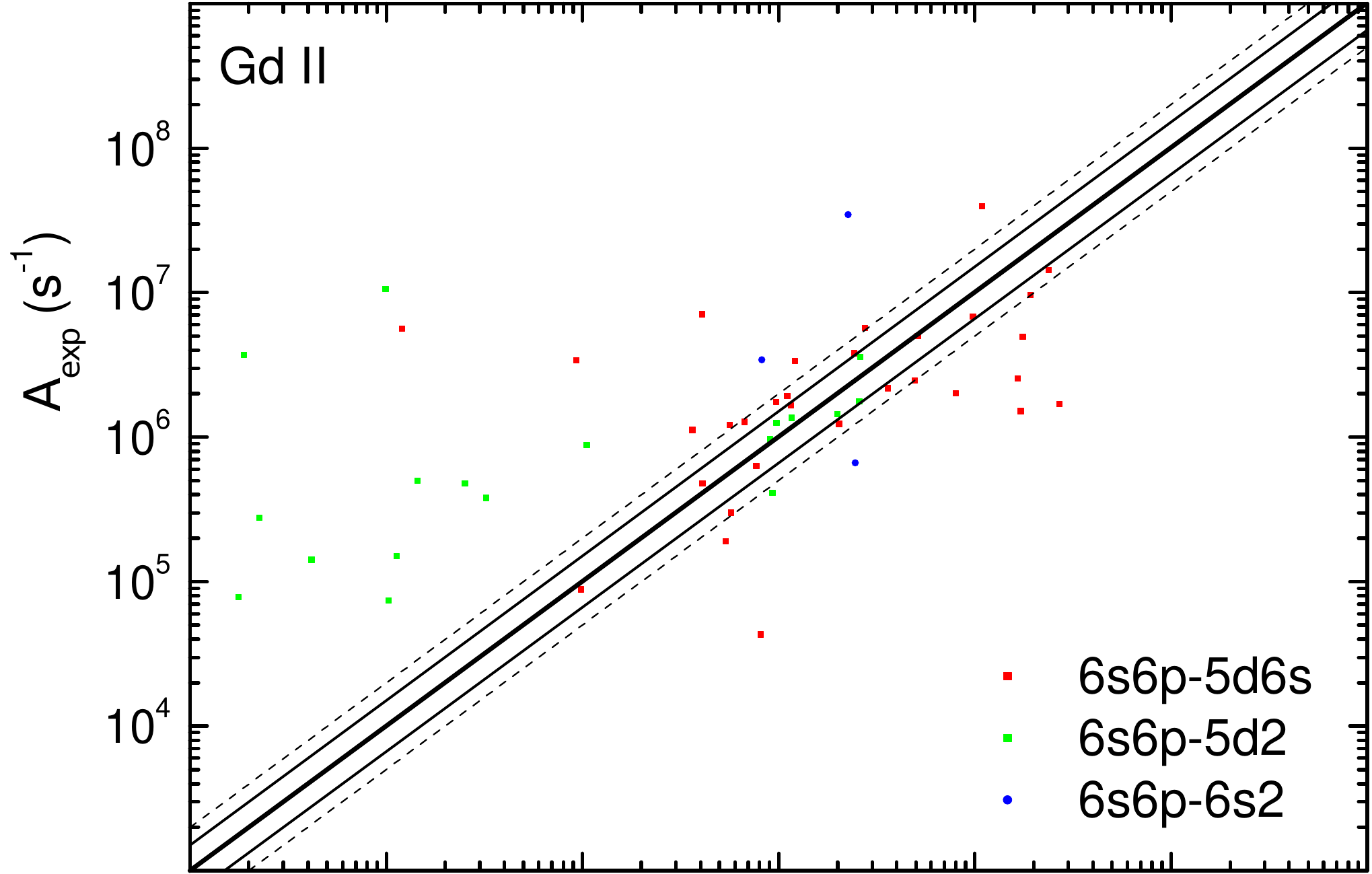}
\includegraphics[width=0.47\textwidth]{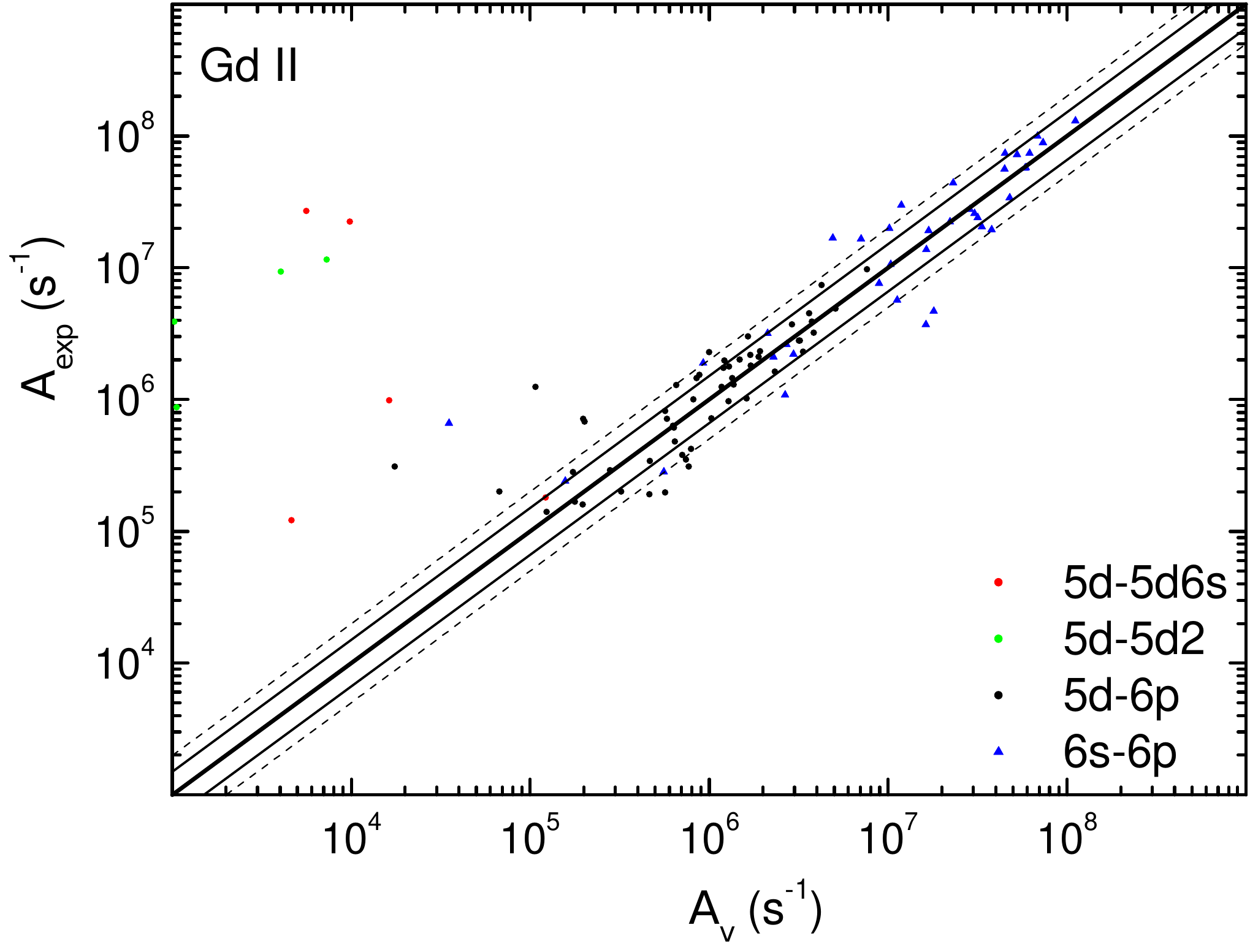}
\caption{Comparison of transition probability of Gd~II between our results ($A_{v}$) and
the results by \citet{Hartog} ($A_{exp}$). 
Three panels are divided according to the configurations involved in the transitions.
The thick line corresponds to perfect agreement, while the thin
solid and dashed lines correspond to deviations by factors of 1.5 and 2.0,
respectively. Transition probabilities are presented in the velocity (Coulomb) form.
	\label{Gd_II_transitions_III}}
\end{figure}

\subsection{Gd~II}

For Gd~II, transitions probabilities are not presented in the NIST database \citep{NIST}.
However, there are several experimental works to address the transition probabilities.
For example, experimental transition probabilities are estimated by \cite{Corliss}.
Also, \citet{Wang_Gd_II_transitions} have experimentally measured branching fractions of 12 levels for Gd~II
using the emission spectrum of a hollow cathode lamp. 
As a results, transition probabilities for 74 lines of Gd~II were derived 
from a combination of the radiative lifetimes reported in the earlier literature 
and newly determined branching fractions. 

More recently, \citet{Hartog} have investigated absolute transition probabilities for 611 lines for Gd~II,
by using combination of LIF radiative lifetime measurements and branching fraction measurements.
Identification of upper and lower energy levels is based on the work by \citet{Martin}. 
In Figure \ref{Gd_II_ambda}, wavelengths of 460 transitions from their experiments are compared with
our calculations. For comparison, we include only the levels with clear identification.
66\% of lines wavelengths are within 10\% agreement range (solid lines) and 12\% of wavelengths have more than 20\% disagreement  (dashed lines).

\begin{deluxetable*}{llrrrrrr}[ht!!]
	\tabletypesize{\footnotesize}
	\setlength{\tabcolsep}{10pt}
	\tablecaption{Transition Energies $\Delta$$E$ (in cm$^{-1}$),
	Transition Wavelengths $\lambda$ (in \AA), 
	Line Strengths $S$ (in a.u.), 
	Weighted Oscillator Strengths $gf$, 
	and Transition Rates $A$ (in s$^{-1}$) for
E1 Transitions of the Pr~II ion. \label{Pr_II_transitions}}
	\tablehead{ \colhead{Lower state} 
	& \colhead{Upper state}  
	& \colhead{$\Delta E$ (cm$^{-1}$)} & \colhead{$\lambda$ (\AA)} & \colhead{$S$} & \colhead{$gf$} & \colhead{$A$ (s$^{-1}$)} & \colhead{$dT$}}
	\startdata
$4f^2 (^3_1H)\,5d^2 (^3_2F)~^5D_{1}$ ~  & $4f^3 (^4_1F)\,5d~^5D_{0}$           ~  &     2769 &        36101 & 1.197D-01 & 1.007D-03 & 5.157D+03 &   0.872\\
$4f^2 (^3_1H)\,5d^2 (^3_2F)~^5F_{1}$ ~  & $4f^3 (^4_1F)\,5d~^5D_{0}$           ~  &      553 &       180550 & 5.998D-02 & 1.009D-04 & 2.065D+01 &   0.997\\
$4f^3 (^4_1F)\,5d~^5D_{0}$           ~  & $4f^2 (^3_1F)\,5d^2 (^3_2F)~^5F_{1}$ ~  &      936 &       106828 & 2.116D-01 & 6.018D-04 & 1.172D+02 &   0.266\\
$4f^3 (^4_1F)\,5d~^5D_{0}$           ~  & $4f^2 (^3_1H)\,5d^2 (^3_2F)~^3D_{1}$ ~  &     1618 &        61775 & 6.203D-03 & 3.050D-05 & 1.777D+01 &   0.972\\
$4f^3 (^4_1F)\,5d~^5D_{0}$           ~  & $4f^2 (^3_1F)\,5d^2 (^3_2F)~^3P_{1}$ ~  &     2826 &        35384 & 4.636D-03 & 3.979D-05 & 7.067D+01 &   0.992\\
$4f^3 (^4_1F)\,5d~^5D_{0}$           ~  & $4f^2 (^3_1F)\,5d^2 (^3_2F)~^5P_{1}$ ~  &     3885 &        25734 & 6.263D-02 & 7.393D-04 & 2.482D+03 &   0.857\\
$4f^3 (^4_1F)\,5d~^5D_{0}$           ~  & $4f^2 (^3_1F)\,5d^2 (^3_2F)~^5D_{1}$ ~  &     4313 &        23184 & 1.093D-02 & 1.432D-04 & 5.927D+02 &   0.915\\
$4f^3 (^4_1F)\,5d~^5D_{0}$           ~  & $4f^2 (^3_1F)\,5d^2 (^3_2F)~^5P_{1}$ ~  &     4739 &        21100 & 5.712D-02 & 8.223D-04 & 4.106D+03 &   0.854\\
$4f^3 (^4_1F)\,5d~^5D_{0}$           ~  & $4f^2 (^3_1F)\,5d^2 (^3_2F)~^3S_{1}$ ~  &     6623 &        15096 & 2.818D-03 & 5.670D-05 & 5.532D+02 &   0.395\\
$4f^3 (^4_1F)\,5d~^5D_{0}$           ~  & $4f^2 (^1_1G)\,5d^2 (^3_2F)~^3D_{1}$ ~  &     7603 &        13151 & 5.877D-03 & 1.357D-04 & 1.745D+03 &   0.883\\
$4f^3 (^4_1F)\,5d~^5D_{0}$           ~  & $4f^2 (^3_1F)\,5d^2 (^3_2F)~^1P_{1}$ ~  &     7933 &        12604 & 3.494D-03 & 8.422D-05 & 1.178D+03 &   0.349\\
$4f^3 (^4_1F)\,5d~^5D_{0}$           ~  & $4f^2 (^3_1F)\,5d^2 (^3_2P)~^5F_{1}$ ~  &     8216 &        12171 & 3.326D-03 & 8.300D-05 & 1.245D+03 &   0.964\\
$4f^3 (^4_1F)\,5d~^5D_{0}$           ~  & $4f^2 (^3_1F)\,5d^2 (^1_2D)~^3P_{1}$ ~  &     8397 &        11908 & 8.419D-03 & 2.147D-04 & 3.367D+03 &   0.148\\
$4f^3 (^4_1F)\,5d~^5D_{0}$           ~  & $4f^2 (^3_1F)\,5d^2 (^3_2F)~^3D_{1}$ ~  &     8690 &        11507 & 4.930D-04 & 1.301D-05 & 2.185D+02 &   0.965\\
$4f^3 (^4_1F)\,5d~^5D_{0}$           ~  & $4f^2 (^1_1G)\,5d^2 (^3_2F)~^3P_{1}$ ~  &    11098 &         9010 & 1.194D-03 & 4.027D-05 & 1.102D+03 &   0.577\\
$4f^3 (^4_1F)\,5d~^5D_{0}$           ~  & $4f^2 (^3_1F)\,5d^2 (^3_2P)~^5D_{1}$ ~  &    12485 &         8009 & 7.617D-03 & 2.888D-04 & 1.001D+04 &   0.750\\
$4f^3 (^4_1F)\,5d~^5D_{0}$           ~  & $4f^2 (^3_1F)\,5d^2 (^1_2G)~^3P_{1}$ ~  &    12982 &         7702 & 3.509D-03 & 1.383D-04 & 5.186D+03 &   0.992\\
$4f^3 (^4_1F)\,5d~^5D_{0}$           ~  & $4f^2 (^3_1H)\,5d^2 (^1_2G)~^3D_{1}$ ~  &    13084 &         7642 & 5.366D-03 & 2.133D-04 & 8.119D+03 &   0.234\\
$4f^3 (^4_1F)\,5d~^5D_{0}$           ~  & $4f^2 (^3_1F)\,5d^2 (^3_2P)~^3D_{1}$ ~  &    14340 &         6973 & 9.093D-05 & 3.961D-06 & 1.811D+02 &   0.991\\
$4f^3 (^4_1F)\,5d~^5D_{0}$           ~  & $4f^2 (^3_1H)\,5d^2 (^1_2G)~^3P_{1}$ ~  &    15904 &         6287 & 4.553D-03 & 2.199D-04 & 1.237D+04 &   0.993\\
				\enddata
\tablecomments{ Table~\ref{Pr_II_transitions} is published in its entirety in the machine-readable format.
All transition data are in length form.
 Part of the values are shown here for guidance regarding its form and content.}
\end{deluxetable*}

As for the transition probabilities, 
we obtain reasonable agreement between our computed values and the LIF measurements
(Figure \ref{Gd_II_transitions_III}, colors of the points represent different configurations). 
In this figure, we include transitions with transition probabilities higher than 10$^3$
from \citet{Hartog}.
At closer look, however, there is disagreement in particular for the two-electron-one-photon transitions 
between states of configurations $4f^7 5d6p$ and $4f^7 6s^2$ and 
$4f^7 6s6p$ and $4f^7 5d^2$. 
Our calculations underestimate the experimental values of these transitions.
These transitions are due to mixing of configurations in the ASFs which allows one-electron-one-photon transitions  
(with one electron jump and $\Delta l\pm 1$).
The calculated values can be changed significantly by a subtle change in degrees of mixing of the allowed configurations in the ASFs. 
On the other hand, agreement with \citet{Hartog} is much better for strong transitions.



\begin{deluxetable*}{llrrrrrr}
	\tabletypesize{\footnotesize}
	\setlength{\tabcolsep}{10pt}
	\tablecaption{Transition Energies $\Delta$$E$ (in cm$^{-1}$),
	Transition Wavelengths $\lambda$ (in \AA), 
	Line Strengths $S$ (in a.u.), 
	Weighted Oscillator Strengths $gf$, 
	and Transition Rates $A$ (in s$^{-1}$) for
E1 Transitions of the Pm~II ion. \label{Pm_II_transitions}}
	\tablehead{ \colhead{Lower state} 
	& \colhead{Upper state}  
	& \colhead{$\Delta E$ (cm$^{-1}$)} & \colhead{$\lambda$ (\AA)} & \colhead{$S$} & \colhead{$gf$} & \colhead{$A$ (s$^{-1}$)} & \colhead{$dT$}}
	\startdata
$4f^5 (^6_0H)~5d~^7F_{0}$           ~  & $4f^4 (^5_1I)~5d^2 (^3_2F)~^7G_{1}$ ~  &     7085 &      14112 & 1.596D-02 & 3.435D-04 & 3.835D+03 &   0.986\\
$4f^5 (^6_0H)~5d~^7F_{0}$           ~  & $4f^4 (^5_1I)~5d^2 (^3_2F)~^7F_{1}$ ~  &    10754 &       9298 & 4.139D-01 & 1.352D-02 & 3.477D+05 &   0.612\\
$4f^5 (^6_0H)~5d~^7F_{0}$           ~  & $4f^4 (^5_1I)~5d^2 (^3_2F)~^5F_{1}$ ~  &    12200 &       8196 & 1.145D-01 & 4.245D-03 & 1.405D+05 &   0.670\\
$4f^5 (^6_0H)~5d~^7F_{0}$           ~  & $4f^4 (^5_1I)~5d^2 (^1_2G)~^5D_{1}$ ~  &    18581 &       5381 & 4.990D-10 & 2.816D-11 & 2.162D-03 &   1.000\\
$4f^5 (^6_0H)~5d~^7F_{0}$           ~  & $4f^4 (^5_0S)~5d^2 (^3_2F)~^7F_{1}$ ~  &    19182 &       5213 & 6.356D-03 & 3.703D-04 & 3.030D+04 &   0.962\\
$4f^5 (^6_0H)~5d~^7F_{0}$           ~  & $4f^4 (^5_1I)~5d^2 (^1_2G)~^5F_{1}$ ~  &    20262 &       4935 & 5.274D-05 & 3.246D-06 & 2.963D+02 &   0.738\\
$4f^5 (^6_0H)~5d~^7F_{0}$           ~  & $4f^4 (^5_1F)~5d^2 (^3_2F)~^7G_{1}$ ~  &    20827 &       4801 & 2.122D-03 & 1.342D-04 & 1.295D+04 &   0.943\\
$4f^5 (^6_0H)~5d~^7F_{0}$           ~  & $4f^4 (^5_1F)~5d^2 (^3_2F)~^7D_{1}$ ~  &    22329 &       4478 & 2.629D-02 & 1.783D-03 & 1.977D+05 &   0.619\\
$4f^5 (^6_0H)~5d~^7F_{0}$           ~  & $4f^4 (^5_1F)~5d^2 (^3_2F)~^7F_{1}$ ~  &    22721 &       4401 & 8.370D-02 & 5.777D-03 & 6.632D+05 &   0.426\\
$4f^5 (^6_0H)~5d~^7F_{0}$           ~  & $4f^4 (^5_1F)~5d^2 (^3_2F)~^5F_{1}$ ~  &    24509 &       4080 & 1.207D-03 & 8.991D-05 & 1.200D+04 &   0.521\\
$4f^5 (^6_0H)~5d~^7F_{0}$           ~  & $4f^4 (^5_1F)~5d^2 (^3_2P)~^7G_{1}$ ~  &    25471 &       3926 & 3.671D-02 & 2.840D-03 & 4.097D+05 &   0.240\\
$4f^5 (^6_0H)~5d~^7F_{0}$           ~  & $4f^4 (^5_0S)~5d^2 (^3_2F)~^5F_{1}$ ~  &    26213 &       3814 & 6.045D-04 & 4.814D-05 & 7.354D+03 &   0.736\\
$4f^5 (^6_0H)~5d~^7F_{0}$           ~  & $4f^4 (^5_1F)~5d^2 (^1_2D)~^5P_{1}$ ~  &    26337 &       3796 & 7.825D-06 & 6.260D-07 & 9.656D+01 &   0.927\\
$4f^5 (^6_0H)~5d~^7F_{0}$           ~  & $4f^4 (^5_1F)~5d^2 (^1_2D)~^5P_{1}$ ~  &    26949 &       3710 & 7.221D-06 & 5.911D-07 & 9.545D+01 &   0.994\\
$4f^5 (^6_0H)~5d~^7F_{0}$           ~  & $4f^4 (^5_1F)~5d^2 (^1_2D)~^5D_{1}$ ~  &    27588 &       3624 & 5.019D-04 & 4.206D-05 & 7.118D+03 &   0.864\\
$4f^5 (^6_0H)~5d~^7F_{0}$           ~  & $4f^4 (^5_1F)~5d^2 (^3_2P)~^7D_{1}$ ~  &    28134 &       3554 & 1.602D-02 & 1.369D-03 & 2.409D+05 &   0.259\\
$4f^5 (^6_0H)~5d~^7F_{0}$           ~  & $4f^4 (^5_1F)~5d^2 (^3_2F)~^5D_{1}$ ~  &    28353 &       3526 & 8.565D-03 & 7.376D-04 & 1.318D+05 &   0.111\\
$4f^5 (^6_0H)~5d~^7F_{0}$           ~  & $4f^4 (^5_1F)~5d^2 (^3_2P)~^7F_{1}$ ~  &    28610 &       3495 & 2.125D-03 & 1.846D-04 & 3.361D+04 &   0.786\\
$4f^5 (^6_0H)~5d~^7F_{0}$           ~  & $4f^4 (^5_1F)~5d^2 (^3_2P)~^7F_{1}$ ~  &    29032 &       3444 & 1.239D-02 & 1.093D-03 & 2.048D+05 &   0.553\\
$4f^5 (^6_0H)~5d~^7F_{0}$           ~  & $4f^4 (^5_1F)~5d^2 (^3_2F)~^5P_{1}$ ~  &    29210 &       3423 & 2.067D-02 & 1.834D-03 & 3.480D+05 &   0.216\\
				\enddata
\tablecomments{ Table~\ref{Pm_II_transitions} is published in its entirety in the machine-readable format.
All transition data are in length form. 
 Part of the values are shown here for guidance regarding its form and content.}
\end{deluxetable*}

\begin{deluxetable*}{llrrrrrr}
	\tabletypesize{\footnotesize}
	\setlength{\tabcolsep}{10pt}
	\tablecaption{Transition Energies $\Delta$$E$ (in cm$^{-1}$),
	Transition Wavelengths $\lambda$ (in \AA), 
	Line Strengths $S$ (in a.u.), 
	Weighted Oscillator Strengths $gf$, 
	and Transition Rates $A$ (in s$^{-1}$) for
E1 Transitions of the Sm~II ion. \label{Sm_II_transitions}}
	\tablehead{ \colhead{Lower state} 
	& \colhead{Upper state}  
	& \colhead{$\Delta E$ (cm$^{-1}$)} & \colhead{$\lambda$ (\AA)} & \colhead{$S$} & \colhead{$gf$} & \colhead{$A$ (s$^{-1}$)} & \colhead{$dT$}}
	\startdata
$4f^6 (^7_0F)~5d~^8H_{11/2}$          ~  & $4f^7 (^6_0I)~^6I_{11/2}$              ~  &    34628 &       2887 & 2.376D-02 & 2.499D-03 & 1.666D+05 &   0.226\\
$4f^6 (^7_0F)~5d~^8H_{11/2}$          ~  & $4f^7 (^6_0F)~^6F_{11/2}$              ~  &    44272 &       2258 & 6.084D-03 & 8.182D-04 & 8.914D+04 &   0.278\\
$4f^6 (^7_0F)~5d~^8H_{11/2}$          ~  & $4f^7 (^6_0F)~^6F_{11/2}$              ~  &    47585 &       2101 & 6.743D-06 & 9.747D-07 & 1.226D+02 &   0.881\\
$4f^6 (^7_0F)~5d~^8H_{11/2}$          ~  & $4f^7 (^4_2H)~^4H_{11/2}$              ~  &    51906 &       1926 & 5.479D-06 & 8.638D-07 & 1.293D+02 &   0.373\\
$4f^6 (^7_0F)~5d~^8H_{11/2}$          ~  & $4f^7 (^4_2H)~^4H_{11/2}$              ~  &    52021 &       1922 & 1.221D-04 & 1.929D-05 & 2.902D+03 &   0.475\\
$4f^6 (^7_0F)~5d~^8H_{11/2}$          ~  & $4f^7 (^4_2H)~^4H_{11/2}$              ~  &    52375 &       1909 & 5.185D-05 & 8.249D-06 & 1.257D+03 &   0.050\\
$4f^6 (^7_0F)~5d~^8H_{11/2}$          ~  & $4f^7 (^4_1K)~^4K_{11/2}$              ~  &    54211 &       1844 & 1.917D-05 & 3.157D-06 & 5.158D+02 &   0.994\\
$4f^6 (^7_0F)~5d~^8H_{11/2}$          ~  & $4f^7 (^4_4H)~^4H_{11/2}$              ~  &    55780 &       1792 & 8.194D-04 & 1.388D-04 & 2.401D+04 &   0.121\\
$4f^6 (^7_0F)~5d~^8H_{11/2}$          ~  & $4f^7 (^4_4H)~^4H_{11/2}$              ~  &    56367 &       1774 & 1.286D-04 & 2.203D-05 & 3.891D+03 &   0.228\\
$4f^6 (^7_0F)~5d~^8H_{11/2}$          ~  & $4f^7 (^4_7G)~^4G_{11/2}$              ~  &    56877 &       1758 & 1.260D-04 & 2.177D-05 & 3.915D+03 &   0.243\\
$4f^6 (^7_0F)~5d~^8H_{11/2}$          ~  & $4f^7 (^4_6G)~^4G_{11/2}$              ~  &    58572 &       1707 & 8.167D-07 & 1.453D-07 & 2.771D+01 &   0.697\\
$4f^6 (^7_0F)~5d~^8H_{11/2}$          ~  & $4f^7 (^4_6G)~^4G_{11/2}$              ~  &    58814 &       1700 & 7.015D-05 & 1.253D-05 & 2.409D+03 &   0.470\\
$4f^6 (^7_0F)~5d~^8H_{11/2}$          ~  & $4f^7 (^4_4I)~^4I_{11/2}$              ~  &    59094 &       1692 & 2.094D-05 & 3.759D-06 & 7.297D+02 &   0.950\\
$4f^6 (^7_0F)~5d~^8H_{11/2}$          ~  & $4f^7 (^4_2I)~^4I_{11/2}$              ~  &    61660 &       1621 & 9.523D-05 & 1.783D-05 & 3.769D+03 &   0.167\\
$4f^6 (^7_0F)~5d~^8H_{11/2}$          ~  & $4f^7 (^4_3K)~^4K_{11/2}$              ~  &    62829 &       1591 & 5.517D-04 & 1.052D-04 & 2.310D+04 &   0.156\\
$4f^6 (^7_0F)~5d~^8D_{11/2}$          ~  & $4f^7 (^6_0I)~^6I_{11/2}$              ~  &    31992 &       3125 & 2.834D-03 & 2.754D-04 & 1.567D+04 &   0.545\\
$4f^6 (^7_0F)~5d~^8D_{11/2}$          ~  & $4f^7 (^6_0F)~^6F_{11/2}$              ~  &    41636 &       2401 & 3.545D-02 & 4.483D-03 & 4.320D+05 &   0.939\\
$4f^6 (^7_0F)~5d~^8D_{11/2}$          ~  & $4f^7 (^6_0F)~^6F_{11/2}$              ~  &    44949 &       2224 & 6.156D-02 & 8.405D-03 & 9.440D+05 &   0.921\\
$4f^6 (^7_0F)~5d~^8D_{11/2}$          ~  & $4f^7 (^4_2H)~^4H_{11/2}$              ~  &    49270 &       2029 & 1.859D-04 & 2.782D-05 & 3.755D+03 &   0.717\\
$4f^6 (^7_0F)~5d~^8D_{11/2}$          ~  & $4f^7 (^4_2H)~^4H_{11/2}$              ~  &    49385 &       2024 & 2.738D-03 & 4.107D-04 & 5.568D+04 &   0.862\\
				\enddata
\tablecomments{ Table~\ref{Sm_II_transitions} is published in its entirety in the machine-readable format.
All transition data are in length form. 
 Part of the values are shown here for guidance regarding its form and content.}
\end{deluxetable*}

\begin{deluxetable*}{llrrrrrr}
	\tabletypesize{\footnotesize}
	\setlength{\tabcolsep}{10pt}
	\tablecaption{Transition Energies $\Delta$$E$ (in cm$^{-1}$),
	Transition Wavelengths $\lambda$ (in \AA), 
	Line Strengths $S$ (in a.u.), 
	Weighted Oscillator Strengths $gf$, 
	and Transition Rates $A$ (in s$^{-1}$) for
E1 Transitions of the Eu~II ion. \label{Eu_II_transitions}}
	\tablehead{ \colhead{Lower state} 
	& \colhead{Upper state}  
	& \colhead{$\Delta E$ (cm$^{-1}$)} & \colhead{$\lambda$ (\AA)} & \colhead{$S$} & \colhead{$gf$} & \colhead{$A$ (s$^{-1}$)} & \colhead{$dT$}}
	\startdata
$4f^7 (^6_0P)\,5d~^7F_{0}$          ~  & $4f^8 (^7_0F)~^7F_{1}$              ~  &     9040 &      11061 & 1.303D-03 & 3.579D-05 & 6.504D+02 &   0.840\\
$4f^7 (^6_0P)\,5d~^7F_{0}$          ~  & $4f^8 (^5_3D)~^5D_{1}$              ~  &    30106 &       3321 & 2.056D-02 & 1.880D-03 & 3.789D+05 &   0.794\\
$4f^7 (^6_0P)\,5d~^7F_{0}$          ~  & $4f^8 (^5_2F)~^5F_{1}$              ~  &    36281 &       2756 & 1.696D-03 & 1.869D-04 & 5.472D+04 &   0.992\\
$4f^7 (^6_0P)\,5d~^7F_{0}$          ~  & $4f^8 (^5_2F)~^5F_{1}$              ~  &    36869 &       2712 & 1.379D-03 & 1.544D-04 & 4.668D+04 &   0.765\\
$4f^7 (^6_0P)\,5d~^7F_{0}$          ~  & $4f^8 (^5_2F)~^5F_{1}$              ~  &    37397 &       2673 & 3.813D-03 & 4.332D-04 & 1.347D+05 &   0.714\\
$4f^7 (^6_0P)\,5d~^5D_{0}$          ~  & $4f^8 (^7_0F)~^7F_{1}$              ~  &     2361 &      42339 & 1.176D-03 & 8.438D-06 & 1.046D+01 &   0.962\\
$4f^7 (^6_0P)\,5d~^5D_{0}$          ~  & $4f^8 (^5_3D)~^5D_{1}$              ~  &    23427 &       4268 & 1.070D-03 & 7.619D-05 & 9.298D+03 &   0.231\\
$4f^7 (^6_0P)\,5d~^5D_{0}$          ~  & $4f^8 (^5_2F)~^5F_{1}$              ~  &    29602 &       3378 & 2.478D-02 & 2.228D-03 & 4.342D+05 &   0.363\\
$4f^7 (^6_0P)\,5d~^5D_{0}$          ~  & $4f^8 (^5_2F)~^5F_{1}$              ~  &    30191 &       3312 & 5.379D-02 & 4.933D-03 & 9.998D+05 &   0.579\\
$4f^7 (^6_0P)\,5d~^5D_{0}$          ~  & $4f^8 (^5_2F)~^5F_{1}$              ~  &    30718 &       3255 & 5.116D-02 & 4.773D-03 & 1.001D+06 &   0.980\\
$4f^8 (^7_0F)~^7F_{1}$              ~  & $4f^7 (^6_0D)\,5d~^7F_{0}$          ~  &     1158 &      86316 & 2.027D-02 & 7.135D-05 & 6.388D+01 &   0.982\\
$4f^7 (^6_0D)\,5d~^7F_{0}$          ~  & $4f^8 (^5_3D)~^5D_{1}$              ~  &    19907 &       5023 & 6.089D-03 & 3.681D-04 & 3.244D+04 &   0.891\\
$4f^7 (^6_0D)\,5d~^7F_{0}$          ~  & $4f^8 (^5_2F)~^5F_{1}$              ~  &    26082 &       3833 & 7.236D-01 & 5.733D-02 & 8.671D+06 &   0.880\\
$4f^7 (^6_0D)\,5d~^7F_{0}$          ~  & $4f^8 (^5_2F)~^5F_{1}$              ~  &    26670 &       3749 & 1.442D-02 & 1.168D-03 & 1.848D+05 &   0.970\\
$4f^7 (^6_0D)\,5d~^7F_{0}$          ~  & $4f^8 (^5_2F)~^5F_{1}$              ~  &    27198 &       3676 & 1.119D-01 & 9.245D-03 & 1.520D+06 &   0.847\\
$4f^8 (^7_0F)~^7F_{1}$              ~  & $4f^7 (^6_0D)\,5d~^5D_{0}$          ~  &     5664 &      17653 & 9.848D-04 & 1.694D-05 & 3.627D+02 &   0.914\\
$4f^7 (^6_0D)\,5d~^5D_{0}$          ~  & $4f^8 (^5_3D)~^5D_{1}$              ~  &    15400 &       6493 & 2.752D-02 & 1.287D-03 & 6.791D+04 &   0.063\\
$4f^7 (^6_0D)\,5d~^5D_{0}$          ~  & $4f^8 (^5_2F)~^5F_{1}$              ~  &    21576 &       4634 & 6.207D-02 & 4.068D-03 & 4.211D+05 &   0.731\\
$4f^7 (^6_0D)\,5d~^5D_{0}$          ~  & $4f^8 (^5_2F)~^5F_{1}$              ~  &    22164 &       4511 & 3.348D-02 & 2.254D-03 & 2.462D+05 &   0.022\\
$4f^7 (^6_0D)\,5d~^5D_{0}$          ~  & $4f^8 (^5_2F)~^5F_{1}$              ~  &    22692 &       4406 & 4.103D-01 & 2.828D-02 & 3.238D+06 &   0.942\\
				\enddata
\tablecomments{ Table~\ref{Eu_II_transitions} is published in its entirety in the machine-readable format.
All transition data are in length form. 
 Part of the values are shown here for guidance regarding its form and content.}
\end{deluxetable*}

\begin{deluxetable*}{llrrrrrr}
	\tabletypesize{\footnotesize}
	\setlength{\tabcolsep}{10pt}
	\tablecaption{Transition Energies $\Delta$$E$ (in cm$^{-1}$),
	Transition Wavelengths $\lambda$ (in \AA), 
	Line Strengths $S$ (in a.u.), 
	Weighted Oscillator Strengths $gf$, 
	and Transition Rates $A$ (in s$^{-1}$) for
E1 Transitions of the Gd~II ion. \label{Gd_II_transitions}}
	\tablehead{ \colhead{Lower state} 
	& \colhead{Upper state}  
	& \colhead{$\Delta E$ (cm$^{-1}$)} & \colhead{$\lambda$ (\AA)} & \colhead{$S$} & \colhead{$gf$} & \colhead{$A$ (s$^{-1}$)} & \colhead{$dT$}}
	\startdata
$4f^7 (^8_0S)\,5d~^9D\,6s~^1D_{11/2}$  ~  & $4f^8 (^7_0F)\,6s~^8F_{11/2}$          ~  &     7731 &      12933 & 1.622D-06 & 3.809D-08 & 1.265D-01 &   0.994\\
$4f^7 (^8_0S)\,5d~^9D\,6s~^1D_{11/2}$  ~  & $4f^8 (^7_0F)\,6s~^6F_{11/2}$          ~  &     8671 &      11531 & 3.131D-07 & 8.247D-09 & 3.447D-02 &   0.998\\
$4f^7 (^8_0S)\,5d~^9D\,6s~^1D_{11/2}$  ~  & $4f^8 (^5_3G)\,6s~^6G_{11/2}$          ~  &    33824 &       2956 & 1.511D-06 & 1.552D-07 & 9.875D+00 &   0.031\\
$4f^7 (^8_0S)\,5d~^9D\,6s~^1D_{11/2}$  ~  & $4f^8 (^5_3G)\,6s~^4G_{11/2}$          ~  &    34708 &       2881 & 1.063D-06 & 1.121D-07 & 7.510D+00 &   0.161\\
$4f^7 (^8_0S)\,5d~^9D\,6s~^1D_{11/2}$  ~  & $4f^8 (^5_0L)\,6s~^6L_{11/2}$          ~  &    35251 &       2836 & 9.906D-09 & 1.060D-09 & 7.327D-02 &   0.508\\
$4f^7 (^8_0S)\,5d~^9D\,6s~^1D_{11/2}$  ~  & $4f^8 (^5_1H)\,6s~^6H_{11/2}$          ~  &    39556 &       2528 & 1.645D-08 & 1.976D-09 & 1.719D-01 &   0.710\\
$4f^7 (^8_0S)\,5d~^9D\,6s~^1D_{11/2}$  ~  & $4f^8 (^5_1H)\,6s~^4H_{11/2}$          ~  &    40715 &       2456 & 6.683D-09 & 8.265D-10 & 7.616D-02 &   0.669\\
$4f^7 (^8_0S)\,5d~^9D\,6s~^1D_{11/2}$  ~  & $4f^8 (^5_2F)\,6s~^6F_{11/2}$          ~  &    41839 &       2390 & 1.807D-07 & 2.297D-08 & 2.235D+00 &   0.803\\
$4f^7 (^8_0S)\,5d~^9D\,6s~^1D_{11/2}$  ~  & $4f^8 (^5_2I)\,6s~^6I_{11/2}$          ~  &    44260 &       2259 & 3.334D-11 & 4.482D-12 & 4.881D-04 &   0.915\\
$4f^7 (^8_0S)\,5d~^9D\,6s~^1D_{11/2}$  ~  & $4f^8 (^5_2I)\,6s~^4I_{11/2}$          ~  &    45531 &       2196 & 1.077D-09 & 1.489D-10 & 1.716D-02 &   0.275\\
$4f^7 (^8_0S)\,5d~^9D\,6s~^1D_{11/2}$  ~  & $4f^8 (^5_0K)\,6s~^6K_{11/2}$          ~  &    47419 &       2108 & 8.083D-08 & 1.164D-08 & 1.455D+00 &   0.426\\
$4f^7 (^8_0S)\,5d~^9D\,6s~^1D_{11/2}$  ~  & $4f^8 (^5_0K)\,6s~^4K_{11/2}$          ~  &    49135 &       2035 & 4.864D-08 & 7.259D-09 & 9.743D-01 &   0.161\\
$4f^7 (^8_0S)\,5d~^9D\,6s~^1D_{11/2}$  ~  & $4f^8 (^5_2G)\,6s~^6G_{11/2}$          ~  &    49860 &       2005 & 1.753D-07 & 2.655D-08 & 3.669D+00 &   0.641\\
$4f^7 (^8_0S)\,5d~^9D\,6s~^1D_{11/2}$  ~  & $4f^8 (^5_0K)\,6s~^4K_{11/2}$          ~  &    50373 &       1985 & 7.274D-08 & 1.113D-08 & 1.569D+00 &   0.924\\
$4f^7 (^8_0S)\,5d~^9D\,6s~^1D_{11/2}$  ~  & $4f^8 (^5_2G)\,6s~^4G_{11/2}$          ~  &    50830 &       1967 & 5.492D-08 & 8.481D-09 & 1.218D+00 &   0.933\\
$4f^7 (^8_0S)\,5d~^9D\,6s~^1D_{11/2}$  ~  & $4f^8 (^5_0K)\,6s~^4K_{11/2}$          ~  &    52462 &       1906 & 1.719D-06 & 2.739D-07 & 4.191D+01 &   0.711\\
$4f^7 (^8_0S)\,5d~^9D\,6s~^1D_{11/2}$  ~  & $4f^8 (^3_5K)\,6s~^4K_{11/2}$          ~  &    54181 &       1845 & 1.796D-10 & 2.956D-11 & 4.824D-03 &   0.992\\
$4f^7 (^8_0S)\,5d~^9D\,6s~^1D_{11/2}$  ~  & $4f^8 (^3_3K)\,6s~^4K_{11/2}$          ~  &    55996 &       1785 & 5.269D-12 & 8.963D-13 & 1.562D-04 &   0.982\\
$4f^7 (^8_0S)\,5d~^9D\,6s~^1D_{11/2}$  ~  & $4f^8 (^3_6G)\,6s~^4G_{11/2}$          ~  &    56703 &       1763 & 6.780D-08 & 1.167D-08 & 2.087D+00 &   0.920\\
$4f^7 (^8_0S)\,5d~^9D\,6s~^1D_{11/2}$  ~  & $4f^8 (^5_2H)\,6s~^6H_{11/2}$          ~  &    57987 &       1724 & 9.638D-09 & 1.697D-09 & 3.173D-01 &   0.574\\
				\enddata
\tablecomments{ Table~\ref{Gd_II_transitions} is published in its entirety in the machine-readable format.
 All transition data are in length form. 
 Part of the values are shown here for guidance regarding its form and content.}
\end{deluxetable*}

\section{Summary}
\label{sec:summary}

We presented {\it ab-initio} atomic calculations of energy levels and E1 transitions from Pr~II to Gd~II ions
based on the strategy developed for the calculations of Nd II \citep{Nd_jonai}.
In total 2~145, 9~774, 8~393, 2~473, and 4~397 levels are presented for Pr~II, Pm~II, Sm~II, Eu~II, and Gd~II, respectively.
Some of the Rydberg states are also included to the computations for Eu~II. 
By comparing with the NIST database and the results by other authors,
we confirmed that our calculations achieve good accuracy. 
For the energy levels, the averaged accuracy compared with the NIST data
are 8\%, 12\%, 6\%, 8\%, and 7\% for Pr~II, Pm~II, Sm~II, Eu~II, and Gd~II, respectively. 
These are the highest accuracies achieved for this kind of complete atomic calculations
needed for opacity calculations. 
There is no clear dependence of accuracy on atomic number $Z$. 
This means that data of lanthanide set can be computed in similar way to the izoelectronic sequence. 
By using the results of atomic structure calculations, 
E1 transitions between levels are computed. 
We provide data for 411~314, 7~104~005, 4~720~626, 467~724, 
and 1 383 694‬ transitions for Pr~II, Pm~II, Sm~II, Eu~II, and Gd~II, respectively. 
Transition probabilities are compared with NIST database as well as the results of other works. 
Our computed E1 type transition probabilities are in good agreement with 
presented in NIST database experimental values, especially in the area of strong transitions. 
\acknowledgments

This research was funded by a grant (No. S-LJB-18-1) from 
the Research Council of Lithuania. This research was also 
supported by the JSPS Bilateral Joint Research Project. D.K. is grateful
for the support by the NINS program of Promoting Research by
Networking among Institutions (grant No. 01411702). 
The computations presented in this paper were performed at the
High Performance Computing Center "HPC Sauletekis" of the
Faculty of Physics at Vilnius University.

\bibliography{reference}



\end{document}